\documentclass[12pt]{article}
\pdfoutput=1

\usepackage{amsmath,amssymb,amsfonts,graphicx,amsfonts}
\usepackage{textcomp,mathcomp}
\usepackage{color}
\usepackage{ascmac}
\usepackage{physics}
\usepackage{dcolumn}
\usepackage{bm}
\usepackage{caption}
\usepackage{subcaption}
\usepackage[utf8]{inputenc}
\usepackage{booktabs}
\usepackage{xcolor}
\usepackage{url}
\usepackage{pifont}

\allowdisplaybreaks[4]

\begin{document}

\begin{center}
    {\Large Gauge Symmetry in Quantum Simulation}\\
    \vspace{0.4cm}
{\Large --- Redundancy Is All You Need --- }\\

\end{center}
\vspace{0.1cm}
\vspace{0.1cm}
\begin{center}
Masanori Hanada$^{\rm a}$, Shunji Matsuura$^{\rm b}$, Andreas Sch\"{a}fer$^{\rm c}$, Jinzhao Sun$^{\rm d}$
\end{center}
\vspace{0.3cm}

\begin{center}
{\small
$^{\rm a}$School of Mathematical Sciences, Queen Mary University of London\\
Mile End Road, London, E1 4NS, United Kingdom\\
$^{\rm a}$qBraid Co., Harper Court 5235, Chicago, IL 60615, United States\\
$^{\rm b}$
Interdisciplinary Theoretical \& Mathematical Science Program (iTHEMS)\\ RIKEN, Wako, Saitama 351-0198, Japan\\
$^{\rm b}$
Department of Electrical and Computer Engineering\\
University of British Columbia, Vancouver, BC V6T 1Z4, Canada\\
$^{\rm c}$ Institute of Theoretical Physics, University of Regensburg\\
Universit\"{a}tsstrasse 31, D-93053 Regensburg, Germany\\
$^{\rm c}$
Department of Physics, National Taiwan University, Taipei, Taiwan 106\\
$^{\rm d}$School of Physical and Chemical Sciences, Queen Mary University of London\\
Mile End Road, London, E1 4NS, United Kingdom
}
\end{center}

\vspace{1.5cm}

\newpage

\begin{center}
  {\bf Abstract}
\end{center}

Quantum simulation of non-Abelian gauge theories requires careful handling of gauge redundancy. We present universal principles for treating gauge symmetry that apply to any quantum simulation approach. Our main result is that enforcing gauge-singlet constraints is conceptually and practically unnecessary for extracting physical observables. Physical states need not be represented by gauge singlets; singlet and non-singlet representations are both valid, with distinct practical trade-offs that we elucidate using analogies to BRST quantization, and the non-singlet approach is likely more efficient toward the continuum limit. As in classical electrodynamics and the path-integral formulation, gauge redundancy simplifies things. We demonstrate these principles within a complete quantum simulation framework based on the orbifold lattice, which enables explicit and efficient circuit constructions relevant to real-world QCD. We show how non-singlet approaches yield gauge-invariant observables through wave packets and string excitations, and that this approach is universal and efficient. For the limited situations where a singlet representation is genuinely required, we introduce a Haar-averaging projection implemented via linear combinations of unitaries and analyze its cost and truncation errors; its post-selection cost makes it a fallback rather than the default. We also introduce a penalty term in the Hamiltonian that eliminates non-singlet states from the low-energy spectrum. Working in temporal gauge, we map lattice Yang-Mills dynamics to Pauli-string Hamiltonians suitable for Trotterization. Classical simulations of small systems validate convergence criteria, quantify truncation and Trotter errors, and yield concrete resource estimates and scalable circuit recipes for SU$(N)$ gauge theories.

\newpage
\tableofcontents
\newpage
\section{Introduction}\label{sec:introduction}
Gauge symmetry is a fundamental feature of many physical theories, including electromagnetism, Yang-Mills theory~\cite{Yang:1954ek,Peskin:1995ev,Weinberg:1996kr}, and Quantum Chromodynamics (QCD)~\cite{Han:1965pf,Fritzsch:1973pi,Gross:1973id,Politzer:1973fx}, and more broadly, the standard model of particle physics~\cite{Glashow:1961tr,Weinberg:1967tq,Salam:1968rm,tHooft:1971qjg,tHooft:1972tcz}. They also lie at the core of the most concrete approaches to quantum gravity through dualities between supersymmetric Yang-Mills theories and string/M-theory~\cite{Maldacena:1997re}. Less widely appreciated is that permutation symmetry in systems of identical bosons~\cite{Bose:1924,Einstein:1924} functions as a gauge symmetry with physical consequences --- the Bose-Einstein condensation~\cite{Einstein:1925,Feynman:1953zz}. Any faithful quantum simulation of such systems must confront the role of gauge symmetry at a fundamental level. While it is often stated that physical states must be gauge-invariant, this statement obscures an important distinction between physical equivalence and representation: gauge symmetry --- or rather, gauge redundancy --- identifies physically equivalent configurations, but does not uniquely prescribe how those configurations should be represented in a quantum simulation. As a result, different representations of physical states --- gauge singlet and non-singlet alike --- can be equally valid, provided they encode the same equivalence class and reproduce gauge-invariant observables. 

The origin of this widespread misconception --- ``physical states must be gauge-invariant'' --- is not clear. The claim that physical states must be gauge-invariant is often attributed to Kogut and Susskind~\cite{Kogut:1974ag}, but their actual statement was \textit{``The physical states are drawn from the space of gauge-invariant states"}, which describes a particular choice of basis convenient for their Hamiltonian lattice formulation. The physical content resides in gauge-equivalence classes; gauge-invariant representatives are one valid choice among many.
In the past, non-singlet descriptions played crucial roles in the understanding of confinement in the large-$N$ gauge theories~\cite{Hanada:2020uvt} and emergent geometry in string theory~\cite{Hanada:2021ipb,Gautam:2024zsj}. See also refs.~\cite{Gautam:2022akq,Fliss:2025kzi} that emphasized the difference between \textit{physical} and \textit{gauge-invariant}.

Recent work established quantum circuit constructions for Yang-Mills theory using extended Hilbert space and non-compact variables~\cite{Buser:2020cvn,Bergner:2024qjl,Halimeh:2024bth,Bergner:2025zkj}, in such a way that the generalization to QCD is straightforward~\cite{Halimeh:2025ivn}. However, crucial questions regarding the treatment of gauge redundancy remained: how to implement gauge-singlet constraints (when necessary), and when to use singlet or non-singlet formulations. Furthermore, it was not very clear how to validate convergence criteria concerning the truncation of the infinite-dimensional Hilbert space associated with bosons. We address these by: (1) establishing universal principles showing that the gauge-singlet constraint is unnecessary for extracting physical observables, which justifies cheap non-singlet simulations; (2) providing, for the limited situations where a singlet representation is genuinely required, a singlet-projection protocol for SU($N$) gauge theory via Haar averaging realized through LCU --- whose post-selection cost is the very reason to avoid it whenever possible; (3) establishing an efficient alternative that removes non-singlets from the low-energy spectrum via a penalty term; and (4) providing systematic numerical validation and complete resource estimates. Our framework completes the quantum simulation toolkit for non-Abelian gauge theories, providing both conceptual clarity and practical implementation protocols with validated scalings.

Yang-Mills theory is a natural generalization of electromagnetism, which is a gauge theory with the U(1) gauge group, to the SU($N$) gauge group. As in electromagnetism, gauge symmetry plays a crucial role in the quantization of the relativistic theory without negative-norm modes in the Hilbert space. To see this, imagine creation operators $\hat{a}^\dagger_\mu$ corresponding to the U(1) gauge field $A_\mu$, where $\mu=0,1,2,3=t,x,y,z$, act on the Lorentz-invariant vacuum $\ket{\rm VAC; inv}$. Then, the natural inner product that is consistent with the Lorentz symmetry is $\bra{\rm VAC; inv}\hat{a}_\mu\hat{a}^\dagger_\nu\ket{\rm VAC; inv}=\eta_{\mu\nu}$, where $\eta_{\mu\nu}=(-1)^{\delta_{\mu,0}}\delta_{\mu\nu}$. This seems to be a problem because $\hat{a}^\dagger_\mu\ket{\rm VAC; inv}$ has a negative norm (specifically, the norm is $-1$ for $\mu=0$), which could conflict with the probability interpretation of quantum mechanics. Gauge symmetry cures this problem because only the transverse modes are physical, and negative-norm modes decouple from physical processes. 

The temporal gauge ($A_t=0$ gauge; see Supplementary Material~\ref{sec:path-integral-and-Hamiltonian}) is convenient for quantum simulations, because the negative-norm modes are absent; if the Hilbert space contains negative-norm states, it is unclear how to map it to qubits. Some caution is required because this gauge condition does not remove gauge redundancy completely. A common option is to restrict the Hilbert space to be invariant under the residual gauge symmetry; we call such a Hilbert space the gauge-invariant Hilbert space, or the singlet Hilbert space, and denote it by $\mathcal{H}_{\rm inv}$.  This is a subspace of the extended Hilbert space $\mathcal{H}_{\rm ext}$ that contains both gauge singlets and non-singlets. This singlet-prescription is, however, not straightforward. If one tries to construct only $\mathcal{H}_{\rm inv}$ on quantum devices, it is difficult to build orthonormal basis, and the operators including the Hamiltonian become complicated. This leads to exponential growth in classical pre-processing cost, and possibly also in the circuit depth, with respect to the number of logical qubits, which destroys any potential quantum advantage~\cite{Hanada:2025yzx}. Therefore, we will use $\mathcal{H}_{\rm ext}$ for quantum simulations. Still, $\mathcal{H}_{\rm inv}$ is realized as a subspace of $\mathcal{H}_{\rm ext}$, and one has the option to use only states in $\mathcal{H}_{\rm inv}$. 

As established in refs.~\cite{Bergner:2024qjl,Halimeh:2024bth,Bergner:2025zkj,Halimeh:2025ivn}, the orbifold lattice framework provides explicit quantum circuit construction for SU($N$) gauge theories in $3+1$ dimensions with polynomial resource scaling, both on quantum devices (the number of gates and circuit depth) and on classical computers (pre-processing cost for designing and compiling quantum circuits). Alternative approaches, despite the significant efforts and success in the Abelian setups (see e.g., refs.~\cite{Zohar:2015hwa,Dalmonte:2016alw,Banuls:2019bmf}), either lack explicit circuits at arbitrary truncation levels or require large classical compilation cost~\cite{Hanada:2025yzx}. Building on this foundation, we now address the issues associated with gauge symmetry. This paper makes two interconnected contributions:\\

\noindent
\textbf{Universal conceptual framework:} We clarify how gauge symmetry should be treated in quantum simulations, resolving some misconceptions and providing principles applicable to any approach. (Sec.~\ref{sec:H_ext-and-H_inv})

\noindent
\textbf{Complete practical implementation:}
We provide explicit protocols, circuits, and resource estimates within the orbifold lattice framework, completing recent advances~\cite{Bergner:2024qjl,Halimeh:2024bth,Bergner:2025zkj,Halimeh:2025ivn} by addressing the treatment of gauge symmetry. To the best of our knowledge, using non-compact variables, as in the orbifold lattice formulation, is the only approach so far, that provides explicit quantum circuits for Yang-Mills theory, which enable scalable simulations towards the continuum limit for any gauge group and dimension without exponential classical compilation costs; see ref.~\cite{Hanada:2025yzx} for the cost analyses for other approaches. (Sec.~\ref{sec:nonsinglet_simulations}, Sec.~\ref{sec:singlet_simulation}, Sec.~\ref{sec:numerical_test}, and Sec.~\ref{sec:outlook})\\

The protocols presented here may appear straightforward to quantum algorithm experts --- \textbf{this simplicity is precisely the achievement}. Through careful physics-based problem design (specifically, temporal gauge quantization, non-compact variables, and the orbifold lattice Hamiltonian), we reduced a seemingly intractable problem to one where textbook quantum algorithms suffice. The difficulty lies not in inventing new algorithms, but in recognizing which physics reformulation enables algorithmic tractability.

Table~\ref{table:then_and_now_comparison} summarizes the key capabilities added in this work compared to the prior orbifold lattice framework (dubbed the ``Universal framework''~\cite{Halimeh:2024bth})~\cite{Bergner:2024qjl,Halimeh:2024bth,Bergner:2025zkj,Halimeh:2025ivn} and generic gauge-invariant approaches.

\begin{center}
\begin{table*}[htbp]
\centering
\renewcommand{\arraystretch}{1.5}
\resizebox{\textwidth}{!}{
\begin{tabular}{|c||c|c|c|}
\hline
 & \textbf{Gauge-invariant approaches} &  \textbf{Orbifold Lattice (prior work)} &  \textbf{Orbifold Lattice (this work)} \\
\hline
\hline
Explicit circuits & \ding{55} &  \ding{51} &  \ding{51} \\
\hline
Singlet constraint required? & Yes (built in) & No & No (optional, local only) \\
\hline
Singlet projection &  N/A & \ding{55} & \ding{51} (expensive, avoid whenever possible)\\
\hline
Penalty for non-singlet&  N/A & \ding{55} & \ding{51} \\
($\mathrm{Tr}\hat{G}^2$ term)  & & & \\
\hline
Non-singlet states & \ding{55} & \ding{55} (not justified) & \ding{51} (justified) \\
\hline
Compilation cost & Exponential in $N^2QV$ & Polynomial in $N$, $Q$, and $V$ & Polynomial in $N$, $Q$, and $V$ \\
\hline
Gate count & Unknown & $O(N^4Q^4V)$ &  $O(N^4Q^4V)$\\
\hline
Circuit depth  & Unknown & $O(N^4Q^4)$ &  $O(N^4Q^4)$\\
\hline
Convergence validation & N/A & \ding{55} & \ding{51} \\
\hline
\end{tabular}
}
\caption{Comparison of generic gauge-singlet approaches (methods attempting to work directly in singlet Hilbert space) and the orbifold approach before and after this work. Here, we consider SU($N$) theory in $2+1$ or $3+1$ dimensions. 
$Q$ is the number of qubits per boson, and $V$ is the number of lattice points. Gate counting and circuit depth can change slightly depending on the details of the implementation.
``Convergence validation" refers to the validation of the convergence when the Hilbert space truncation is removed. (Ref.~\cite{Bergner:2025zkj} provided the convergence validation without Hilbert space truncation, using the Euclidean path integral.) 
Circuit depth and Gate Depth are for Hamiltonian time evolution. 
The singlet projection protocol (Sec.~\ref{sec:singlet_projection}) involves expensive post-selections. 
}\label{table:then_and_now_comparison}
\end{table*}
\end{center}
\subsection*{Outline of the paper}
In this work, we explain how quantum simulations can be conducted respecting gauge symmetry. 
Sec.~\ref{sec:H_ext-and-H_inv} constitutes the conceptual essence of this paper; the remaining sections serve to provide the technical implementation protocols and concrete validations.

In Sec.~\ref{sec:H_ext-and-H_inv}, we start with emphasizing a simple fact: using gauge singlets is only one of many ways to represent physical states. We can choose a convenient option depending on the device and the problem in our hands. We provide analogies with a well-known example --- the Becchi-Rouet-Stora-Tyutin (BRST) quantization~\cite{Kugo:1979gm,Peskin:1995ev,Weinberg:1996kr} --- that provides us with a better intuition for the meaning of ``physical states", and a few examples of non-singlet descriptions to illustrate the important points. 

Whether one uses singlets or non-singlets for quantum simulations, it is convenient if one can project arbitrary non-singlet states to the corresponding singlet states. In Sec.~\ref{sec:singlet_projection}, we provide such a procedure for SU($N$) gauge theory for any $N$ and any dimension. The projection naturally adopts a form of Linear Combinations of Unitaries (LCU)~\cite{Childs:2012gwh} or Quantum Singular Value Transformation (QSVT)~\cite{Yoder:2014sls,Gilyen:2018khw} and thus can be implemented on a quantum computer. To do so, the use of non-compact variables is a key ingredient. This projection makes simulations with singlet states technically straightforward. 
As another option, we show how to include a penalty term that removes non-singlet states from the low-energy spectrum, keeping the efficiency of the quantum circuits, in Sec.~\ref{sec:penalty_G^2}.
As simulations with non-singlet states, we consider two options in Sec.~\ref{sec:nonsinglet_simulations}: non-singlet excitations of a gauge-invariant state and singlet excitations of a wave packet.

Finally, Sec.~\ref{sec:numerical_test} provides crucial numerical validation of convergence criteria. Specifically, it presents systematic numerical experiments on simplified $\mathrm{S}^n$ models (bosons constrained to $n$-spheres) that capture the essential physics of the scalar mass terms in the orbifold lattice. These experiments on classical computers quantify how truncation error depends on the discretization scale $\delta_x$ and mass $m$, establishing that $\delta_x\lesssim 1/\sqrt{m}$ is required for accurate low-energy physics. As a bonus, this section clarifies how the state $\ket{\mathcal{G}}$ used for the singlet projection protocol introduced in Sec.~\ref{sec:singlet_simulation} can be constructed. 
These also demonstrate that Trotter step sizes are determined by physical energy scales (like $m$), not by the ultraviolet cutoff $\Lambda$, confirming that high-energy modes don't spoil the simulation efficiency. 

Appendices~\ref{sec:Kogut_Susskind} and \ref{sec:orbifold_lattice} provide parallel technical reviews of the two main Hamiltonian formulations used throughout the paper. Appendix~\ref{sec:Kogut_Susskind} covers the Kogut-Susskind formulation, which uses unitary link variables and serves as the standard lattice gauge theory Hamiltonian. It details the operators and their commutation relations, the structure of the Hamiltonian with its electric and magnetic terms, and crucially, how string states (Wilson loops and lines) behave under the Hamiltonian evolution, including their splitting and joining interactions. A technical subsection compares left-acting versus right-acting electric field operators. Appendix~\ref{sec:orbifold_lattice} provides analogous material for the orbifold lattice formulation, which uses complex (non-compact) link variables instead of unitary ones. This formulation is central to the paper's practical implementation because it allows straightforward truncation to finite-dimensional Hilbert spaces and natural expression in terms of Pauli strings, making it far more suitable for quantum circuit design than the Kogut-Susskind approach.

Supplementary Material offers technical details and review materials that make the paper self-contained. 
\section{Extended Hilbert space, gauge-invariant Hilbert space, and physical states}\label{sec:H_ext-and-H_inv}

\subsection{Hilbert space}\label{sec:Hilbert_space}
We start with discussing what are the `Hilbert space' and `physical states' in gauge theories, and how they should be treated in quantum simulations. These principles apply to any approach, not just the Kogut-Susskind Hamiltonian and the orbifold lattice Hamiltonian, which will be used as concrete examples below. (See Appendices~\ref{sec:Kogut_Susskind} and~\ref{sec:orbifold_lattice} for reviews of these Hamiltonians.)

The Kogut-Susskind Hamiltonian uses unitary link variables $U_{j,\vec{n}}$ as dynamical degrees of freedom, while the orbifold lattice Hamiltonian uses the complex link variables $Z_{j,\vec{n}}$. Here, $j$ and $\vec{n}$ labels spatial directions and lattice points, respectively. $U_{j,\vec{n}}$ and $Z_{j,\vec{n}}$ live on a link connecting points $\vec{n}$ and $\vec{n}+\hat{j}$, where $\hat{j}$ represents a unit vector along the $j$-direction and not an operator.

The extended Hilbert space corresponds to all possible field configurations that are compatible with the gauge condition $A_t=0$. It is defined by 
\begin{align}
\mathcal{H}_{\rm ext}
=
\otimes_{j,\vec{n}}{\cal H}_{j,\vec{n}}
=
\otimes_{j,\vec{n}}
\textrm{Span}\left\{
|U\rangle_{j,\vec{n}}
\ |\ U\in{\rm SU}(N)
\right\}\, , 
\end{align}
where
\begin{align}
\hat{U}_{j,\vec{n}}
|U\rangle_{j,\vec{n}}
=
U|U\rangle_{j,\vec{n}}
\qquad 
U\in{\rm SU}(N)\, ,  
\end{align}
for the Kogut-Susskind formulation and 
\begin{align}
\mathcal{H}_{\rm ext}
=
\otimes_{j,\vec{n}}{\cal H}_{j,\vec{n}}
=
\otimes_{j,\vec{n}}
\textrm{Span}\left\{
|Z\rangle_{j,\vec{n}}
\ |\ Z\in\mathbb{C}^{N\times N}
\right\}\, ,
\end{align}
where
\begin{align}
\hat{Z}_{j,\vec{n}}
|Z\rangle_{j,\vec{n}}
=
Z|Z\rangle_{j,\vec{n}}
\qquad 
Z\in\mathbb{C}^{N\times N}\, , 
\end{align}
for the orbifold lattice. 
More precisely, we consider only the Hilbert space of square-integrable wave functions. 

The residual gauge transformation (time-independent gauge transformation) by $\Omega\in\mathcal{G}=\prod_{\vec{n}}[\mathrm{SU}(N)]_{\vec{n}}$ is 
\begin{align}
\hat{\Omega}\ket{U}=|U^{(\Omega)}\rangle\, , 
\qquad
U_{j,\vec{n}}^{(\Omega)}\equiv \Omega_{\vec{n}}U_{j,\vec{n}}\Omega^{-1}_{\vec{n}+\hat{j}}
\end{align}
and 
\begin{align}
\hat{\Omega}\ket{Z}=|Z^{(\Omega)}\rangle\, , 
\qquad
Z_{j,\vec{n}}^{(\Omega)}\equiv \Omega_{\vec{n}}Z_{j,\vec{n}}\Omega^{-1}_{\vec{n}+\hat{j}}\, , 
\end{align}
respectively. States connected to each other by gauge transformation are physically equivalent. In this sense, `gauge symmetry' is gauge \textit{redundancy}. A large fraction of the redundancy has been already removed by taking the $A_t=0$ gauge. One way to remove the residual redundancy is to use the gauge-invariant Hilbert space:
\begin{align}
    \mathcal{H}_{\rm inv}
    =
    \left\{
 \ket{\Phi}\in\mathcal{H}_{\rm ext}
 \ \Big|\ 
 \hat{\Omega}\ket{\Phi}=\ket{\Phi}
 \ \textrm{for}\ \forall \Omega\in\mathcal{G}
    \right\}\, . 
\end{align}

Note that, unless the time direction is noncompact or a special boundary condition is taken, 
the condition $A_t=0$ cannot be imposed in the literal sense; we can only push $A_t$ to one point in time. As shown in Supplementary Material~\ref{sec:path-integral-and-Hamiltonian}, the integration of this remaining piece of $A_t$ leads to the projection to $\mathcal{H}_{\rm inv}$, 
\begin{align}
\hat{\mathcal{P}}\equiv\frac{1}{{\rm vol}(\mathcal{G})}\int_{g \in \mathcal{G}} dg\hat{g}\, .     
\end{align}
Here, the integration is performed over the Haar measure on $\mathcal{G}$. 

For later purposes, let us rephrase the gauge invariance in terms of the generators of the gauge transformation. On $\mathcal{H}_{\rm ext}$, the link variables and their canonical conjugates (the electric field $\hat{E}$ in the Kogut-Susskind formulation and $\hat{P}$ in the orbifold lattice formulation; see Appendices~\ref{sec:Kogut_Susskind} and~\ref{sec:orbifold_lattice}) are defined as operators. From them, we can construct the generators of the gauge transformation at each lattice point $\vec{n}$, which we denote by $\hat{G}^\alpha_{\vec{n}}$; see \eqref{eq:gauge_generator_orbifold} for the explicit form in the orbifold lattice formulation. Writing the gauge-transformation parameter as $\Omega_{\vec{n}}=\exp\left(\mathrm{i}\sum_\alpha\theta^\alpha_{\vec{n}}\tau_\alpha\right)$, where $\tau_\alpha$ are the generators of SU($N$), we have
\begin{align}
\hat{\Omega}
=
\exp\left(\mathrm{i}\sum_{\vec{n}}\sum_\alpha\theta^\alpha_{\vec{n}}\hat{G}^\alpha_{\vec{n}}\right)\, . 
\end{align}
Because any $\Omega\in\mathcal{G}$ can be written in this form, a state $\ket{\Phi}$ is in $\mathcal{H}_{\rm inv}$ if and only if $\hat{G}^\alpha_{\vec{n}}\ket{\Phi}=0$ for all $\alpha$ and $\vec{n}$. We call this condition the singlet constraint. The Hamiltonian commutes with $\hat{G}^\alpha_{\vec{n}}$, and hence, with $\hat{\Omega}$ and $\hat{\mathcal{P}}$ as well. Note also that, because $\mathcal{G}$ is compact, $\hat{\mathcal{P}}$ is a well-defined projection operator on the Hilbert space of square-integrable wave functions, which satisfies $\hat{\mathcal{P}}^2=\hat{\mathcal{P}}=\hat{\mathcal{P}}^\dagger$ and $\hat{\Omega}\hat{\mathcal{P}}=\hat{\mathcal{P}}\hat{\Omega}=\hat{\mathcal{P}}$. Therefore, gauge-invariant states are normalizable, and $\mathcal{H}_{\rm inv}=\hat{\mathcal{P}}\mathcal{H}_{\rm ext}$ is a genuine subspace of $\mathcal{H}_{\rm ext}$.

Up to a normalization factor, $\hat{G}^\alpha_{\vec{n}}$ is the lattice counterpart of the covariant divergence of the electric field (when matter fields are added, their color charge density is also included). For this reason, $\hat{G}^\alpha_{\vec{n}}$ are often called the Gauss-law operators, and the singlet constraint is often called the Gauss law constraint. For non-Abelian gauge groups, however, this terminology is misleading. The Gauss law in classical electromagnetism states that the electric flux does not emerge from nothing; it can start or end only on charges. This is a gauge-invariant statement because the electric field is gauge-invariant, and for the U(1) gauge group, it is equivalent to the singlet constraint. In non-Abelian gauge theory, on the other hand, neither the electric field nor $\hat{G}^\alpha_{\vec{n}}$ is gauge-invariant; they transform in the adjoint representation. As we will see around \eqref{eq:open_Wilson_line} and \eqref{eq:nonsinglet_open_string} in Sec.~\ref{sec:singlet_vs_non-singlet}, the Gauss law in the above sense corresponds only to the part of the singlet constraint associated with the Cartan subgroup U(1)$^{N-1}$ of SU($N$). The rest of the singlet constraint concerns the color orientation and has no counterpart in electromagnetism. In the following, when we refer to the Gauss law for quantum states, we mean it in the above sense.

\subsection{``Physical states"}\label{sec:physical_states}
There is a parallel between the quantization in $A_t=0$ gauge and the Lorentz-covariant quantization that uses BRST cohomology~\cite{Kugo:1979gm,Peskin:1995ev,Weinberg:1996kr} (Table~\ref{table:A_t=0-vs-BRST}). In our case, setting $A_t=0$ is analogous to requiring the BRST-closed condition, $\hat{Q}_{\rm BRST}\ket{\rm physical}=0$ --- both eliminate the negative norm modes. That is, $\mathcal{H}_{\rm ext}$ corresponds to $\mathrm{Ker}(\hat{Q}_{\rm BRST})$. Both $\mathcal{H}_{\rm ext}$ and $\mathrm{Ker}(\hat{Q}_{\rm BRST})$ are redundant and physically equivalent states appear multiple times. To remove the redundancy, in our case, we identify the states that differ only by a gauge transformation, while in covariant quantization, states that differ only by BRST-exact states are identified. In more formal terminologies, the true physical Hilbert space without redundancy is the gauge orbit $\mathcal{H}_{\rm ext}/\mathrm{Ker}(\hat{\mathcal{P}})$ in our case and the BRST cohomology $\mathrm{Ker}(\hat{Q}_{\rm BRST})/\mathrm{Im}(\hat{Q}_{\rm BRST})$ for covariant quantization. 

It is important to note that the word ``physical" is used loosely in a few different ways. In the context of the BRST quantization, that the state is ``physical" often means that the norm is not negative, which is guaranteed by requiring $\hat{Q}_{\rm BRST}\ket{\rm physical}=0$. In this weaker sense, all states in $\mathcal{H}_{\rm ext}$ are physical. A stronger, and ultimately more useful, meaning of ``physical'' is the equivalence class obtained after removing redundancies; that is, elements of the BRST cohomology or, in the temporal gauge, elements of $\mathcal{H}_{\rm ext}/\mathrm{Ker}(\hat{\mathcal{P}})$ obtained by quotienting by gauge orbits. When, in the context of the $A_t=0$ gauge, people say ``physical states are gauge-invariant," the latter, stronger version is assumed. More precisely, what is meant there is \textit{how to fix residual gauge symmetry}, which is analogous to the ambiguity associated with the BRST-exact states. Two states connected by a residual gauge transformation are equivalent, just as two BRST-closed states are equivalent if the difference is BRST-exact. Therefore, ``physical states are gauge-invariant" is a misleading statement. A more precise statement is that \textit{one way to remove the residual gauge redundancy in the $A_t=0$ gauge is to use the gauge-invariant Hilbert space.} 

BRST-exact states are    ``unphysical" or ``redundant". In this paper, we adopt ``redundant" to emphasize that it is associated with the redundancy of $\mathrm{Im}(\hat{Q}_{\rm BRST})$. In the $A_t=0$ gauge, $\mathrm{Ker}(\hat{\mathcal{P}})$ is redundant in the same sense. Note, however, there is an important difference: while $\mathrm{Im}(\hat{Q}_{\rm BRST})$ consists of zero-norm states, $\mathrm{Ker}(\hat{\mathcal{P}})$ consists of positive-norm states. As we will see in Sec.~\ref{sec:nonsinglet_simulations}, `redundant' states in $\mathrm{Ker}(\hat{\mathcal{P}})$ can sometimes be perfectly fine for practical computations, in the sense that the same expectation values are obtained.

\begin{table}[htbp]
\centering
\renewcommand{\arraystretch}{1.0}
\begin{tabular}{|c|c|}
\hline
\textbf{$A_t = 0$ Gauge Quantization} & \textbf{Covariant Quantization} \\
\hline
\hline
$A_t = 0$ gauge  & BRST-closed condition \\
\hline
$\mathcal{H}_{\rm ext}$ & $\mathrm{Ker}(\hat{Q}_{\rm BRST})$ \\
\hline
$\mathrm{Ker}(\hat{\mathcal{P}})$ &  $\mathrm{Im}(\hat{Q}_{\rm BRST})$ \\
\hline
Gauge orbits & BRST cohomology \\
$\mathcal{H}_{\rm ext}/\mathrm{Ker}(\hat{\mathcal{P}})$& $\mathrm{Ker}(\hat{Q}_{\rm BRST})/\mathrm{Im}(\hat{Q}_{\rm BRST})$ \\
\hline
\end{tabular}
\caption{Comparison of $A_t = 0$ gauge quantization and Lorentz-covariant BRST quantization.}\label{table:A_t=0-vs-BRST}
\end{table}

\subsection{Singlet vs non-singlet}\label{sec:singlet_vs_non-singlet}
There are many ways to describe a gauge orbit: one can either choose a point on the orbit (non-singlet state) or average over the orbit (singlet state). Below, we will see that both singlet and non-singlet states can describe gauge-invariant physics. Depending on the problems and simulation architectures, different approaches may be appropriate. 

Let us consider classical electromagnetism as an example that shows the importance of the use of the extended Hilbert space and non-singlet states. 
In classical electromagnetism, classical field configurations are not gauge invariant. This is not a problem because physics is determined by field strength (equivalently, electric field and magnetic field), which is gauge invariant. For explicit computations, it is convenient to fix the gauge and remove the redundancy. Here, let us take the temporal gauge condition $A_t=0$.  This condition does not remove the redundancy completely because time-independent gauge transformations preserve it.

Such classical field configurations correspond to low-energy wave packets in quantum field theory that are localized around the classical value of the field, $A_j^{\rm (cl.)}$. Clearly, the extended Hilbert space $\mathcal{H}_{\rm ext}$ is used. All wave packets related by the gauge transformation are physically equivalent. In principle, one could use the gauge-invariant Hilbert space $\mathcal{H}_{\rm inv}$ using the projector $\hat{\mathcal{P}}$, but no one does that because the same result is obtained anyway. 

Let us elaborate on the above. Let $\ket{\Phi}$ be a wave packet around a field configuration $A_j^{\rm (cl.)}$. A gauge transformation by $g\in\mathcal{G}$ maps $\ket{\Phi}$ to $\hat{g}\ket{\Phi}$, which is a wave packet localized around $A_j^{\rm (cl.)}-g^{-1}\partial_jg$.  Any wave packet on the gauge orbit $\{\hat{g}\ket{\Phi}\, |\, g\in\mathcal{G}\}$ is equally fine for gauge-invariant operators because they give the same expectation values:
\begin{align}
\bra{\Phi}\hat{\mathcal{O}}\ket{\Phi}
=
(\bra{\Phi}\hat{g}^{-1})\, \hat{\mathcal{O}}\, (\hat{g}\ket{\Phi})\, 
\qquad
\mathrm{if}\quad
\hat{g}^{-1}\hat{\mathcal{O}}\hat{g}
=
\hat{\mathcal{O}}\, . 
\end{align}
In principle, one can use a gauge-invariant state\footnote{
Sec.~\ref{sec:singlet_simulation} discusses an implementation of this projection on digital quantum computer.
} 
\begin{align}
\ket{\Phi; {\rm inv}}\equiv\mathcal{N}^{-1/2}\int_{\mathcal{G}}dg\hat{g}\ket{\Phi}\, ,
\end{align}
where the normalization factor is 
\begin{align}
\mathcal{N}={\rm Vol}(\mathcal{G})\int_{\mathcal{G}}dg\bra{\Phi}\hat{g}\ket{\Phi}\, . 
\end{align}
However, we obtain the same result anyway. To see this, note that the inner product of $\ket{\Phi}$ and $\hat{g}\ket{\Phi}$ is essentially the delta function $\delta(\hat g-\textbf{1})$ because we are considering a well-localized wave packet whose width is much smaller than the typical values of $A_j^{\rm (cl.)}$. This is particularly so in physically meaningful setups of large lattice volume (which is the case in the continuum limit, even at fixed physical volume), because even a small transformation at each link accumulate to significant change that leads to vanishing overlap. 
It is worth elaborating further to address some plausible concerns. First, wave packets cannot have a nontrivial stabilizer; any non-constant $g$ moves all wave packets. With the standard boundary condition $\hat g\to\mathbf{1}$, the stabilizer group can only be $\{\mathbf{1}\}$.
Second, the value of $\|\hat{g}-\textbf{1}\|$ typically increases with the number of lattice points (and also with $N^2-1$ in the case of SU($N$) theory). An $O(1)$ value (of $\|\hat{g}-\textbf{1}\|$) already suppresses the overlap to a sufficiently small value.

Furthermore, we are interested in `small' operators, say a polynomial of field strength of a finite degree, which does not alter the wave packet significantly. To see this point in concrete terms, let $x$ be the deviation of the coordinate variable $A_j$ from its value at the center of the wave packet. The wave function is sharply peaked around $x=0$. Suppose $\hat{f}$ consists of several coordinate operators, say $\hat{x}^k$. Then, the wave function changes from $\Phi(x)$ to $x^k\Phi(x)$. Assuming that $\Phi(x)$ is localized (approximately) as $\Phi(x)\sim e^{-x^2/2\sigma^2}$ ($\sigma\ll 1$), and assuming $k$ is finite (which is the meaning of `small' here), $x^k\Phi(x)$ is also localized. If $\hat{f}$ is also gauge-invariant,
\begin{align}
\bra{\Phi; {\rm inv}}
\hat{f}
\ket{\Phi; {\rm inv}}
&\simeq
\mathcal{N}^{-1}
{\rm Vol}(\mathcal{G})
f(A^{\rm cl.})
\int_{\mathcal{G}}dg
\bra{\Phi}
\hat{g}\ket{\Phi}
\nonumber\\
&=
f(A^{\rm cl.})\, . 
\label{eq:expectation_value_singlet_wave_packet}
\end{align}
For the same reason, 
\begin{align}
\bra{\Phi}
\hat{f}
\ket{\Phi}
\simeq
f(A^{\rm cl.})\, . 
\label{eq:expectation_value_nonsinglet_wave_packet}
\end{align}
Therefore, we can use either the singlets or non-singlets; this is merely a matter of taste. In practice, however, it is common to use  non-singlet states because it is more convenient to do so. 

To make the point even clearer, it is instructive to consider a system of two identical, non-interacting bosons. As explained in Supplementary Material~\ref{sec:gauge_theory_review}, this system has a gauge symmetry $\mathrm{S}_2$. Suppose that the state is obtained from two one-particle wave functions $\phi_1$ and $\phi_2$ with negligible overlap. Then, whether we use a non-singlet wave function $\phi_1(x_1)\phi_2(x_2)$ or singlet wave function $[\phi_1(x_1)\phi_2(x_2)+\phi_1(x_2)\phi_2(x_1)]/\sqrt{2}$, we obtain the same expectation value for singlet operators such as $\hat{x}_1^2+\hat{x}_2^2$. This illustrates that physical content resides in gauge-equivalence classes, not in any particular representative. From this viewpoint, it is clear why we do not need to consider a fully symmetrized wave function of all identical bosons in the universe.

At the classical level, the equation of motion for $A_t$ leads to the Gauss law constraint. Naively, the quantum counterpart to using the equation of motion for $A_t$ would be to integrate out $A_t$, which, in turn, is equivalent to acting with a projector $\hat{\mathcal{P}}$ on quantum states (Supplementary Material~\ref{sec:path-integral-and-Hamiltonian}). In the operator language of Sec.~\ref{sec:Hilbert_space}, $\hat{\mathcal{P}}$ is the projector onto the states that satisfy the singlet constraint, $\hat{G}^\alpha_{\vec{n}}\ket{\Phi}=0$. Among such states, the ground state in $\mathcal{H}_{\rm inv}$, which we denote by $\ket{\rm VAC; inv}$, plays a special role in the following. By definition, it is annihilated by all the generators of the gauge transformation, i.e., $\hat{G}^\alpha_{\vec{n}}\ket{\rm VAC; inv}=0$ for all $\alpha$ and $\vec{n}$, or equivalently, $\hat{\Omega}\ket{\rm VAC; inv}=\ket{\rm VAC; inv}$ for all $\Omega\in\mathcal{G}$ and $\hat{\mathcal{P}}\ket{\rm VAC; inv}=\ket{\rm VAC; inv}$. The statements about string states in the rest of this section rely on this property. To see why the Gauss law matters, let us consider an open Wilson line connecting two spatial points $\vec{x}$ and $\vec{x}'$, denoted by $\hat{W}(\vec{x},\vec{x}')$ acting on the gauge-invariant ground state $\ket{\rm VAC; inv}$:
\begin{align}
    \hat{W}(\vec{x},\vec{x}')\ket{\rm VAC; inv}\, . 
\label{eq:open_Wilson_line}
\end{align}
There is no classical counterpart of such a state because it would violate the Gauss law. 
Here, by the Gauss law, we mean that the electric flux does not emerge from nothing, as in classical electromagnetism: in \eqref{eq:open_Wilson_line}, the flux carried by the Wilson line emerges from nothing at one end point and disappears into nothing at the other. In terms of the gauge transformation, the Gauss law in this sense is the invariance under the gauge transformations with diagonal $\Omega_{\vec{n}}$, which form the Cartan subgroup U(1)$^{N-1}$ of SU($N$) at each lattice point (U(1)$^{N}$ for U($N$)). Each U(1) factor is the same as the gauge symmetry of electromagnetism, and the invariance under it means that the flux of each color does not emerge from nothing. This is the part of the singlet constraint associated with the Cartan subgroup. Each component of \eqref{eq:open_Wilson_line}, $\hat{W}_{a,b}(\vec{x},\vec{x}')\ket{\rm VAC; inv}$, does not satisfy it; it acquires a nontrivial phase under the diagonal gauge transformations at $\vec{x}$ or $\vec{x}'$.
Such a state is projected to zero because $\hat{\mathcal{P}}\hat{\Omega}=\hat{\mathcal{P}}$:
\begin{align}
    \hat{\mathcal{P}}\hat{W}(\vec{x},\vec{x}')\ket{\rm VAC; inv}
    =
    0\, . 
\end{align}
Therefore, such a state is redundant. Of course, we can add fermions at the end points so that the Gauss law is satisfied. Namely, we can consider
\begin{align}
    \hat{\bar{\psi}}(\vec{x})\hat{W}(\vec{x},\vec{x}')\hat{\psi}(\vec{x}')\ket{\rm VAC; inv}\, . 
\end{align}
This state is gauge invariant, and the classical counterpart satisfies the Gauss law. Hence, one might think the Gauss law constraint in the classical theory should translate into the gauge invariance in the quantum theory; but that is not the case, as we can see from the following example~\cite{Hanada:2020uvt} in the SU($N$) theory with fermions in the fundamental representation:
\begin{align}
    \hat{\bar{\psi}}_a(\vec{x})\hat{W}_{a,a'}(\vec{x},\vec{x}')\hat{\psi}_{a'}(\vec{x}')\ket{\rm VAC; inv}\, . 
\label{eq:nonsinglet_open_string}
\end{align}
Here, $a,a'=1,2,\cdots,N$ are color indices, which are \underline{not} summed over. Such a state is not gauge-invariant, but it satisfies the Gauss law in the sense that the color flux ends at sources (i.e., it is invariant under the diagonal gauge transformations at every lattice point, unlike the components of \eqref{eq:open_Wilson_line}), and it is not projected to zero:
\begin{align}
&\hat{\mathcal{P}}\left(
    \hat{\bar{\psi}}_a(\vec{x})\hat{W}_{a,a'}(\vec{x},\vec{x}')\hat{\psi}_{a'}(\vec{x}')\ket{\rm VAC; inv}
    \right)
    \nonumber\\
   &\qquad =
\frac{1}{N^2}\sum_{b,c=1}^N\hat{\bar{\psi}}_b(\vec{x})\hat{W}_{b,c}(\vec{x},\vec{x}')\hat{\psi}_{c}(\vec{x}')\ket{\rm VAC; inv}\, . 
\label{eq:projection_of_nonsinglet_open_string}
\end{align}
The right hand side $\frac{1}{N^2}\sum_{b,c=1}^N\hat{\bar{\psi}}_b(\vec{x})\hat{W}_{b,c}(\vec{x},\vec{x}')\hat{\psi}_{c}(\vec{x}')\ket{\rm VAC; inv}$ is obtained by averaging over the gauge orbit. $\hat{\bar{\psi}}_a(\vec{x})\hat{W}_{a,a'}(\vec{x},\vec{x}')\hat{\psi}_{a'}(\vec{x}')\ket{\rm VAC; inv}$ is a particular choice of a representative element of this gauge orbit. 
Specifically, because $\ket{\rm VAC; inv}$ is SU($N$) invariant, the SU($N$) transformation is 
\begin{align}
&
    (\hat{\bar{\psi}}\Omega^{-1})_a(\vec{x})(\Omega\hat{W}\Omega^{-1})_{a,a'}(\vec{x},\vec{x}')(\Omega\hat{\psi})_{a'}(\vec{x}')\ket{\rm VAC; inv}
    \nonumber\\
    &\qquad=
    \sum_{b,b',c,c'} \hat{\bar{\psi}}_b(\vec{x})\Omega^{-1}_{ba}(\vec{x})\Omega_{ab'}(\vec{x})\hat{W}_{b'c}(\vec{x},\vec{x}')\Omega^{-1}_{ca'}(\vec{x}')\Omega_{a'c'}(\vec{x}')\hat{\psi}_{c'}(\vec{x}')\ket{\rm VAC; inv}\, . 
\end{align}
By taking the average over Haar measure, we obtain $\delta_{bb'}/N$ from $\Omega^{-1}_{ba}(\vec{x})\Omega_{ab'}(\vec{x})$ and $\delta_{cc'}/N$ from $\Omega^{-1}_{ca'}(\vec{x})\Omega_{a'c'}(\vec{x})$, and hence \eqref{eq:projection_of_nonsinglet_open_string} follows. 
To specify the orbit, we can use either the orbit itself or a representative element --- there is no difference in the information in our hands. 

\begin{figure}[htbp]
\begin{center}
  \includegraphics[width=8cm]{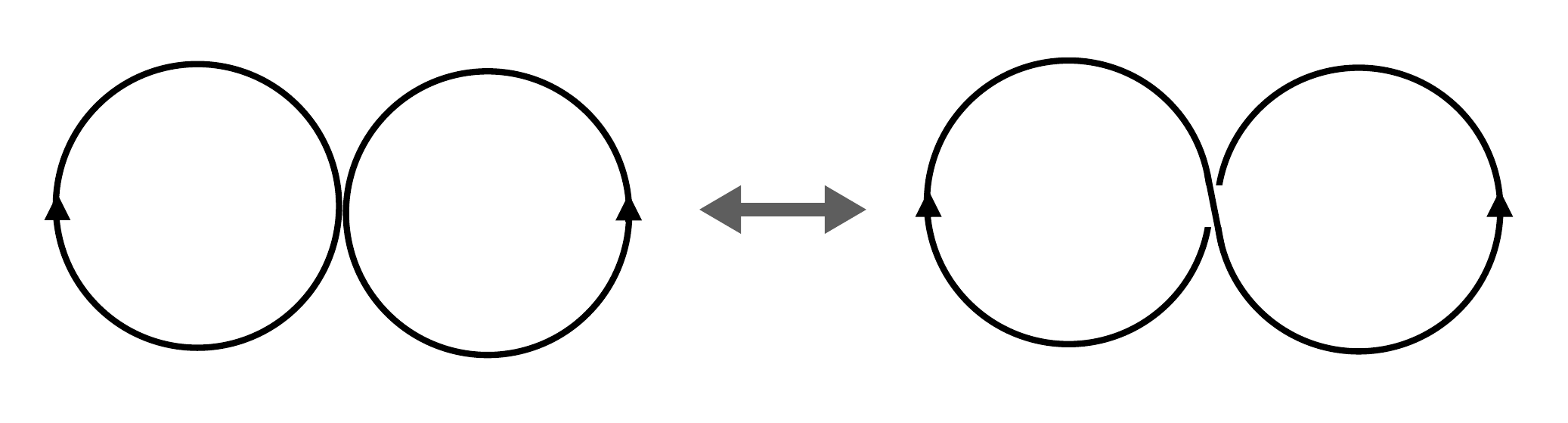}
\end{center}  
  \caption{Joining/splitting interactions of strings. }
  \label{fig:join-and-split}
\end{figure}

The same holds for closed strings. Namely, 
\begin{align}
\hat{W}^{\rm (singlet)}_{\rm closed}\equiv{\rm Tr}(\hat{Z}_{j_1,\vec{x}}\hat{Z}_{j_2,\vec{x}+\hat{j}_1}\cdots\hat{Z}_{j_l,\vec{x}-\hat{j}_l})
\label{sec;closed_string_singlet}
\end{align}
and 
\begin{align}
\hat{W}^{\rm (non\mathchar`-singlet)}_{\rm closed}\equiv N^l\hat{Z}_{j_1,\vec{x};a_1a_2}\hat{Z}_{j_2,\vec{x}+\hat{j}_1;a_2a_3}\cdots\hat{Z}_{j_l,\vec{x}-\hat{j}_l;a_la_1}
\label{sec;closed_string_nonsinglet}
\end{align}
carry the same information. Note that the joining/splitting interactions between strings (Fig.~\ref{fig:join-and-split}) are described in the same manner in both singlet and non-singlet descriptions, as we show in Appendix~\ref{sec:string_interaction}. 

Furthermore, note that there is no contradiction with the Elitzur's theorem~\cite{Elitzur:1989nr} that states the impossibility of  ``spontaneous breaking" of gauge symmetry. 
If, instead of \eqref{sec;closed_string_nonsinglet}, we consider 
\begin{align}
\hat{Z}_{j_1,\vec{x};a_1b_1}\hat{Z}_{j_2,\vec{x}+\hat{j}_1;a_2b_2}\cdots\hat{Z}_{j_l,\vec{x}-\hat{j}_l;a_lb_l}\, , 
\end{align}
the expectation value is proportional to $\delta_{b_1a_2}\cdots\delta_{b_la_1}$, which is invariant under the local SU($N$) transformation and hence does not break the gauge symmetry spontaneously. 

\begin{figure}[htbp]
\begin{center}
  \includegraphics[width=8cm]{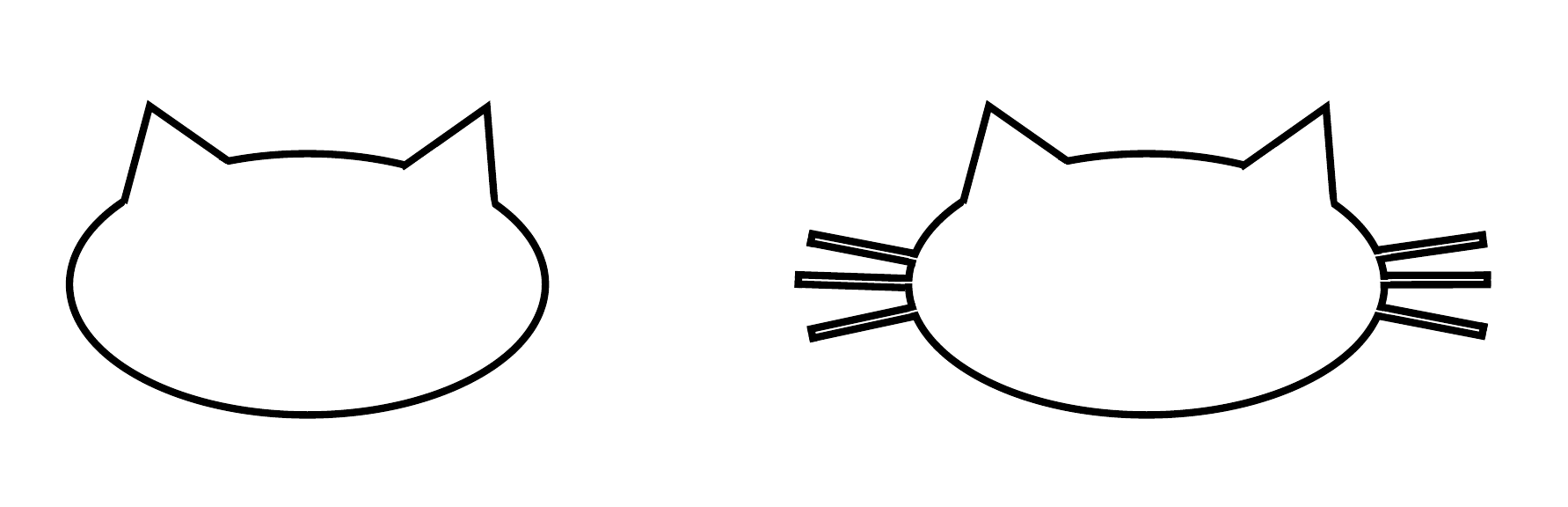}
\end{center}  
  \caption{
  Wilson loops with and without whiskers. For Kogut-Susskind, whiskers disappear (in other words, they become $\textbf{1}$) after the singlet projection. The difference between $\hat{W}^{\rm (non\mathchar`-singlet)}$ with and without whiskers is in $\mathrm{Ker}(\hat{\mathcal{P}})$ and thus redundant. With or without whiskers, a cat is a cat!
  }\label{fig:whiskers}
\end{figure}

We can also create a redundant state, which is analogous to BRST-exact states, by taking a difference between two elements on the orbit, e.g., $\hat{\bar{\psi}}_1(\vec{x})\hat{W}_{1,1}(\vec{x},\vec{x}')\hat{\psi}_{1}(\vec{x}')\ket{\rm VAC; inv}-\hat{\bar{\psi}}_2(\vec{x})\hat{W}_{2,2}(\vec{x},\vec{x}')\hat{\psi}_{2}(\vec{x}')\ket{\rm VAC; inv}$. For unitary link variables, `whiskers' such as $\hat{U}_{j,\vec{n}}^{ab}\hat{U}^{\dagger bc}_{j,\vec{n}}$ are projected to the identity matrix, and hence $\hat{W}^{\rm (non\mathchar`-singlet)}$ with and without whiskers (see Figure~\ref{fig:whiskers}) are physically equivalent. In this case, it is more reasonable to use a loop without whiskers. 

When one considers thermodynamics, it would be good to use singlet states to avoid over-counting. If one is interested in a microstate, such as the Hamiltonian time evolution of a specific initial condition, then either the singlet or the non-singlet can work, depending on the details of the task. In fact, even a redundant state --- an element of $\mathrm{Ker}(\hat{\mathcal{P}})$ --- may be fine. For a well-localized wave packet $\ket{\Phi}$, we can define a redundant state $(\ket{\Phi}-\hat{g}\ket{\Phi})/\sqrt{2}$. If $\ket{\Phi}$ and $\hat{g}\ket{\Phi}$ are sufficiently separated and a gauge-invariant operator $\hat{O}$ does not mix them, then the expectation value of $\hat{O}$ is the same whether we use $\ket{\Phi}$, $\hat{g}\ket{\Phi}$, or $(\ket{\Phi}-\hat{g}\ket{\Phi})/\sqrt{2}$. 

To summarize, a singlet representation is actually needed only in two situations: (i) thermodynamics and other ensemble quantities, where non-singlet states would be over-counted, and (ii) when non-singlet operators act on non-singlet states, so that gauge invariance must be restored locally. In every other case considered in this paper --- ground states and wave packets probed by gauge-invariant operators, and spatially separated string excitations on a gauge-invariant state --- the non-singlet description gives identical expectation values at lower cost, and the projection of Sec.~\ref{sec:singlet_projection} should simply not be used.

\section{Quantum simulations without singlet constraint}\label{sec:nonsinglet_simulations}
In Sec.~\ref{sec:H_ext-and-H_inv}, we emphasized the importance of a non-singlet description of gauge-invariant physics, taking electromagnetism as an example. Specifically, a `physical state' can be described using either a singlet or a non-singlet, and certain simple types of the latter, such as a wave packet, are conceptually and technically easier. Here, we are talking about a  ``wave packet" in the field space, i.e., gauge field $A_\mu(x)$ is localized at each spatial point $x$. In the lattice theory, it is a wave packet on the group manifold. This is a rather fundamental class; for example, the ground state is a wave packet around $U=1$ (and also $W=1$, for the case of the orbifold lattice), up to gauge transformation. More precisely, such a wave packet is a non-singlet representative of the gauge-invariant ground state $\ket{\rm VAC; inv}$ used in Sec.~\ref{sec:singlet_vs_non-singlet}; see Sec.~\ref{sec:wave_packet}.

In this section, we focus on the practical utility of the non-singlet description in quantum simulations. We consider two options: non-singlet-string excitation on a gauge-invariant state, and a wave packet. The use of the non-singlet description can be useful to avoid the cost of the singlet projection introduced in Sec.~\ref{sec:singlet_simulation} whenever possible. 
\subsection{Non-singlet strings}
Suppose that the gauge-invariant ground state $\ket{\rm VAC; inv}$ is already prepared. We can act with non-singlet operators on it to obtain non-singlet states. For example, we consider closed strings (Wilson loops) and open strings (Wilson lines). For many cases of interest, we obtain the same result whether we use singlets or non-singlets. 

To understand the concepts, we can use the vanilla Kogut-Susskind or orbifold lattice formulation; The details do not matter, unless one considers the circuit implementation, which is much more efficient for the orbifold formulation. Below, we use the orbifold lattice for concreteness.

Gauge-invariant states in $\mathcal{H}_{\rm inv}$ can be obtained by acting with Wilson loop operators (often called closed strings) $\hat{W}_{\rm closed}$ defined by \eqref{sec;closed_string_singlet}
on $\ket{\rm VAC; inv}$. 
The number of terms in the sum increases exponentially with the length of the loop $l$ as $N^l$. When $N$ or $l$ is large, it is better to avoid such an increase. 

A simple way to avoid such an increase is to use non-singlet states~\footnote{
See Supplementary Material~\ref{Sec:algorithms} for an alternative approach.
}. 
Suppose that no loop intersects with other loops or with itself. Then, such an operator is obtained by symmetrizing $\hat{W}^{\rm (non\mathchar`-singlet)}_{\rm closed}$ defined by \eqref{sec;closed_string_nonsinglet}.
Note that, in this case, symmetrization over SU($N$) is the same as taking the average over all possible choices of $a_1,\cdots,a_l$. 
Using $\hat{W}^{\rm (singlet)}$ or $\hat{W}^{\rm (non\mathchar`-singlet)}$ does not make a difference in the following case. Suppose that $\ket{\rm VAC; inv}$ is gauge invariant. Let us consider a correlation function of multiple loops $\hat{W}^{\rm (singlet)}_1,\cdots,\hat{W}^{\rm (singlet)}_n$. 
If all loops are spatially separated,
\begin{align}
&
\bra{\rm VAC; inv}
 \hat{W}^{\rm (singlet)}_1 e^{-\mathrm{i}t_1\hat{H}}
 \hat{W}^{\rm (singlet)}_2 e^{-\mathrm{i}t_2\hat{H}}
 \cdots 
 \nonumber\\
 &\qquad\qquad
 \cdots
 e^{-\mathrm{i}t_{n-1}\hat{H}}
 \hat{W}^{\rm (singlet)}_n
 \ket{\rm VAC; inv}
 \nonumber\\
 &\quad=
 \bra{\rm VAC; inv}
 \hat{W}^{\rm (non\mathchar`-singlet)}_1 e^{-\mathrm{i}t_1\hat{H}}
 \hat{W}^{\rm (non\mathchar`-singlet)}_2 e^{-\mathrm{i}t_2\hat{H}}
 \cdots 
 \nonumber\\
 &\qquad\qquad\quad
 \cdots
 e^{-\mathrm{i}t_{n-1}\hat{H}}
 \hat{W}^{\rm (non\mathchar`-singlet)}_n
 \ket{\rm VAC; inv}\, . 
\label{eq:loop_correlators_singlet_vs_nonsinglet}
\end{align}
(This can be seen using $\mathcal{P}\ket{\rm VAC; inv}=\ket{\rm VAC; inv}$, $\bra{\rm VAC; inv}\mathcal{P}=\bra{\rm VAC; inv}$, and projection properties explained in Sec.~\ref{sec:singlet_vs_non-singlet}.)
As in the case of \eqref{eq:projection_of_nonsinglet_open_string}, this equality relies on the gauge invariance of $\ket{\rm VAC; inv}$ and $\hat{H}$: the gauge transformation acts only on the loop operators, and because the loops do not share lattice points, the Haar average can be taken for each loop separately.
In this way, singlet and non-singlet descriptions can lead to the same result.

Note that the same holds for the lattice Monte Carlo simulation on classical computers: if configurations are generated without gauge fixing (analogous to the use of a gauge-invariant ground state $\ket{\rm VAC; inv}$), we obtain the same expectation values for $\hat{W}^{\rm (singlet)}$ and $\hat{W}^{\rm (non\mathchar`-singlet)}$. 

Although the same expectation value is guaranteed, more measurements might be needed to estimate the expectation value precisely when the non-singlet states are used. Still, the non-singlet description can simplify the circuit implementations, and also, it can provide some conceptual advantages when certain problems are considered, as mentioned at the beginning of this section.

\subsubsection*{Preparation of string states}
Let us consider closed strings. (Open strings are similar.)
Whether we use the singlet version like $\hat{W}^{\rm (singlet)}$ or non-singlet version like $\hat{W}^{\rm (non\mathchar`-singlet)}$, they are written as a sum of Pauli strings made of only $\sigma_z$s with length $l$, because each $\hat{Z}$ is a sum of $\sigma_z$s. The number of Pauli strings scales as $(2Q)^l$ for $\hat{W}^{\rm (non\mathchar`-singlet)}$ and $(2NQ)^l$ for $\hat{W}^{\rm (singlet)}$. (The relative factor $N^l$ gives us a strong motivation to use non-singlet states. In case we prefer singlet states, we can compute traces explicitly, or use the projection method in Sec.~\ref{sec:singlet_simulation}, or use yet another method in Supplementary Material~\ref{Sec:algorithms}.) That we have a simple Pauli string expansion makes state preparation straightforward. 

Let us consider a method based on LCU (Supplementary Material~\ref{Sec:algorithms}). The task is, for a given state $\ket{\Phi}$ (say the ground state), to obtain $\frac{\hat{O}\ket{\Phi}}{||\hat{O}\ket{\Phi}||}$, where $\hat{O}$ is $\hat{W}^{\rm (singlet)}$ or $\hat{W}^{\rm (non\mathchar`-singlet)}$. For this purpose, we can apply the LCU method (as reviewed in Supplementary Material~\ref{Sec:algorithms}) directly without modification, because the Pauli-string expansion of the following form is known explicitly:
\begin{align}
    \hat{O}
    =
    \sum_{i}
    \alpha_i\hat{\Pi}_i\, . 
\end{align}
Note that we can take all $\alpha_i$ real and non-negative, attributing the phase to $\hat{\Pi}_i$. As long as the loop consists of a finite number of links, $\sum_i |\alpha_i|$ is bounded. 
\subsection{Wave packet}\label{sec:wave_packet}
Let us use a wave packet as discussed in Sec.~\ref{sec:H_ext-and-H_inv}. If we consider the ground state at weak coupling~\footnote{We can even take $g^2\to 0$ at fixed lattice spacing $a$, which is a weaker coupling than the continuum limit. Such a trick can be useful when we use adiabatic state preparation.}, a wave packet localized around $W_{j,\vec{n}}=\textbf{1}_N$ and $U_{j,\vec{n}}=\textbf{1}_N$, which we denote by $\ket{\Phi}$, describes the ground state. Such a wave packet is not gauge invariant; any wave packet equivalent to $\ket{\Phi}$ up to a gauge transformation has the same property. A gauge transformation by $\Omega\in\mathcal{G}$ maps $\ket{\Phi}$ to $\hat{\Omega}\ket{\Phi}$, which is a wave packet localized around $W_{j,\vec{n}}=\textbf{1}_N$ and $U_{j,\vec{n}}=\Omega_{\vec{n}}\Omega^{-1}_{\vec{n}+\hat{j}}$. 
Hence $\ket{\Phi}$ can be invariant only under the global SU($N$) transformation. The gauge-invariant ground state is obtained by projection, as we saw in Sec.~\ref{sec:H_ext-and-H_inv}. Namely, $\ket{\rm VAC; inv}$ used in Sec.~\ref{sec:singlet_vs_non-singlet} and in the previous subsection is $\hat{\mathcal{P}}\ket{\Phi}$ up to a normalization factor, and $\ket{\Phi}$ is a non-singlet representative of the same physical state (a point on the gauge orbit). Note that $\ket{\Phi}$ is not annihilated by the generators of the gauge transformation $\hat{G}^\alpha_{\vec{n}}$. Therefore, the relations that rely on the gauge invariance of $\ket{\rm VAC; inv}$, such as \eqref{eq:projection_of_nonsinglet_open_string} and \eqref{eq:loop_correlators_singlet_vs_nonsinglet}, do not hold if $\ket{\rm VAC; inv}$ is replaced with $\ket{\Phi}$. Instead, the equivalence between $\ket{\Phi}$ and $\ket{\rm VAC; inv}$ for gauge-invariant operators follows from a different mechanism --- the suppression of the overlap between $\ket{\Phi}$ and $\hat{\Omega}\ket{\Phi}$ --- as explained around \eqref{eq:expectation_value_singlet_wave_packet} and \eqref{eq:expectation_value_nonsinglet_wave_packet}.

To take the continuum limit, the lattice spacing $a$ and coupling constant $g$ are sent to zero, and the lattice size is sent to infinity. In this limit, $\bra{\Phi}\hat{\Omega}\ket{\Phi}$ is essentially a product of delta functions, $\prod_{\vec{n}}\delta(\Omega_{\vec{n}}\Omega^{-1}_{\vec{n}+\hat{j}}-\textbf{1}_N)$. 
This is because $\ket{\Phi}$ and $\hat{\Omega}\ket{\Phi}$ can have a non-negligible overlap only when the overlap is almost perfect at all sites (i.e., $\Omega_{\vec{n}}\Omega^{-1}_{\vec{n}+\hat{j}}$ is very close to $\textbf{1}_N$ at all $\vec{n}$.) 

A natural strategy for quantum simulations is to act with gauge-singlet operators on this non-singlet state, as we discussed for classical electromagnetism in Sec.~\ref{sec:H_ext-and-H_inv}. 

To obtain a wave packet localized around $W=\textbf{1}_N$ and $U=\textbf{1}_N$, we can add a penalty term proportional to 
\begin{align}
\sum_{\vec{n}}\sum_{j=1}^3
{\rm Tr}
\left|\langle\hat{Z}_{j,\vec{n}}\rangle -\sqrt{\frac{a}{2g^2}}\cdot\textbf{1}_N\right|^2\, . 
\end{align}

Alternatively, we might be able to use the adiabatic state preparation method, starting with a special point in the orbifold lattice where the ground state can be obtained analytically (e.g., the weak-coupling limit) and then changing the parameters adiabatically. Whether this approach works depends on the structure of the phase diagram, which can be studied using the standard Euclidean path integral. 

When using LCU and Taylor expansion, imaginary time evolution is straightforward. 

\section{Quantum simulations with singlet constraint}\label{sec:singlet_simulation}

\subsection{Projection to gauge singlet}\label{sec:singlet_projection}
We stress at the outset that the protocols of this section are \emph{not} the recommended default: as summarized in Sec.~\ref{sec:when_to_project}, the singlet projection carries a post-selection cost and should be used only locally and sparingly, in the situations where a singlet representation is genuinely required. The default, cheaper strategy is the non-singlet approach of Sec.~\ref{sec:nonsinglet_simulations}. We nevertheless present the projection in full, because a complete toolkit must cover the exceptional cases as well.

From here on, we consider circuit implementations. We will use the orbifold lattice Hamiltonian (Appendix~\ref{sec:orbifold_lattice}) and the universal framework~\cite{Halimeh:2024bth} (see also Supplementary Material~\ref{sec:universal_framework}) that allows us to construct quantum circuits explicitly. Combined with our previous work~\cite{Halimeh:2024bth,Halimeh:2025ivn}, this section completes the efficient quantum simulation method for Yang-Mills theory and QCD with gauge-singlet constraint. 

Conceptually, the simplest way to achieve quantum simulation with the gauge-singlet constraint is to explicitly construct the projection procedure from $\mathcal{H}_{\rm ext}$ to $\mathcal{H}_{\rm inv}$~\footnote{Refs.~\cite{Than:2024zaj,Chakraborty:2025veu} introduced a singlet-projection protocol for QCD in $1+1$ dimensions, utilizing a special property specific to $1+1$ dimensions to make the implementation tractable. We do not use any properties specific to certain dimensions, gauge group, or matter content. This makes our approach readily applicable to real-world QCD, and at the same time, more resource-intensive. 
}. Such a procedure is characterized by its action on the coordinate eigenstate. Specifically, it must transform $\ket{Z}$ into $\int d\Omega\ket{Z^{(\Omega)}}$, where the integral is taken over the Haar measure. Note that such a procedure is not realizable by a unitary operator acting on the Hilbert space, because many non-singlets correspond to one singlet. 

Our protocol consists of the following steps:
\begin{itemize}
    \item 
    \textbf{Step 1.}
    Create operators $\hat{\xi}_{\vec{n}}$ and states $\ket{\xi}=\otimes_{\vec{n}}\ket{\xi_{\vec{n}}}$ satisfying $\hat{\xi}_{\vec{n}}\ket{\xi_{\vec{n}}}=\xi_{\vec{n}}\ket{\xi_{\vec{n}}}$, where $\xi_{\vec{n}}\in\mathbb{C}^{N\times N}$.

    \item 
    \textbf{Step 2.}
    Produce $\ket{Z^{(\xi)}}\ket{\xi}$ from $\ket{Z}\ket{\xi}$, where $Z^{(\xi)}_{j,\vec{n}}=\xi_{\vec{n}} Z_{j,\vec{n}}\bar{\xi}_{\vec{n}+\hat{j}}$.

    \item 
    \textbf{Step 3.}    Take an average over $\xi_{\vec{n}}\in\mathrm{SU}(N)\subset\mathbb{C}^{N\times N}$. 
\end{itemize}

\noindent
\textbf{Step 1.}
We can introduce a complex matrix $\xi_{\vec{n}}\in\mathbb{C}^{N\times N}$ and ancilla states $\ket{\xi}=\otimes_{\vec{n}}\ket{\xi_{\vec{n}}}$ just by using the techniques in the orbifold lattice formulation. We assign $Q$ qubits for each bosonic degree of freedom. (In principle, we can assign different number than $Q$.)\\

\noindent
\textbf{Step 2.}
To shift $\ket{Z}$ to $\ket{Z^{(\xi)}}$, we can use
\begin{align}
    \bra{P}\ket{Z}
    =
    \exp\left(
-\mathrm{i}\sum_{j,\vec{n}}\mathrm{Tr}\left[P_{j,\vec{n}}\bar{Z}_{j,\vec{n}}
    +
    \bar{P}_{j,\vec{n}}Z_{j,\vec{n}}\right]
    \right)
\end{align}
and a similar equation for $\ket{Z^{(\xi)}}$, which give us
\begin{align}
    |Z^{(\xi)}\rangle 
    &= \int dP \ket{P}\bra{P}\ket{Z}
    \nonumber\\
    &\times
    \exp\Biggl(
\mathrm{i}\sum_{j,\vec{n}}\mathrm{Tr}\Bigl[P_{j,\vec{n}}(\bar{Z}_{j,\vec{n}}-\bar{Z}^{(\xi)}_{j,\vec{n}})
    \nonumber\\
    &\qquad
    +
    \bar{P}_{j,\vec{n}}(Z_{j,\vec{n}}
    -Z^{(\xi)}_{j,\vec{n}})\Bigl]
    \Biggl)\, . 
\end{align}
Note also that, by using 
\begin{align}
    \hat{Z}^{(\xi)}_{j,\vec{n}}
    \equiv
    \hat{\xi}_{\vec{n}}
    \hat{Z}_{j,\vec{n}}
    \hat{\bar{\xi}}_{\vec{n}+\hat{j}}\, , 
\end{align}
we have 
\begin{align}
    \hat{Z}^{(\xi)}\ket{Z}\ket{\xi}
    =
    Z^{(\xi)}\ket{Z}\ket{\xi}\, . 
\end{align}
To produce $\ket{Z^{(\xi)}}$ using this operator, firstly we create $\int\mathrm{d}ZF(Z)\ket{Z}_1\ket{Z}_2\ket{\xi}$ from $\int\mathrm{d}ZF(Z)\ket{Z}_1\ket{\xi}$. 
Here, we used a subscript to distinguish two copies. This is straightforward because $\ket{Z}$ is binary; we just tensor $\ket{0,0,\cdots,0}_2$ and act CNOT gates. Now, we define a new unitary operator:
\begin{align}
\hat{\mathcal{T}}^{(\xi)}
\equiv
    \exp\left(
\mathrm{i}\mathrm{Tr}\left[\hat{P}_1(\hat{\bar{Z}}_2-\hat{\bar{Z}}_2^{(\xi)})
    +
    \hat{\bar{P}}_1(\hat{Z}_2-\hat{Z}^{(\xi)}_2)\right]
    \right)\, . 
    \label{eq:singlet_projection_T_operator}
\end{align}
Here, $\hat{P}_1$ and $\hat{\bar{P}}_1$ acts on $\ket{Z}_1$, $\hat{Z}_2$ and $\hat{\bar{Z}}_2$ acts on $\ket{Z}_2$, and $\hat{Z}_2^{(\xi)}$ is made of $\hat{Z}_2$ and $\hat{\xi}$. 
(We did not write $j,\vec{n}$ explicitly just to avoid cluttered notations.)
Note that all operators inside the exponential commute with each other and hence there is no ordering ambiguity. Using this operator, we can easily see that~\footnote{
If $\ket{P=0}$ exists in the truncated Hilbert space, we could introduce $\ket{P=0}_2$ and act  
$\exp\left(-\mathrm{i}\mathrm{Tr}\left(\hat{P}_2\hat{\bar{Z}}_1^{(\xi)}+\hat{\bar{P}}_2\hat{Z}^{(\xi)}_1)\right)\right)$ to get 
$\ket{Z}_1\ket{Z^{(\xi)}}_2\ket{\xi}$. 
(In our specific truncation, exact zero-momentum mode does not exist, though it can be introduced slightly changing the details.)
Then, by relabeling $1$ and $2$, the rest goes the same. 
} 
\begin{align}
    \hat{\mathcal{T}}^{(\xi)}
    \ket{Z}_1\ket{Z}_2\ket{\xi}
= 
    |Z^{(\xi)}\rangle_1\ket{Z}_2\ket{\xi}\, . 
\end{align}
By seeing $\mathrm{Tr}\left[\hat{P}_1(\hat{\bar{Z}}_2-\hat{\bar{Z}}_2^{(\xi)})+\hat{\bar{P}}_1(\hat{Z}_2-\hat{Z}^{(\xi)}_2)\right]$ as `Hamiltonian', we can regard $\hat{\mathcal{T}}^{(\xi)}$ as `Hamiltonian time evolution'. 
By performing the Quantum Fourier Transform to $\ket{Z}_1$ and going to the momentum basis, $\hat{P}_1$ and $\hat{\bar{P}}_1$ becomes a sum of Pauli $\sigma_z$s. The other operators ($\hat{Z}_2$, $\hat{\bar{Z}}_2$, $\hat{Z}_2^{(\xi)}$, $\hat{\bar{Z}}_2^{(\xi)}$) are also written using Pauli $\sigma_z$s. Therefore, we can use the universal framework~\cite{Halimeh:2024bth} (see also Supplementary Material~\ref{sec:Trotter_cost_piece}). 

Now we need to eliminate $\ket{Z}_2$. An expensive but precise method is to project it to $\frac{1}{2^{N^2Q}}\left(\sum_{Z'}\ket{Z'}_2\right)$. We can simply perform measurement in $\ket{\pm}=\frac{\ket{0}\pm\ket{1}}{\sqrt{2}}$ basis or use amplitude amplification. Either way, the cost scaling is exponential in $N^2Q$. A better method is to act another operator, 
\begin{align}
\hat{\mathcal{T}}^{\prime(\xi)}
\equiv
    \exp\left(
\mathrm{i}\mathrm{Tr}\left[\hat{P}_2\hat{\bar{Z}}_1^{(\bar{\xi})}
    +
    \hat{\bar{P}}_2\hat{Z}^{(\bar{\xi})}_1\right]
    \right)\, . 
    \label{eq:singlet_projection_T_prime_operator}
\end{align}
Then, $\hat{Z}^{(\bar{\xi})}_1$ acting on $\ket{Z^{(\xi)}}_1$ picks up $Z^{(\bar{\xi}\xi)}$. 
In the infinite mass limit, $\xi$ is unitary, and $\bar{\xi}$ is $\xi^{-1}$, and hence, $Z^{(\bar{\xi}\xi)}=Z$. Therefore, by acting $\hat{\mathcal{T}}^{\prime(\xi)}$, $\ket{Z}_2$ is approximately sent to $\ket{Z=0}_2$, which allows us more reliable projection with a higher probability. 
Either way, we use $\left(\hat{\mathcal{T}}^{(\xi)}\ket{Z}_1\ket{Z}_2\ket{\mathcal{G}}\right)_{\rm proj.}$ to denote the state after projecting out $\ket{Z}_2$. 
Note that, even if we improve the success probability of post selection in this way, repeated post-selection leads to exponentially large cost in general.

\noindent
\textbf{Step 3.}
We prepare a state $\ket{\mathcal{G}}=\int d\xi\varphi(\xi)\ket{\xi}$ in such a way that $\ket{\mathcal{G}}$ is a linear combination of $\xi_{\vec{n}}\in\mathrm{SU}(N)$ with Haar measure. 
Equivalently, $\varphi(\xi)$ is SU($N$)-invariant and vanishes quickly as $\xi$ departs from $\mathrm{SU}(N)\in\mathbb{C}^{N\times N}$. 
There exist many ways to build $\ket{\mathcal{G}}$. One way is to multiply $\exp(-C\, \mathrm{Tr}(\hat{\xi}\hat{\bar{\xi}}-\textbf{1})^2)$ and $\exp(-C'|\det\hat{\xi}-1|^2)$ to a state proportional to $\sum_\xi\ket{\xi}$, where $C$ and $C'$ are large positive numbers. As discussed in Supplementary Material~\ref{Supplementary_Material:singlet_simulation}, this is straightforward by using the Linear Combination of Unitaries (LCU; see Supplementary Material~\ref{Sec:algorithms}) or Quantum Singular Value Transform (QSVT).
Another natural option would be to minimize a Hamiltonian
\begin{align}
\hat{H}_\xi=
\mathrm{Tr}\hat{P}_\xi\hat{\bar{P}}_\xi
+
c_1\cdot
\mathrm{Tr}\left(\hat{\xi}\hat{\bar{\xi}}-1\right)^2
+
c_2\cdot\left|\det\hat{\xi}-1\right|^2
\label{preparation_xi_Hamiltonian}
\end{align}
with large coefficients $c_1$ and $c_2$. In the limit of $c_1\to\infty$, $\xi$ is forced to be unitary, and in the limit of $c_2\to\infty$, determinant becomes 1. The momentum part makes the distribution uniform along the SU($N$) group manifold. See Sec.~\ref{sec:numerical_test} for numerical demonstration of this approach. 
By using $\ket{\mathcal{G}}$, we can realize the average over SU($N$):
\begin{align}
\bra{\mathcal{G}}
\left(\hat{\mathcal{T}}^{(\xi)}
    \ket{Z}_1\ket{Z}_2\ket{\mathcal{G}}\right)_{\rm proj.}
    &=
    \int_{\mathrm{SU}(N)}\mathrm{d}\Omega\, |Z^{(\Omega)}\rangle\, .
\end{align}
This projection involves a post-selection, which, again, can lead to exponentially large cost if the projection is repeated multiple times.

Supplementary Material~\ref{Supplementary_Material:singlet_simulation} provides more details of the computations in this section.

\subsection{Alternative way by adding a penalty term}\label{sec:penalty_G^2}
A simple way to remove gauge non-singlets from the low-energy spectrum is to add a penalty to the non-singlet states. This idea has been discussed extensively in the past; see e.g.~\cite{Banerjee:2012pg}. A nontrivial question is whether we can incorporate this into the efficient circuits constructed based on the noncompact variables ---  and the answer is yes. 

Using non-compact variables, the generators of U($N$) transformation acting on the link variable from left and right are defined as 
\begin{align}
\hat{\mathcal{E}}_{j,\vec{n}}^{(\mathrm{L})ab}
&=
\sum_c
\left(
\mathrm{i}\hat{Z}_{j,\vec{n}}^{ac}\cdot\hat{\bar{P}}^{cb}_{j,\vec{n}}
-
\mathrm{i}\hat{\bar{Z}}^{cb}_{j,\vec{n}}\cdot\hat{P}^{ac}_{j,\vec{n}}
\right)
\end{align} 
and
\begin{align}
\hat{\mathcal{E}}_{j,\vec{n}}^{(\mathrm{R})ab}
&=
\sum_c
\left(
\mathrm{i}\hat{Z}^{cb}_{j,\vec{n}}\cdot\hat{\bar{P}}^{ac}_{j,\vec{n}}
-
\mathrm{i}\hat{\bar{Z}}^{ac}_{j,\vec{n}}\cdot\hat{P}^{cb}_{j,\vec{n}}
\right)\, . 
\end{align} 
We can use U($N$) generators $\tau_\alpha$ ($\alpha=1,\cdots,N^2$) to extract the generators of U($N$) as 
\begin{align}
    \hat{\mathcal{E}}^{(\mathrm{L})\alpha}
    =
    \mathrm{Tr}\left(
    \tau_\alpha
    \hat{\mathcal{E}}^{(\mathrm{L})}
    \right)\, , 
    \qquad
        \hat{\mathcal{E}}^{(\mathrm{R})\alpha}
    =
    \mathrm{Tr}\left(
    \tau_\alpha
    \hat{\mathcal{E}}^{(\mathrm{R})}
    \right)\, .      
\end{align}
The generators of SU($N$) gauge transformation is 
\begin{align}
\hat{G}^\alpha_{\vec{n}}
=
\sum_{j=1}^3
\left(
-\hat{\mathcal{E}}^{(\mathrm{L})\alpha}_{j,\vec{n}}
+
\hat{\mathcal{E}}^{(\mathrm{R})\alpha}_{j,\vec{n}-\hat{j}}
\right)\, . 
\label{eq:gauge_generator_orbifold}
\end{align}
By adding a penalty term proportional to $\sum_{\alpha,\vec{n}}\left(\hat{G}_{\vec{n}}^\alpha\right)^2$ to the Hamiltonian, we can remove gauge non-singlets. 

As explained in Supplementary Materials~\ref{sec:gauge_penalty_implementation}, this term can be implemented to the Hamiltonian time evolution by slightly generalizing the universal framework. 
If we can prepare low-energy states of the theory with such a penalty term, the singlet-constraint can be imposed. One way to achieve this would be to take the energy out of the system by attaching it to an external system analogous to the heat bath.

\subsection{When is the singlet projection actually needed?}\label{sec:when_to_project}
Adding projections does not change the gate-count scaling of Sec.~\ref{sec:outlook} provided they are not used excessively. The cost for the post-selection in the projection can be significant, and if we perform projection to the entire lattice, exponential increase of the post-selection cost looks inevitable. To tame the cost for the projection, what we should keep in mind are:
\begin{itemize}
\item
We should use non-singlet description when possible. (Sec.~\ref{sec:nonsinglet_simulations})

\item
Perform singlet-projection only to a small fraction of the space, and do not repeat it too many times. 

\end{itemize}
From the results in Sec.~\ref{sec:nonsinglet_simulations}, we see that complications may arise only when non-singlet operators acts on a non-singlet state; therefore, we should perform the singlet projection only for such a case, in a small fraction of the lattice where such complications can arise. The use of non-singlet initial condition and singlet operators, avoiding any post-selection where possible, seems to be the optimal strategy. In other words, within our framework the singlet projection is a \emph{local, occasional subroutine} applied to a few links when gauge invariance must be restored --- not a global constraint imposed on the simulation.

\section{Validation of Convergence Criteria}\label{sec:numerical_test}
We validate our framework through systematic numerical studies on classical computers. This approach allows exact verification of algorithm correctness, convergence criteria, and scaling properties at system sizes where analytical results or exact diagonalization provide ground truth. Classical validation is the appropriate methodology for algorithm verification; quantum hardware becomes relevant when system sizes exceed classical simulation capabilities, where our validated scaling ensures quantum advantage.

In this section, we will focus on the truncation effect and Trotter scaling for the mass terms that enforce unitarity constraints. We first provide a theoretical analysis that holds regardless of the details of the interactions, dimensions, and lattice size, and then we provide numerical demonstrations on classical computers using small systems. The validation conducted in this section is relevant both for the singlet-projection protocol introduced in Sec.~\ref{sec:singlet_simulation} and other parts of the universal simulation protocol~\cite{Halimeh:2024bth} (see also Supplementary Material~\ref{sec:universal_framework}). 

To write an efficient quantum circuit for gauge theory, the use of non-compact variables (complex link variable $Z_{j,\vec{n}}$, and a complex matrix $\xi$ to express the SU($N$) gauge transformation in Sec.~\ref{sec:singlet_simulation}) was crucial. Furthermore, the state $\ket{\mathcal{G}}$, which plays the crucial role in the singlet-projection protocol introduced in Sec.~\ref{sec:singlet_simulation}, could be described as the ground state of the Hamiltonian \eqref{preparation_xi_Hamiltonian}, which is actually an embedding of strongly-coupled lattice gauge theory into noncompact variables, as we will see below. Natural questions associated with this approach are: how many qubits do we need to express unitary variables precisely? How large a mass do we need if we want to remove the scalars so that the orbifold lattice Hamiltonian reduces to the Kogut-Susskind Hamiltonian? Does a large mass in such a limit affect the simulation cost? In this section, we address these questions by numerical experiments on classical computers.

Numerically demonstrating convergence criteria using classical computers is challenging --- that is the very reason why we do need a quantum computer! To reduce the computational cost for a demonstration, we consider the counterpart of the strong-coupling limit of the Kogut-Susskind Hamiltonian. Specifically, we drop the second and third lines in the orbifold lattice Hamiltonian \eqref{eq:orbifold_Hamiltonian}, keeping the kinetic term $\sum_{j,\vec{n}}\mathrm{Tr}\hat{P}_{j,\vec{n}}\hat{\bar{P}}_{j,\vec{n}}$ and the mass term $\hat{H}_{\rm mass}$ defined by \eqref{eq:orbifold_mass_term}. Then, there is no interaction between different sets of $(j,\vec{n})$. Therefore, we can focus on a single link and consider the following Hamiltonian:
\begin{align}
    \hat{H}
    =
    \mathrm{Tr}\hat{P}\hat{\bar{P}}
    +
    \hat{H}^{\rm (mass)}\, , 
    \label{eq:orbifold_Hamiltonian_for_strong_copuling}
\end{align}
\begin{align}
    \hat{H}^{\rm (mass)}
    =
    \mathrm{Tr}
    \left(
    m^2\left(\hat{Z}\hat{\bar{Z}}-\mathbf{1}_N\right)^2
    +
    m_{\rm U(1)}^2|\det\hat{Z}-1|^2
    \right)\, . 
\end{align}
Here, to simplify the expression, we set $g^2$ and $a$ to 1.

We want to see the behavior of this model in the infinite-mass limit. This is still a challenging task because the SU($N$) group manifold is embedded into $\mathbb{C}^{N^2}=\mathbb{R}^{2N^2}$ and the dimension of the truncated Hilbert space is $\Lambda^{2N^2}$. Here, $\Lambda$ is the truncation level for each boson. It is easy to see that the dimension of the truncated Hilbert space grows fast even for $N=2$. Hence, let us simplify the problem without losing the essence, in the following manner. 

For $N=2$, the constraint imposed by the mass term can be written as $\vec{v}_i^\dagger\vec{v}_j=\delta_{ij}$, where $v_i$ are column vectors (i.e., $Z=(\vec{v}_1,\vec{v}_2)$), and $\det Z=1$. Among them, let us first impose $\vec{v}_1^\dagger\vec{v}_2=0$, $|\vec{v}_1|^2=|\vec{v}_2|^2$, and $\det Z\in\mathbb{R}$. This can be achieved by arranging the mass term appropriately and sending the mass parameter corresponding to these constraints to large values. Then, the complex matrix $Z$ becomes
\begin{align}
    Z
    =
    \left(
    \begin{array}{cc}
         \alpha & -\beta^\ast \\
         \beta & \alpha^\ast
    \end{array}
    \right)\,  
    \qquad\alpha,\beta\in\mathbb{C}\, . 
\end{align}
Note that $Z\in\mathrm{SU}(2)$ when the remaining constraint $|\alpha|^2+|\beta|^2=1$ is imposed.
This gives the embedding of $SU(2)=\mathrm{S}^3$ into $\mathbb{R}^4$. 
By writing this $Z$ as $Z=e^\phi U$, we can interpret $\phi$ and $U$ as a scalar and SU(2) link variable, respectively. To get SU(2), we remove $\phi$ by taking its mass large. 

This is a special case of the standard embedding of S$^n$ into $\mathbb{R}^{n+1}$. As a counterpart of \eqref{eq:orbifold_Hamiltonian_for_strong_copuling}, we will study the following Hamiltonian consisting of $n+1$ bosons:
\begin{align}
\hat{H}
=
\frac{1}{2}\sum_{a=1}^{n+1}\hat{p}_a^2
+
\frac{m^2}{8}\left(\sum_{a=1}^{n+1}\hat{x}_a^2-1\right)^2\, . 
\label{n-sphere-model}
\end{align}
This corresponds to dropping the second and third lines of \eqref{eq:orbifold_Hamiltonian}, which is the counterpart of the strong-coupling limit of the Kogut-Susskind Hamiltonian.
Interpreting the radial coordinate as a scalar $\phi$, i.e., $\sum_{a=1}^{n+1}x_a^2=e^{2\phi}$, the second term is the scalar mass term $\frac{m^2}{2}\phi^2$. When mass is large, $\phi$ vanishes, and the low-energy states are restricted to the unit sphere S$^n$. This is the same mechanism as that for the SU($N$) embedded into $\mathbb{R}^{2N^2}$. For $n=3$, we obtain the example of the orbifold lattice discussed above. For $n=1$, we obtain U(1) embedded into $\mathbb{C}=\mathbb{R}^2$. Note that, for SU(2), this setup (four bosons per link) is likely to be more useful for quantum simulations than the original formulation (eight bosons per link) because the number of qubits and quantum gates necessary for quantum simulations can be reduced significantly. 

This setup makes the numerical test tractable with a laptop for $n=1,2$, and $3$. As we show in Supplementary Material~\ref{Supplementary_Material:singlet_simulation}, this gives a testbed for the singlet-projection protocol discussed in Sec.~\ref{sec:singlet_simulation}, too.

We take $m^2$ large and focus on the low-energy states. Then, $\phi$ behaves as a harmonic oscillator decoupled from the S$^n$ part, as in the case of scalars in the orbifold lattice Hamiltonian. The ground-state wave function is proportional to $\exp(-\frac{m\phi^2}{2})$, whose width scales as $1/\sqrt{m}$ with $m$. Namely, the sphere created by the scalar mass term has a thickness of order $\frac{1}{\sqrt{m}}$. For a good approximation of the low-energy modes, we need sufficiently many points in this width, and hence, 
\begin{align}
\delta_x=\frac{2R}{\Lambda}\lesssim\frac{1}{\sqrt{m}}
\end{align}
is needed. We can derive the same convergence criterion for the orbifold lattice, when the Kogut-Susskind limit is considered taking $m$ larger than $a^{-1}$. 
Note that this estimate holds regardless of the details of the interactions, dimensions, and lattice size. 

The numerical studies below confirm this theoretical scaling and demonstrate that the convergence is rapid and controllable in practice. 
\subsection{Ground-state wave function}
\begin{figure}[htbp]
  \includegraphics[width=8cm]{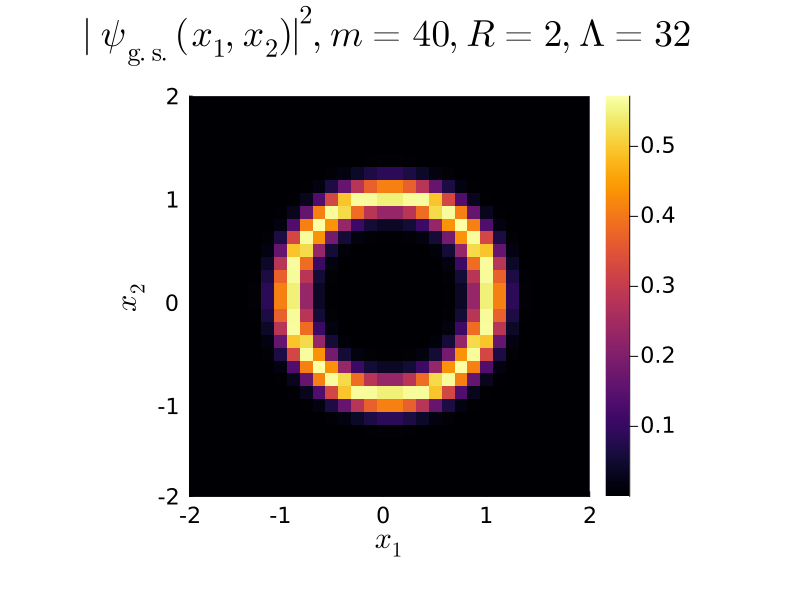}
  \includegraphics[width=8cm]{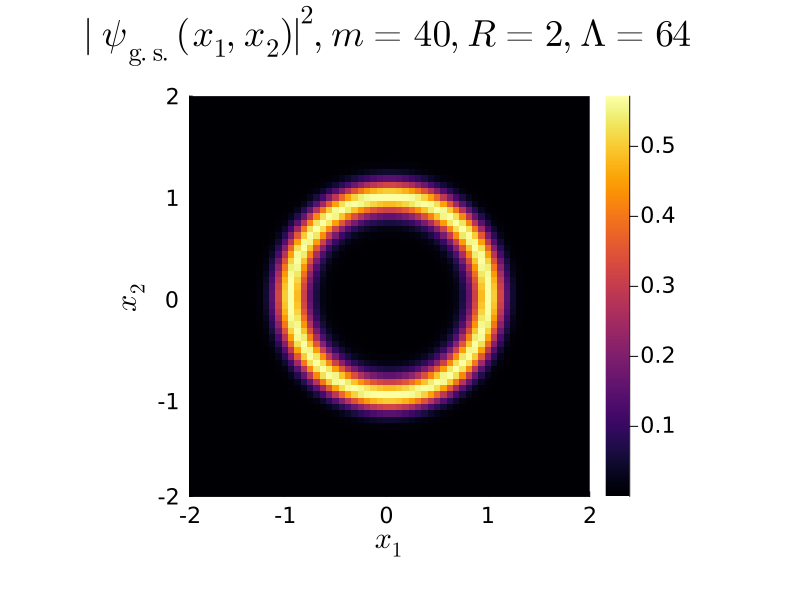}
  \caption{The square of the ground-state wave function for the $n=1$ model, $|\psi_{\rm g.s.}(x_1,x_2)|^2$. We used $m=40$, $R=2$, and $\Lambda=32,64$. 
  }
  \label{fig:ground_state_heatmap_n=1}
\end{figure}

Let $\psi_{\rm g.s.}(x_1,x_2)$ be the ground-state wave function. 
In Fig.~\ref{fig:ground_state_heatmap_n=1}, $|\psi_{\rm g.s.}(x_1,x_2)|^2$ is shown for $m=40$, $R=2$, and $\Lambda=32,64$. The wave function is normalized to match the limit of $\Lambda\to\infty$, dividing by $\delta_x^2$. The distribution is a thin ring close to S$^1$, as it should be. 

\subsection{Energy spectrum}\label{Supplementary_Material:benchmark_energy}
We study the energy spectrum varying the scalar mass $m^2$ and truncation level $\Lambda$, while fixing $R$ to be $2.0$.\\

\noindent
\underline{\textbf{$n=1$ model.}}
In the energy spectrum, we can see non-degenerate eigenvalues that correspond to zero modes along the U(1) direction and two-fold-degenerate eigenvalues that correspond to nonzero momentum along the U(1) direction. 
In Table~\ref{table:n=1_energy}, we show the first and second non-degenerate energy levels, with $R=2.0$ and $\Lambda=64$, for $m=10, 20, 30$, and $40$.
The first one is always $E_0$, as expected. We can see the approach to the harmonic oscillator values, $\frac{m}{2}$ and $\frac{3m}{2}$. 

\begin{figure}[htbp]
\begin{center}
  \includegraphics[width=8cm]{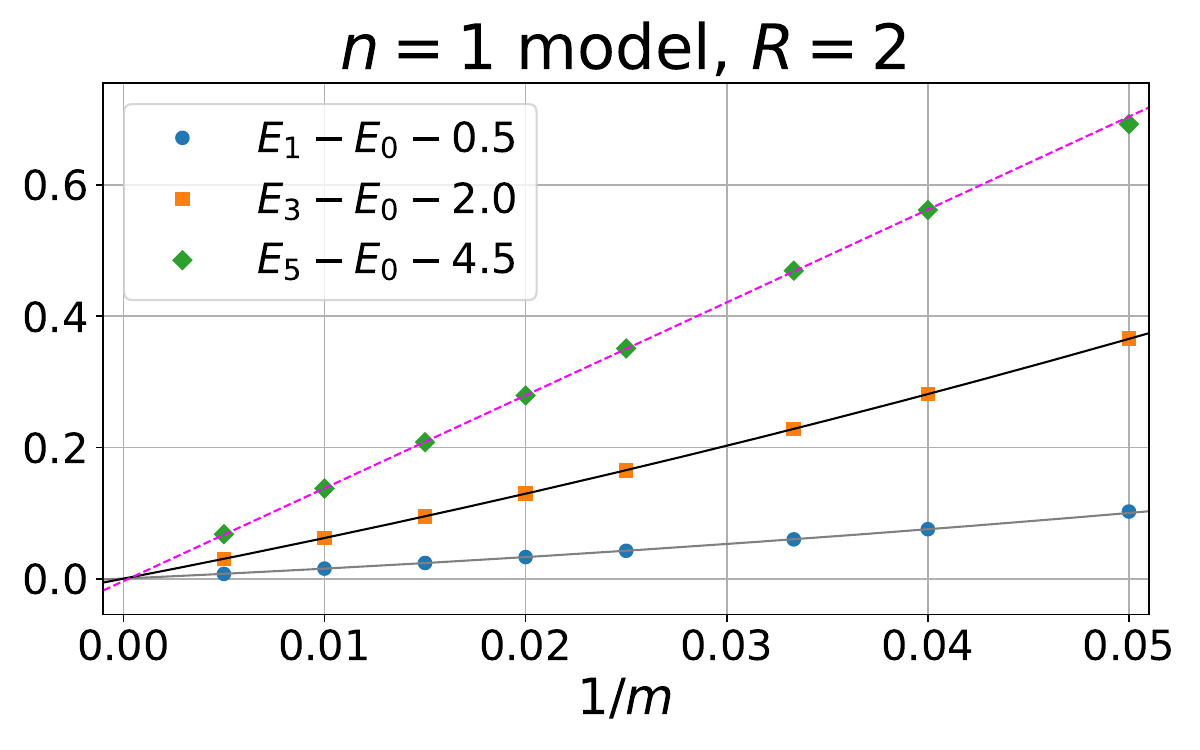}
\end{center}
  \caption{For the $n=1$ model, the difference between $E_0$ and two-fold degenerate states corresponding to momentum $p=\pm 1$, $\pm 2$, and $\pm 3$ along the U(1) direction should be $\frac{p^2}{2}$ in the infinite-mass limit. Specifically, $E_1-E_0-0.5$, $E_3-E_0-2.0$, and $E_5-E_0-4.5$ should converge to zero. We plotted these numbers for several values of $m$ between 20 and 200. The lines are quadratic fit with respect to $1/m$. We used $R=2$, and $\Lambda$ was taken sufficiently large so that the error is less than $10^{-5}$. 
  }\label{fig:mass_dependence_n=1}
\end{figure}

For sufficiently large $m$, low-lying modes correspond to the excitations along the U(1) direction. $E_1=E_2$, $E_3=E_4$, and $E_5=E_6$ correspond to momentum $p=\pm 1$, $\pm 2$, and $\pm 3$, respectively. The difference from $E_0$ should be $\frac{p^2}{2}$ in the infinite-mass limit. We demonstrated this in Fig.~\ref{fig:mass_dependence_n=1}. 

\begin{figure}[htbp]
\begin{center}
  \includegraphics[width=8cm]{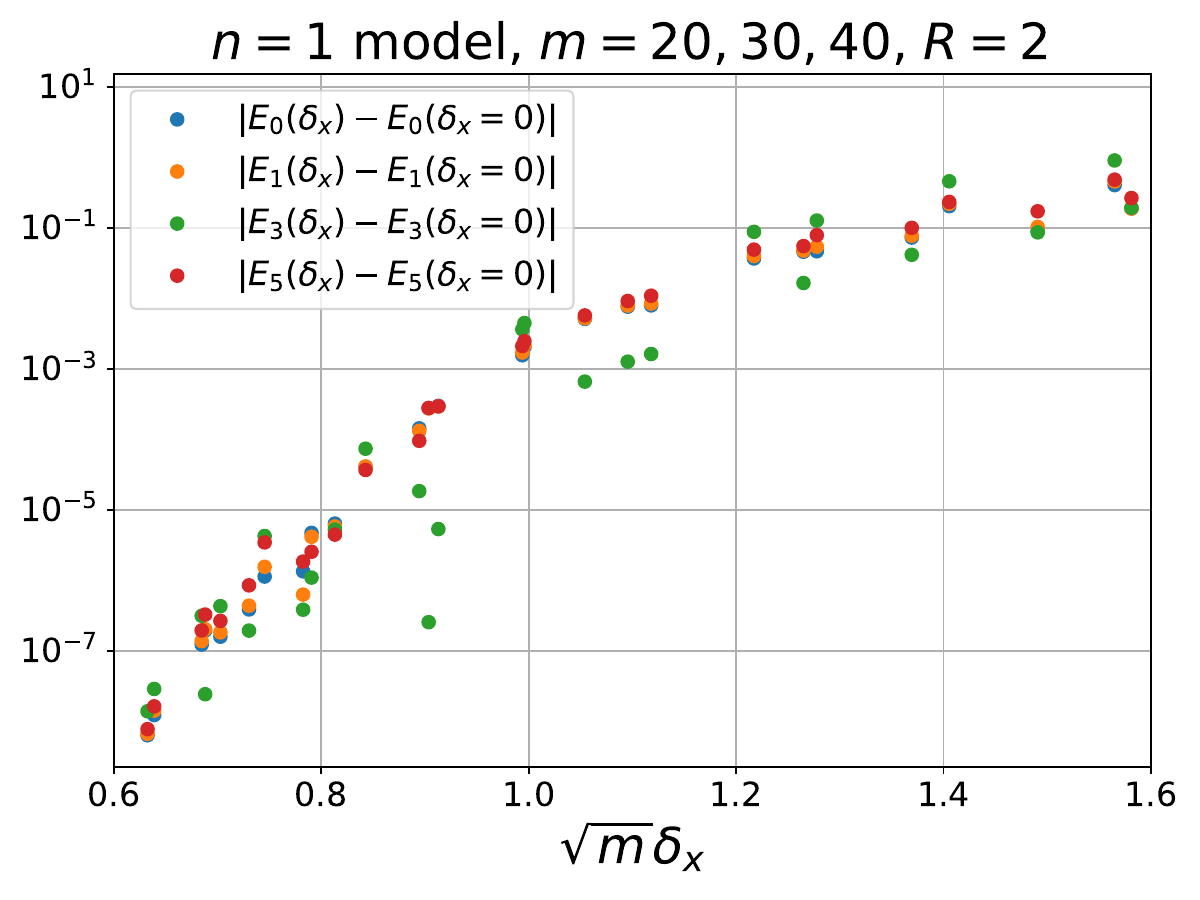}
  \end{center}
  \caption{For the $n=1$ model at $R=2$ and $m=20, 30$, and $40$, the low-lying spectrum was calculated varying $\Lambda$ (and hence $\delta_x=2R/\Lambda$), and the difference from the values at $\delta_x=0$ was plotted. The energy eigenvalues at $\delta_x=0$ were estimated using sufficiently small $\delta_x$ where so that the error is smaller than $10^{-10}$. By using $\sqrt{m}\delta_x$ as the horizontal axis, results from different mass line up and show the same exponential decay. This means we should take $\delta_x\lesssim\frac{1}{\sqrt{m}}$ to approximate the spectrum well. Note that we plotted the absolute value $|E_\ell(\delta_x)-E_\ell(\delta_x=0)|$ because the sign oscillates.   
  }\label{fig:spectrum_delta_x_dependence_n=1}
\end{figure}

In Fig.~\ref{fig:spectrum_delta_x_dependence_n=1}, we showed how the low-energy spectrum depends on $\delta_x$. We observe roughly the same amount of corrections at $\sqrt{m}\delta_x$. Therefore, as expected, we can control the corrections by taking $\delta_x\lesssim\frac{1}{\sqrt{m}}$. 
\\

\begin{table}[htbp]
\centering
\renewcommand{\arraystretch}{1.0}
\begin{tabular}{|c||c|c|}
\hline
$m$ & $E_0$ & 2nd non-degenerate level\\
\hline
\hline
10 & 4.503 & 12.437\\
\hline
20 & 9.582 & 27.712 \\
\hline
30 & 14.599 & 42.903\\
\hline
40 & 19.606 & 57.971\\
\hline
\end{tabular}
\caption{
The first and second non-degenerate energy levels, with $R=2.0$ and $\Lambda=64$, for $m=10, 20, 30$, and $40$.
The first one is always $E_0$, as expected. We can see the approach to the harmonic oscillator values, $\frac{m}{2}$ and $\frac{3m}{2}$. The $n=1$ model. 
}\label{table:n=1_energy}
\end{table}

\begin{figure}[htbp]
\begin{center}
  \includegraphics[width=8cm]{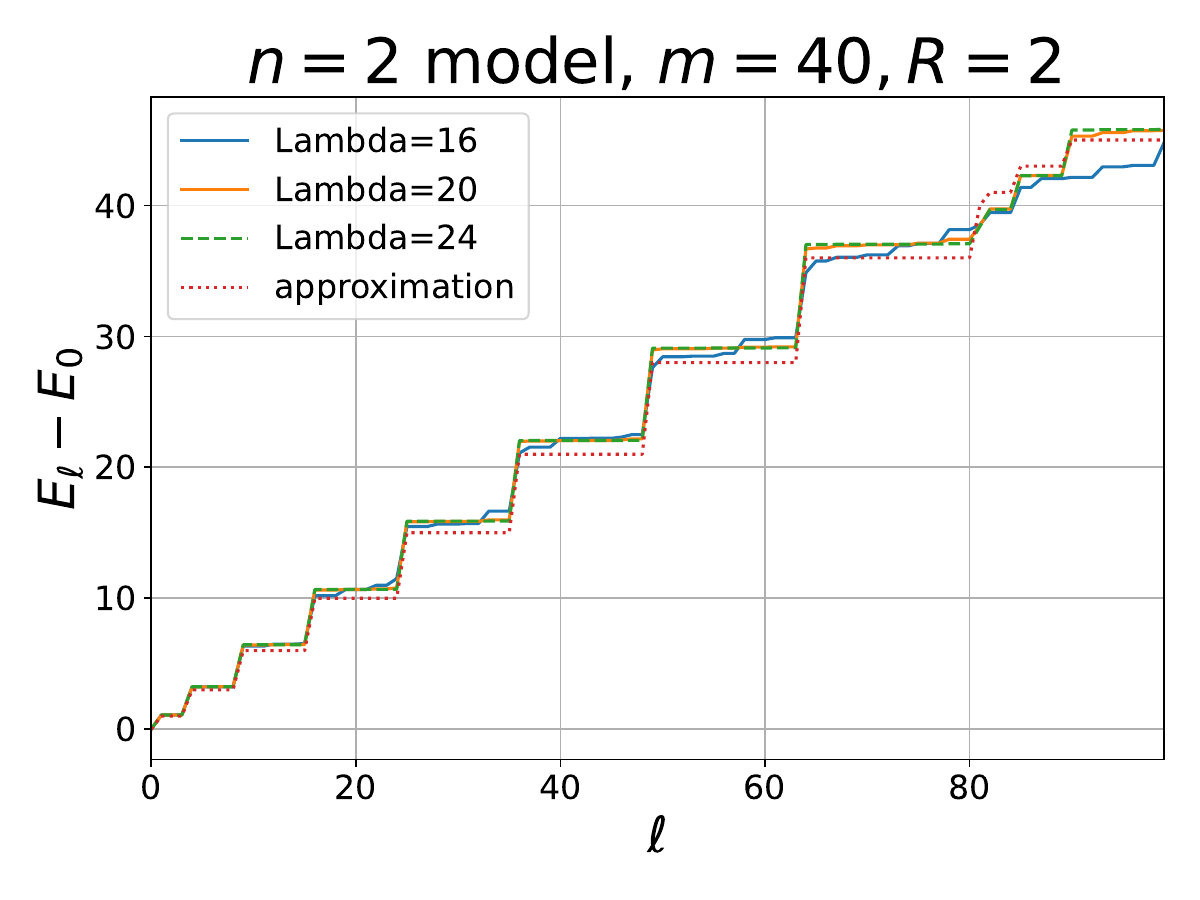}
\end{center}  
  \caption{$E_\ell-E_0$ ($0\le \ell<100$) for the $n=2$ model, $m=40$, $R=2$. An `approximation' is obtained assuming that the spectrum is a sum of the contribution from the Laplacian on S$^2$ and a harmonic oscillator describing the radial coordinate, without interaction. Such an approximation is good when $m$ is large. 
  }
  \label{fig:spectrum_n=2}
\end{figure}

\noindent
\underline{\textbf{$n=2$ model.}}
When the mass is large, the energy can be approximated by the contributions from the Laplacian on S$^2$ and the harmonic oscillator describing a small fluctuation of the scalar. The former provides $\frac{1}{2}j(j+1)$, $j=0,1,2,\cdots$, with degeneracy $2j+1$. The latter adds $\left(\frac{1}{2}+k\right)m$, $k=0,1,2,\cdots$. In Fig.~\ref{fig:spectrum_n=2}, we took $m=40$ and $R=2$, and plotted $E_\ell-E_0$ for $0\le\ell<100$, taking the truncation level $\Lambda=16, 20$, and $24$. We also show the `approximation' explained above. We can see a quick convergence as $\Lambda$ increases, and the large-$\Lambda$ result is close to the `approximation'. 

\begin{figure}[htbp]
\begin{center}
  \includegraphics[width=8cm]{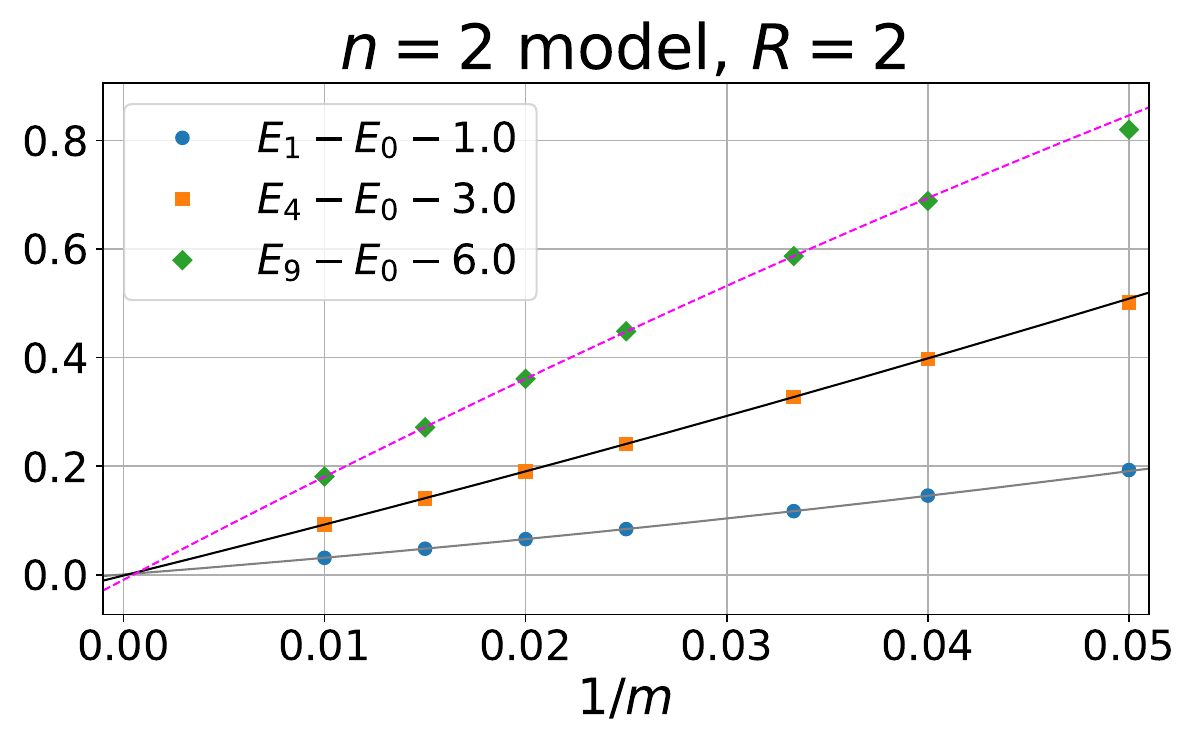}
\end{center}
  \caption{For the $n=2$ model, $E_1-E_0-1$, $E_4-E_0-3$, $E_9-E_0-6$, which should converge to zero as $m\to\infty$, are plotted. We plotted these numbers for several values of $m$ between 20 and 100. The lines are quadratic fit with respect to $1/m$. We used $R=2$, and $\Lambda$ was taken sufficiently large so that the error is smaller than $10^{-3}$.
  }\label{fig:mass_dependence_n=2}
\end{figure}

Fig.~\ref{fig:mass_dependence_n=2} shows that the Laplacian on S$^2$ is obtained precisely in the infinite-mass limit. Specifically, we showed that the differences between the first three groups of degenerate eigenvalues and $E_0$ ($E_1-E_0$, $E_4-E_0$, and $E_9-E_0$) converge to the values consistent with the values obtained from the Laplacian on S$^2$ (1.0, 3.0, and 6.0).

\begin{figure}[htbp]
\begin{center}
  \includegraphics[width=8cm]{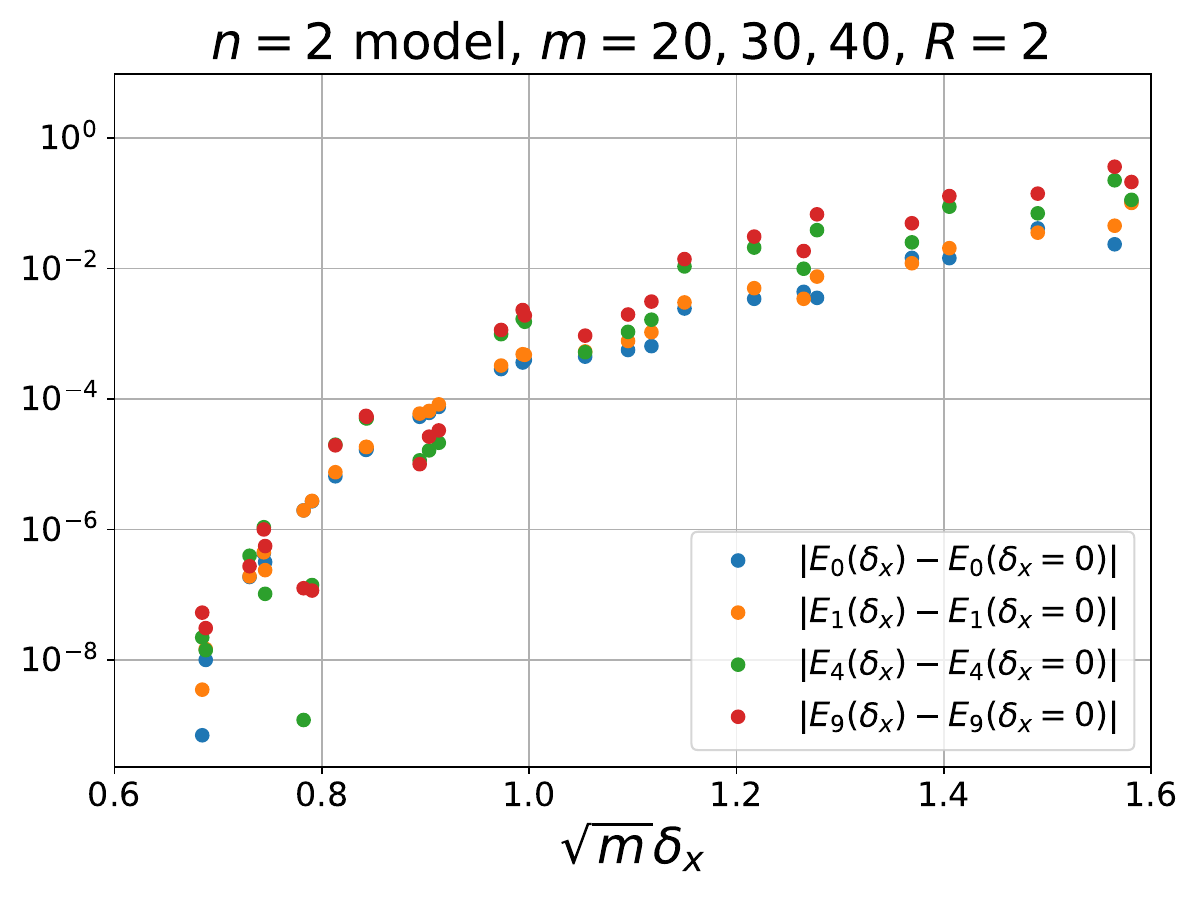}
\end{center}
  \caption{For the $n=2$ model at $R=2$ and $m=20, 30$, and $40$, the low-lying spectrum was calculated varying $\Lambda$ (and hence $\delta_x=2R/\Lambda$), and the difference from the values at $\delta_x=0$ was plotted. The energy eigenvalues at $\delta_x=0$ were estimated using sufficiently small $\delta_x$ so that the error is smaller than $10^{-10}$. By using $\sqrt{m}\delta_x$ as the horizontal axis, results from different mass line up and show the same exponential decay. This means we should take $\delta_x\lesssim\frac{1}{\sqrt{m}}$ to approximate the spectrum well. Note that we plotted the absolute value $|E_\ell(\delta_x)-E_\ell(\delta_x=0)|$ because the sign oscillates. 
}\label{fig:spectrum_delta_x_dependence_n=2}
\end{figure}

In Fig.~\ref{fig:spectrum_delta_x_dependence_n=2}, we showed how the low-energy spectrum depends on $\delta_x$. We observe roughly the same amount of corrections at $\sqrt{m}\delta_x$. Therefore, as expected, we can control the corrections by taking $\delta_x\lesssim\frac{1}{\sqrt{m}}$. 
\\

\begin{figure}[htbp]
\begin{center}
  \includegraphics[width=8cm]{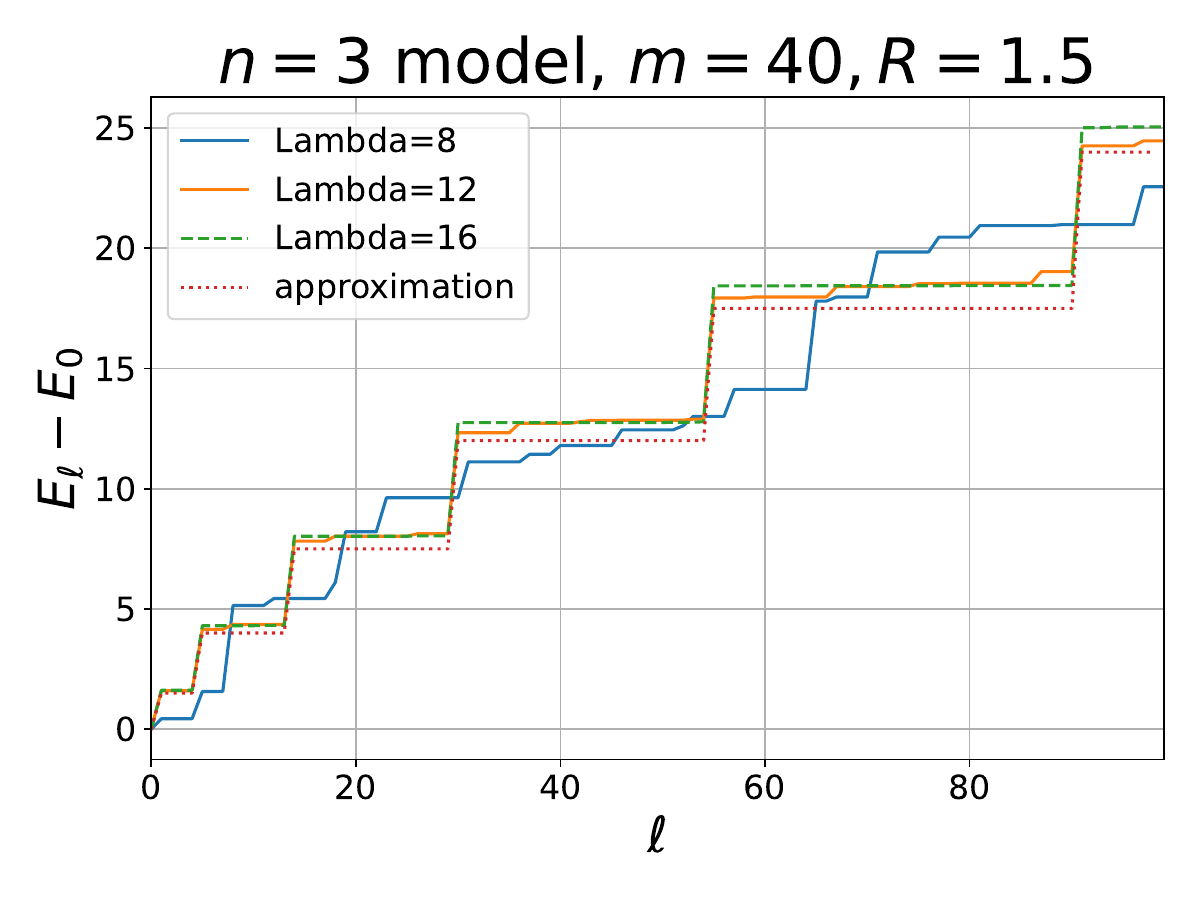}
\end{center}  
  \caption{$E_\ell-E_0$ ($0\le \ell<100$) for the $n=3$ model, $m=40$, $R=1.5$. An `approximation' is obtained assuming the spectrum to be the sum of the contribution from the Laplacian on $\mathrm{S}^3=\mathrm{SU}(2)$ and harmonic oscillator describing the radial coordinate, without interaction. Such an approximation is good when $m$ is large.
  }
  \label{fig:spectrum_n=3}
\end{figure}

\noindent
\underline{\textbf{$n=3$ model.}}
The Laplacian on $\mathrm{S}^3=\mathrm{SU}(2)$ provides $\frac{1}{2}j(j+2)$, $j=0,1,2,\cdots$, with degeneracy $(j+1)^2$. The latter adds $\left(\frac{1}{2}+k\right)m$, $k=0,1,2,\cdots$. In Fig.~\ref{fig:spectrum_n=3}, we took $m=40$ and $R=1.5$, and plotted $E_\ell-E_0$ for $0\le\ell<100$, taking the truncation level $\Lambda=8, 12$, and $16$. We also show the `approximation' explained above. Note that the excited modes of the scalar ($k\ge 1$) do not appear in this range. We can see a quick convergence as $\Lambda$ increases, and that the large-$\Lambda$ result is close to the `approximation'. The correct degeneracy pattern shows that the structure of the SU(2) group manifold is properly captured. Therefore, we conclude that the ground state serves as the state $\ket{\mathcal{G}}$, which plays the crucial role in the singlet-projection protocol introduced in Sec.~\ref{sec:singlet_simulation}, and can be systematically improved by taking $m$ and $\Lambda$ larger.

\subsection{Numerical demonstration for singlet projection}\label{sec:singlet_projection_demonstration}
Next, we study how the Hilbert space truncation affects the singlet projection. (Supplementary Material~\ref{sec:projection_numerical_test_U1} gives further details.)
A reasonable measure of the truncation effect is
\begin{align}
    \bra{Z}_{\rm proj.}
    \hat{G}^2_{\rm truncated}
    \ket{Z}_{\rm proj.}\, , 
\end{align}
where $\ket{Z}_{\rm proj.}$ is the projected version of the singlet-projected state and $\hat{G}_{\rm truncated}$ is the truncated version of the SU($N$) generators. 

Numerical evaluation of this quantity requires a large cost. As a tractable case, let us consider U$(1)$ embedded into $\mathbb{C}=\mathbb{R}^2$, and just one link.   
The Hilbert space is a truncated version of $\mathrm{Span}\{\ket{z}\, |\, z\in\mathbb{C}\}$. We take 
{$z=(x+\mathrm{i}y)/\sqrt{2}$}
, and restrict 
{$x$ and $y$}
to $\pm\delta_x/2,\pm 3\delta_x/2,\cdots,(\Lambda-1)\delta_x/2$. The dimension of the truncated Hilbert space is $\Lambda^2$. In the same way, we introduce $\xi\in\mathbb{C}$, so that $z$ is transformed to $z^{(\xi)}\equiv\xi z$. 

\begin{figure}[htbp]
\begin{center}
  \includegraphics[width=8cm]{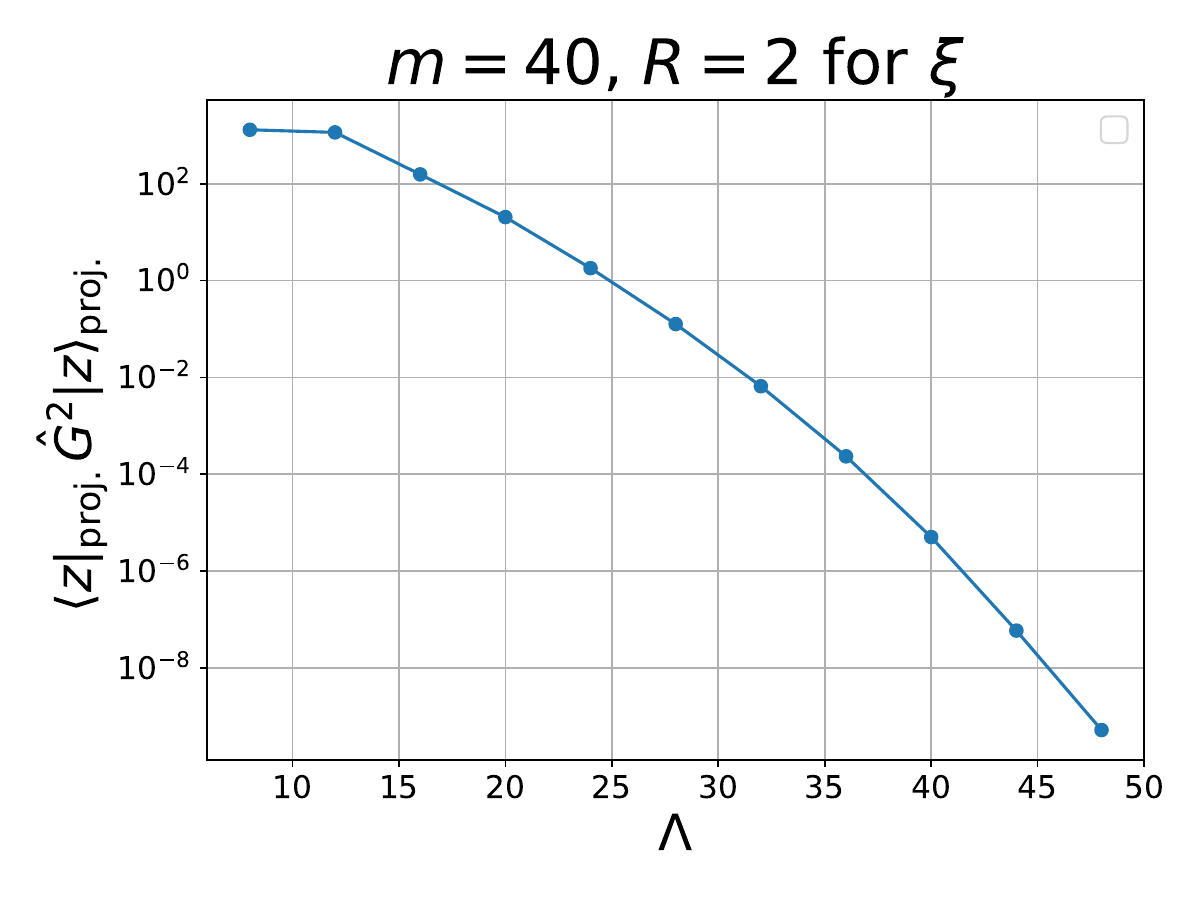}
\end{center}  
  \caption{$\bra{z}_{\rm proj.}\hat{G}^2\ket{z}_{\rm proj.}$ described in the text for $\Lambda=8,12,\cdots,48$.}
  \label{fig:G2_convergence}
\end{figure}

The U(1) generator is
\begin{align}
\hat{G}
=
\hat{x}\hat{p}_y-\hat{y}\hat{p}_x\, . 
\end{align} 
We choose $\varphi(\xi)$ in such a way that $\sum_\xi\varphi(\xi)\ket{\xi}$ gives the ground state of the Hamiltonian similar to \eqref{n-sphere-model}, specifically
\begin{align}
\hat{H}_\xi
=
\frac{1}{2}\hat{p}_{\xi^{\rm (R)}}^2
+
\frac{1}{2}\hat{p}_{\xi^{\rm (I)}}^2
+
\frac{m^2}{8}\left((\hat{\xi}^{\rm (R)})^2+(\hat{\xi}^{\rm (I)})^2-1\right)^2\, , 
\end{align}
where $\xi=\xi^{\rm (R)}+\mathrm{i}\xi^{\rm (I)}$, $[\hat{\xi}^{\rm (R)},\hat{p}_{\xi^{\rm (R)}}]=[\hat{\xi}^{\rm (I)},\hat{p}_{\xi^{\rm (I)}}]=\mathrm{i}$,
with $m=40$, $R=2$ and $\delta_x=2R/\Lambda=4/\Lambda$.
We truncate $\xi^{\rm (R)}$ and $\xi^{\rm (I)}$ such that they take the values $\pm\delta_x/2,\pm 3\delta_x/2,\cdots,(\Lambda-1)\delta_x/2$. Then, we can obtain $\sum_\xi\varphi(\xi)\ket{\xi}$ and $\ket{z}_{\rm proj.}$ numerically for a given $\ket{z}$. As a concrete test case, we took $\Lambda$ to be even, and take $x=\frac{\delta_x}{2}\left(\frac{\Lambda}{2}-1\right)=1-\frac{2}{\Lambda}$ and $y=0$. The result is shown in Fig.~\ref{fig:G2_convergence}. We can see that, as $\Lambda$ increases, $\bra{z}_{\rm proj.}\hat{G}^2\ket{z}_{\rm proj.}$ quickly approaches zero. 

The projection protocol converges exponentially with truncation level. As a result, the residual gauge-symmetry violation at $\Lambda=48$ is already far below the noise floor of current quantum hardware, so the truncation will not be the limiting factor in near-term demonstrations. This rigorous classical verification ensures the protocol will work correctly when implemented on quantum hardware at larger scales. For the complete SU($N$) projection, the same principles apply with additional complexity from the larger group structure. The U(1) case captures the essential physics of truncation effects 
and validates the Haar averaging mechanism. 

As discussed in Appendix~\ref{appendix:near-term-demonstration}, we can simplify the setup further for U(1) and $\mathbb{Z}_N$ theories, because we do not have to introduce the scalar to describe the Hilbert space in a simple manner. Refs.~\cite{Stryker:2018efp,Rajput:2021trn} considered the singlet projection in these cases. What is new in our protocol is that essentially the same method can be applied to non-Abelian gauge theories, and this is a reflection of the universal structure of the orbifold lattice. 

\subsection{Numerical test for Trotter decomposition}\label{sec:Trotter_error_numerical_test}
In Supplementary Material~\ref{sec:universal_framework}, the simulation cost for Hamiltonian time evolution via Trotter decomposition is estimated using standard operator-norm bounds on the error. While rigorous, this approach is conservative: the operator norm $\|\hat{H}\|$ increases with the truncation level $\Lambda$, potentially suggesting that higher cutoffs worsen simulation efficiency. However, our interest lies in \textit{low-energy physics} --- the relevant energy scales are independent of $\Lambda$, and we evolve only linear combinations of low-energy modes. In this regime, the Trotter step size should be determined by \textit{physical} energy scales --- the lattice spacing, gauge coupling, and in our setup, the scalar mass $m$ that sets the effective cutoff --- rather than by the formal UV cutoff $\Lambda$. 

\begin{figure}[htbp]
\begin{center}
  \includegraphics[width=8cm]{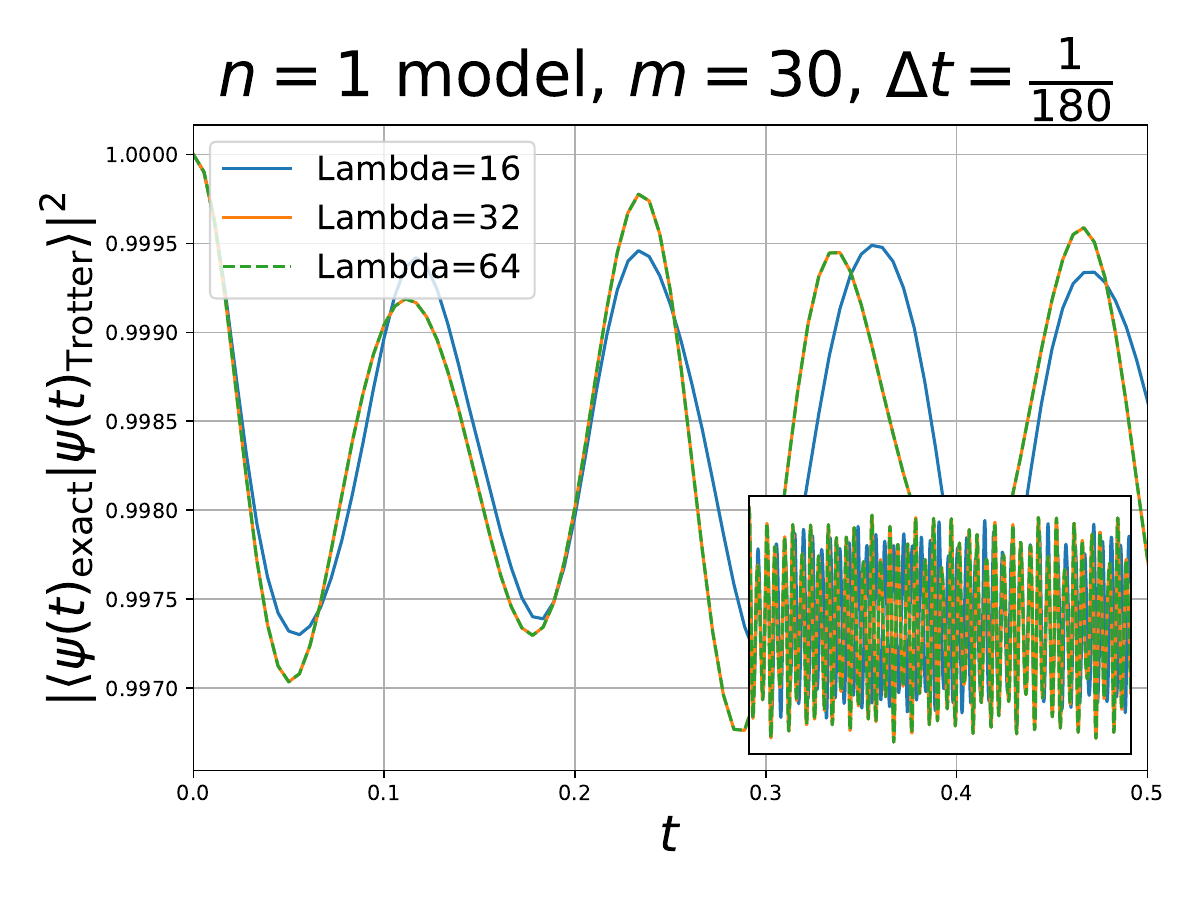}
\end{center}  
  \caption{Fidelity $|\bra{\psi(t)_{\rm exact}}\ket{\psi(t)_{\rm Trotter}}|^2$, $n=1$, mass $m=30$, step size $\Delta t=\frac{1}{180}$. We took the initial state to be the ground state of the truncated Hamiltonian. The results from $\Lambda=32$ and $\Lambda=64$ are indistinguishable. The inset is a plot from $t=0$ to $t=5$.}
  \label{fig:Fidelity_n=1_1}
\end{figure}

\begin{figure}[htbp]
\begin{center}
  \includegraphics[width=8cm]{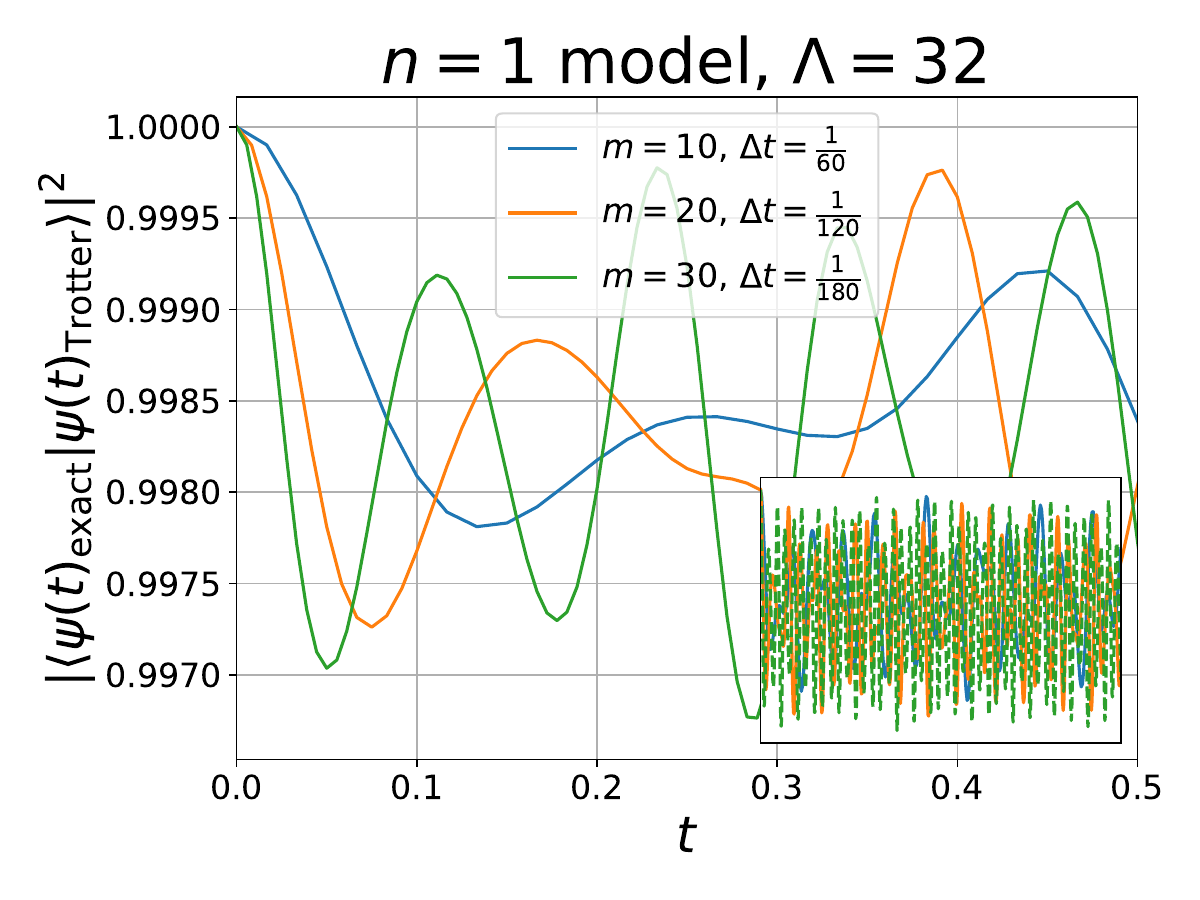}
\end{center}
  \caption{Fidelity $|\bra{\psi(t)_{\rm exact}}\ket{\psi(t)_{\rm Trotter}}|^2$, $n=1$,  $\Lambda=32$, and $m\Delta t=\frac{1}{6}$. We took the initial state to be the ground state of the truncated Hamiltonian. We can see that, when $m\Delta t$ is fixed, fidelity does not change with $m$ (note the scale, the fidelity always stays very close to 1).
  Note that the ground state energy $E_0$ is proportional to $m$ up to the truncation errors. The inset is a plot from $t=0$ to $t=5$. 
  }
  \label{fig:Fidelity_n=1_2}
\end{figure}

\begin{figure}[htbp]
\begin{center}
  \includegraphics[width=8cm]{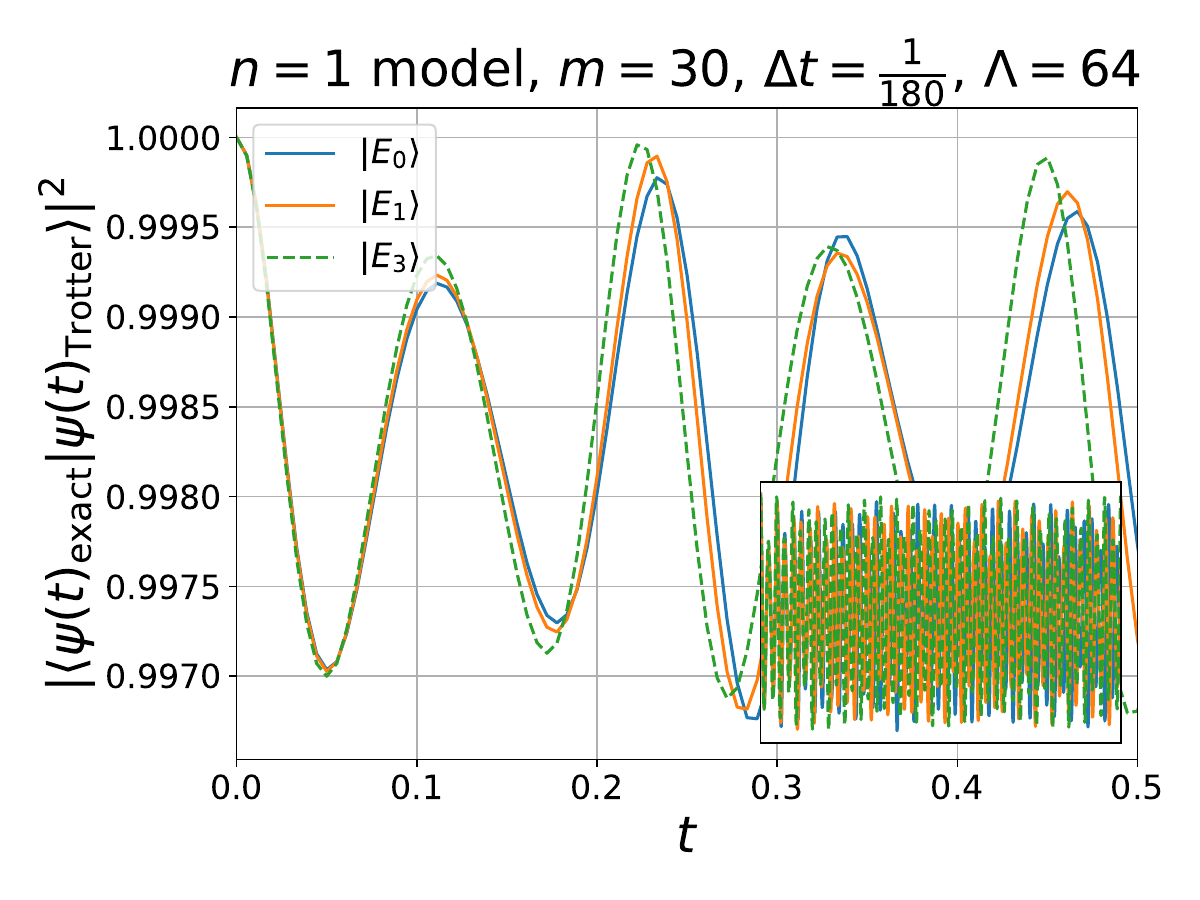}
\end{center}  
  \caption{Fidelity $|\bra{\psi(t)_{\rm exact}}\ket{\psi(t)_{\rm Trotter}}|^2$, $n=1$, mass $m=30$, truncation level $\Lambda=64$, step size $\Delta t=\frac{1}{180}$. We took the initial state to be the ground state $\ket{E_0}$, the first excited mode $\ket{E_1}$, and the third excited mode $\ket{E_3}$. The inset is a plot from $t=0$ to $t=5$. 
  }
  \label{fig:Fidelity_n=1_3}
\end{figure}

For the $n=1$ model, we studied the fidelity $|\bra{\psi(t)_{\rm exact}}\ket{\psi(t)_{\rm Trotter}}|^2$; see Fig.~\ref{fig:Fidelity_n=1_1}, Fig.~\ref{fig:Fidelity_n=1_2}, and Fig.~\ref{fig:Fidelity_n=1_3}. 
From Fig.~\ref{fig:Fidelity_n=1_1}, we can see that the divergence of the largest energy eigenvalue as $\Lambda\to\infty$ does not affect the necessary Trotter step size for low-energy modes, as expected. On the other hand, the energy of the low-lying modes increases with $m$, and hence, we have to use a smaller Trotter step for larger $m$. In Fig.~\ref{fig:Fidelity_n=1_2}, we can see that the error remains approximately the same when $m\Delta t$ is fixed. Fig.~\ref{fig:Fidelity_n=1_3} shows that the same step size can be used for the first few low-lying modes; this is expected because the energy is dominated by the scalar part, which is common for all low-lying modes. 

These results confirm the expectation stated at the beginning of this subsection: Trotter step sizes are governed by physical energy scales, not by the formal UV cutoff, ensuring that simulation efficiency is maintained even as $\Lambda$ increases to improve truncation accuracy.

\section{Resource estimates }\label{sec:outlook}

\subsection{Resource scaling}
Building on the orbifold lattice circuits of refs.~\cite{Bergner:2024qjl,Halimeh:2024bth,Bergner:2025zkj,Halimeh:2025ivn}, we now provided a complete framework for gauge symmetry treatment in quantum simulation: singlet projection protocol, universal principles for choosing singlet and non-singlet formulations, and convergence validation. Our framework enables concrete resource estimates --- qubit counts, gate counts, circuit depths, and compilation costs --- for Yang-Mills theory and QCD on fault-tolerant quantum computers.\\ 

\noindent
\textbf{Qubit counting:} 
Assigning $Q$ qubits for each boson, each link variable requires $2N^2Q$ qubits. For $(d+1)$-dimensional Yang-Mills theory on $V$ lattice points, there are $dV$ links, and hence, $2dN^2VQ$ qubits are needed. 

For SU(2), this number can be cut down to $4Q$ per boson using the embedding into $\mathbb{R}^4$. For two spatial dimensions and $Q=4$ (which corresponds to $\Lambda=2^Q=16$, which is promising based on the spectrum shown in Fig.~\ref{fig:spectrum_n=3}), we need $32V$ qubits, which is 512 for a $4^2$ lattice and 3200 for a $10^2$ lattice. For three spatial dimensions and $Q=4$, we need $48V$ qubits, which is 3072 qubits for a $4^3$ lattice and 48000 qubits for a $10^3$ lattice. 

When Creutz~\cite{Creutz:1980zw} performed the first lattice simulation of the SU(2) Yang-Mills theory that motivated many followers, he used very small lattice sizes $4^4$ to $10^4$ which resulted in large systematic errors which one could overcome only later. For quantum computing, if a fault-tolerant digital quantum computer with a few thousands of logical qubits becomes available, it will provide a much more solid basis and, therefore, should create even more excitement. Major quantum computer companies expect to produce machines with $O(10^4)$ logical qubits in the mid-2030s, which would allow for physics-wise relevant simulations in 3+1 dimensions, which is only about 10 years from the present.

This qubit counting does not change significantly even if the singlet-projection protocol introduced in Sec.~\ref{sec:singlet_simulation} is used. It would be reasonable to assign the same $Q$ to $\ket{\xi_{\vec{n}}}$ and $\ket{Z_{j,\vec{n}}}$, though it is possible to use finer mesh for $\xi$. Because gauge transformation is defined locally, we can perform the SU($N$)-averaging either simultaneously over the entire lattice, or one-by-one at each lattice point. (Of course, it is possible to perform averaging simultaneously at more than one, but not all, lattice point.) For the former approach, we need $2N^2VQ$ qubits to describe $\ket{\xi_{\vec{n}}}$ for all points $\vec{n}$. For the latter, we can use the same set of $\ket{\xi}$ at each time, and hence, we need to add only $2N^2Q$ qubits. Either way, this is at most the same order as the qubit requirement for the complex link fields (total $2dN^2VQ$ for $d$ spatial dimensions). The same argument holds when we use ancilla states $\{\ket{Z}_2\}$, too. 

Practically, we would not need the projection over the entire lattice; when we act a local non-singlet operator to a gauge-invariant state, we need only a local projection procedure to restore the gauge invariance. \\

\noindent
\textbf{Gate complexity:} For the Hamiltonian time evolution, the simulation cost scales polynomially with $Q$. If we do not explicitly remove the U(1) part (which is not an issue for Yang-Mills theory because U(1) and SU($N$) sectors decouple), each Trotter step requires $O(N^2Q^2V)$ gates for Fourier transforms and $\mathcal{O}(N^4Q^4V)$ gates for quartic interactions, giving a total gate count per step of $\mathcal{O}(N^4Q^4V)$. Circuit depth scales as $\mathcal{O}(N^4Q^4)$ when parallelization is optimized. For evolution time $t$ with target accuracy $\epsilon$, the number of steps in the second-order Trotterization is bounded by $\mathcal{O}(t^{3/2}E^{3/2}/\epsilon^{1/2})$ where $E$ is the characteristic energy scale (which is not the truncation level); see Supplementary Material~\ref{sec:universal_framework}.

The addition of projections to the circuit does not affect this scaling if we avoid excessive, unnecessary use of the projections. The cost for the post-selection in the projection can be significant, and if we perform projection to the entire lattice, exponential increase of the post-selection cost looks inevitable. See Sec.~\ref{sec:when_to_project} to see when we should use the singlet projection. \\

\noindent
\textbf{Circuit design:}
Our simulation protocols are simple enough so that the quantum circuits can be designed mostly by hand~\cite{Halimeh:2024bth}. The only part that requires a compiler is the explicit construction of one-qubit rotations from one-qubit gates, whose cost is negligible. (Of course, further optimization could be achieved by using a compiler.) This is in stark contrast with other approaches; for example, if we use a naive gauge-invariant approach, the compilation cost could increase exponentially with the total number of qubits~\cite{Hanada:2025yzx}.  
\subsection{Validation of convergence criteria}
We validated our framework on S$^n$ benchmark systems (Sec.~\ref{sec:numerical_test}) that reproduce the essential physics of scalar mass in the orbifold formulation, including the strong-coupling limit of SU(2) theory. Key findings include:
\begin{itemize}
    \item 
    Truncation requirement: $\delta_x\lesssim\frac{1}{\sqrt{m}}$ 
    
    \item 
    Trotter step scaling: Determined by physical energy scales such as $m$, not the cutoff $\Lambda$ or $\delta_x$ 
     \item 
     Convergence rate: Rapid as $\Lambda$ increases for low-energy states 

\end{itemize}

For validating convergence with full gauge interactions, a classical sampling algorithm~\cite{Hanada:2022pps} can be employed --- yet another advantage of non-compact variables. While such validation could provide even more precise resource estimates before fault-tolerant quantum computers arrive, it is computationally expensive and fundamentally unnecessary: our rigorous error bounds already guarantee convergence.
\section{Outlook}
Our framework is complete and ready for implementation. The truncated Hamiltonian and efficient quantum circuits are given explicitly, using analytic methods. This enables us to pursue various research directions. The remaining work is optimization and experimental implementation, not 
algorithmic development. Specifically, immediate next steps are circuit optimization for specific architectures and specific physical process, NISQ-era demonstrations on small lattices, and classical simulation of noisy circuits. For demonstration purposes, we could try SU(2) or U(1) theory on a very small lattice --- even one or two links --- that is already a perfect setup to demonstrate the singlet-projection protocol --- which, we stress, targets the \emph{local} projection subroutine identified in Sec.~\ref{sec:when_to_project}, i.e., the one place where the projection is actually used in a large-scale simulation, and is not a step toward globally constrained simulations; see Appendix~\ref{appendix:near-term-demonstration}. Note that our protocol is straightforward and scalable given more logical qubits; even a very small-scale demonstration does not involve a simplification for the sake of demonstration that spoils the scalability.

For experimental implementation, we could consider the scattering of glueballs or hadrons. To construct a wave packet of gluons, we can take an appropriate linear combination of plaquettes, whose Pauli-string expansion can be determined analytically using only Pauli $\sigma_z$ gates. Implementation and cost-estimate of such an operator are straightforward when using block encoding via LCU. Other important problems include ground-state preparation, for which the simplicity of the coordinate basis in the orbifold lattice may play a prominent role. 

Concerning experiments on real quantum devices, the important point is scalability: we must study something relevant for future simulations on fault-tolerant devices and \textit{not} study a toy model crafted for the sake of small-scale experiments. As we have seen above, $(2+1)$-dimensional SU(2) Yang-Mills theory on a $4^2$ lattice requires 512 qubits. To make quantum simulations tractable, we should also simplify the circuits. For this purpose, it is important to notice that a few terms in the orbifold lattice Hamiltonian can be dropped without affecting the infinite-mass limit~\cite{Bergner:2024qjl}. By dropping a few terms, the number of gates needed for one Trotter step can be reduced, while the number of qubits needed for sufficiently precise descriptions may or may not change. The point we really want to make in this paper is that the combination of the orbifold lattice formulation, the rapid increase in quantum computer resources  and the various optimization steps we and others suggest could imply that practical quantum computing for lattice gauge theory is substantially closer than one might fear. 

The quantum simulation of non-Abelian gauge theories has advanced from theoretical possibility to practical readiness. Our framework provides everything needed for implementation: explicit quantum circuits with polynomial scaling, protocols for 
gauge-singlet projection when desired, validated convergence criteria, and complete resource estimates. By working in an extended Hilbert space and using the orbifold lattice formulation, we avoid the exponential classical preprocessing that plagues alternative approaches. The universal principles we establish --- clarifying when and why singlet and non-singlet formulations work --- apply beyond our specific implementation to any quantum simulation approach. Our numerical validations demonstrate that truncation errors are controllable and that Trotter step sizes are determined by physical energy scales rather than formal cutoffs, confirming practical efficiency. For SU(2) Yang-Mills theory in 2+1 dimensions, the number of logical qubits needed for small lattices aligns with quantum computing roadmaps for the mid-2030s. For demonstration purposes, we could try SU(2) or U(1) theory on a very small lattice, even one or two links; see Appendix~\ref{appendix:near-term-demonstration} for concrete qubit requirements and scalability analysis.

The path from current algorithms to future quantum advantage in gauge theory simulation is now clear and concrete. What remains is not algorithmic development, but hardware maturation and experimental optimization --- the transition from if to when.
\section*{Acknowledgement}
We thank Masaki Tezuka for discussions and support for some computations in Sec.~\ref{sec:numerical_test}. We thank Georg Bergner, Yuta Kikuchi, Lento Nagano, Shinsuke Nishigaki, Enrico Rinaldi, and Yuki Sato for discussions and comments. 
M.~H.~thanks the STFC for the support through the consolidated grant ST/Z001072/1. Sec.~\ref{sec:H_ext-and-H_inv} and Sec.~\ref{sec:nonsinglet_simulations} are largely based on materials prepared for M.~H.'s talk at ``Quantum Simulation of Strong Interactions (QuaSI) Workshop 1 : Theoretical Strategies for Gauge Theories" at InQubator for Quantum Simulation, the University of Washington, in April 2021. They were not presented in the main sessions due to time limitations, but discussions with the participants were certainly useful. 

\appendix
\section{Review of Kogut-Susskind formulation}\label{sec:Kogut_Susskind}

The Kogut-Susskind Hamiltonian~\cite{Kogut:1974ag} (see e.g., ref.~\cite{Zohar:2015hwa} for a review) is a Hamiltonian counterpart of Wilson's lattice action~\cite{Wilson:1974sk}. A crucial point is to introduce a lattice regularization that preserves important symmetries of the continuum theory in such a way that the radiative corrections are under control. At the lattice level, the Kogut-Susskind Hamiltonian has discrete translation (translation by a lattice unit), 90-degree rotations, parity, charge conjugation, and gauge symmetry. 
\subsection{Operators and commutation relations}
To realize the lattice version of gauge symmetry, unitary link variables that correspond to $\exp(\mathrm{i}agA_j)$ are used. Here, $a$ and $g$ are the lattice spacing and coupling constant, respectively. The link variable $U_{j,\vec{n}}$ lives on a link connecting a point $\vec{n}$ and the neighboring point $\vec{n}+\hat{j}$, and the gauge transformation is defined by $U_{j,\vec{n}}\to\Omega_{\vec{n}}U_{j,\vec{n}}\Omega^{-1}_{\vec{n}+\hat{j}}$. As mentioned above, the electric field $E_j$ is the conjugate momentum of $A_j$. Therefore, we want something like $[\hat{E}_{j,\vec{n}},\hat{A}_{j',\vec{n}'}]=-\mathrm{i}\delta_{jj'}\delta_{\vec{n}\vec{n}'}/a^3$. Note that $\delta_{\vec{n}\vec{n}'}/a^3$ is the lattice version of delta function. 

Let $\tau_\alpha$ be the generator of the group ($\alpha=1,\cdots,N^2$ for U($N$) and $\alpha=1,\cdots,N^2-1$ for SU($N$)), which is normalized as $\mathrm{Tr}(\tau_\alpha\tau_\beta)=\delta_{\alpha\beta}$. We write an $N\times N$ matrix $E_j$ as $E_j=\sum_\alpha E_j^\alpha\tau_\alpha$. Then, $U_{j,\vec{n}}=\exp(\mathrm{i}agA_{j,\vec{n}}^\alpha\tau_\alpha)$, and hence, as an analog of $[\hat{E}_{j,\vec{n}},\hat{A}_{j',\vec{n}'}]=-\mathrm{i}\delta_{jj'}\delta_{\vec{n}\vec{n}'}/a^3$ we can take the commutation relations 
\begin{eqnarray}
\left[
\hat{E}_{j,\vec{n}}^\alpha,
\hat{U}_{k,\vec{n}'}^{pq}
\right]
=
a^{-2}g\delta_{jk}\delta_{\vec{n}\vec{n}'}(\tau_\alpha\hat{U}_{k,\vec{n}'})^{pq}\, , 
\end{eqnarray}
\begin{eqnarray}
\left[
\hat{E}_{j,\vec{n}}^\alpha,
\hat{U}^{\dagger pq}_{k,\vec{n}'}
\right]
=
-a^{-2}g\delta_{jk}\delta_{\vec{n}\vec{n}'}(\hat{U}_{k,\vec{n}'}^\dagger\tau_\alpha)^{pq}\, .   
\end{eqnarray}
From the Jacobi identity, we also have
\begin{eqnarray}
\left[
\hat{E}_{j,\vec{n}}^\alpha,
\hat{E}_{k,\vec{n}'}^\beta
\right]
=
-\mathrm{i}a^{-2}g\sum_\gamma f^{\alpha\beta\gamma}\delta_{jk}\delta_{\vec{n}\vec{n}'}\hat{E}^\gamma_{k,\vec{n}'}\, ,
\end{eqnarray}
where $f^{\alpha\beta\gamma}$ is the structure constant:
\begin{align}
[\tau_\alpha,\tau_\beta]=\mathrm{i}\sum_\gamma f_{\alpha\beta\gamma}\tau_\gamma\, . 
\end{align}

Here, we defined the electric field operator $\hat{E}_j$ using an infinitesimal transformation acting from the left of the link variable. We can also use an infinitesimal transformation acting from the right of the link variable, but as we will see later, using only the former is enough. 

We can choose the generators in such a way that 
\begin{eqnarray}
\sum_{\alpha=1}^{N^2}\tau^{\alpha}_{pq}\tau^{\alpha}_{rs}
=
\delta_{ps}\delta_{qr}\, , 
\qquad
\sum_{\alpha=1}^{N^2}\left(\tau^{\alpha}\tau^\alpha\right)_{pq}=N\delta_{pq} 
\end{eqnarray}
are satisfied for U($N$) and 
\begin{align}
&
\sum_\alpha\tau_\alpha^{pq}\tau_\alpha^{rs}=\delta^{ps}\delta^{qr}-\frac{\delta^{pq}\delta^{rs}}{N}\, , 
\nonumber\\
&
\sum_{\alpha=1}^{N^2}\left(\tau^{\alpha}\tau^\alpha\right)_{pq}=\left(1-\frac{1}{N^2}\right)N\delta_{pq}
\end{align}
are satisfied for SU($N$). 

Unlike $\hat{E}$'s, $\hat{U}$'s commute with each other:
\begin{eqnarray}
\left[
\hat{U},
\hat{U}
\right]
=
\left[
\hat{U},
\hat{U}^\dagger
\right]
=
\left[
\hat{U}^\dagger,
\hat{U}^\dagger
\right]
=
0\, .   
\end{eqnarray}
\subsection{Kogut-Susskind Hamiltonian}
The Hamiltonian consists of electric and magnetic terms, 
\begin{eqnarray}
\hat{H}
=
\hat{H}_{\rm E}
+
\hat{H}_{\rm B}\, .  
\end{eqnarray}
The electric part is
\begin{eqnarray}
\hat{H}_{\rm E}
=
\frac{a^3}{2}
\sum_{\vec{n}}\sum_{j=1}^3\sum_{\alpha=1}^{N^2}\left(\hat{E}_{j,\vec{n}}^{\alpha}\right)^2\, . 
\end{eqnarray}
The magnetic part is 
\begin{eqnarray}
\hat{H}_{\rm B}
=
\frac{1}{2ag^2}
\sum_{\vec{n}}\sum_{j\neq k}
\mathrm{Tr}\left(
\textbf{1}
-
\hat{U}_{j,\vec{n}}
\hat{U}_{k,\vec{n}+\hat{j}}
\hat{U}^\dagger_{j,\vec{n}+\hat{k}}
\hat{U}^\dagger_{k,\vec{n}}
\right)\, . 
\nonumber\\
\end{eqnarray}
The magnetic term (plaquette) is dropped in the strong coupling limit, $g^2\to\infty$~\cite{Susskind:1979up}. Note that this ``strong coupling limit" is different from the continuum limit, which is obtained by scaling the coupling constant and lattice spacing to zero properly.  

The ground state of this strong-coupling lattice gauge theory Hamiltonian $|{\rm g.s.}\rangle$ satisfies $\hat{E}_{j,\vec{n}}^\alpha|{\rm g.s.}\rangle = 0$ for any $\alpha,j$ and $\vec{n}$. Hence, let us use the notation $|E=0\rangle$ to denote the ground state.

The operator $\hat{U}_{j,\vec{n}}$ is interpreted as the coordinate of the group manifold $G=\mathrm{U}(N)$ or SU($N$)  for the link variable on the site $\vec{n}$ in the $j$-direction. The operators $\hat{U}$ and $\hat{E}$ are defined on the extended Hilbert space $\mathcal{H}_{\rm ext}$ that contains gauge non-singlet modes. 
A convenient basis of $\mathcal{H}_{\rm ext}$ is the coordinate representation, 
\begin{eqnarray}
{\cal H}_{\rm ext}
=
\otimes_{j,\vec{n}}{\cal H}_{j,\vec{n}}
\sim
\otimes_{j,\vec{n}}
\left(
\oplus_{g\in G}
|g\rangle_{j,\vec{n}}
\right)\, ,
\end{eqnarray}
 where
 \begin{eqnarray}
\hat{U}_{j,\vec{n}}
|g\rangle_{j,\vec{n}}
=
g|g\rangle_{j,\vec{n}}
\qquad 
g\in G
\end{eqnarray}
More precisely, we will consider only the Hilbert space of square-integrable wave functions. The ground state is the constant mode, 
\begin{align}
&
\ket{E=0}
=
\otimes_{\mu,\vec{n}}
\ket{E=0}_{j,\vec{n}}\, , 
\nonumber\\
&
\ket{E=0}_{j,\vec{n}}
=
\frac{1}{\sqrt{{\rm vol}G}}\int_{G} dg
\ket{g}_{j,\vec{n}}\, . 
\end{align}
Namely, the wave function $\langle g|E=0\rangle$ is constant. 

Let $G_{\vec{n}}$ be the group $G$ living on a point $\vec{n}$, and $\mathcal{G}=\prod_{\vec{n}}G_{\vec{n}}$ be the group of all local gauge transformations. 
Gauge transformation by $\hat{\Omega}=\otimes_{\vec{n}}\hat{\Omega}_{\vec{n}}$ is defined by 
\begin{align}
\hat{\Omega}
\left(
\otimes_{j,\vec{n}}
\ket{g}_{j,\vec{n}}
\right)
=
\otimes_{j,\vec{n}}
\left(
\hat{\Omega}_{\vec{n}}|g\rangle_{j,\vec{n}}
\right)
=
\otimes_{j,\vec{n}}
\ket{\Omega_{\vec{n}}g\Omega_{\vec{n}+\hat{j}}^{-1}}_{j,\vec{n}}. 
\end{align}
Note that the strong-coupling ground state is gauge-invariant:
\begin{align}
\hat{\Omega}|E=0\rangle= \ket{E=0}\, .  
\end{align}
We can define the projection operator to the gauge-invariant Hilbert space $\mathcal{H}_{\rm inv}$ as
\begin{align}
\hat{\mathcal{P}}
=
\frac{1}{{\rm Vol}\mathcal{G}}\int_{\mathcal{G}}d\Omega \hat{\Omega}, 
\end{align}
where the integral is taken by using the Haar measure. By using this projection operator, the canonical partition function at temperature $T$ can be written in two ways as~\cite{Hanada:2020uvt}
\begin{align}
Z(T)
=
{\rm Tr}_{\mathcal{H}_{\rm inv}}\left(e^{-\hat{H}/T}\right)
\end{align}
and
\begin{align}
Z(T)
=
{\rm Tr}_{\mathcal{H}_{\rm ext}}\left(\hat{\mathcal{P}}e^{-\hat{H}/T}\right). 
\label{YM_partition_function}
\end{align}
The latter expression is directly related to the path-integral formalism with gauge field $A_t$, as shown in Supplementary Material~\ref{sec:path-integral-and-Hamiltonian}. The insertion of the projector $\hat{\mathcal{P}}$ corresponds to the integration of $A_t$, and it means that we should identify the states connected by a gauge transformation. This does not mean we must use gauge singlet; in the path-integral formulation, field configurations related by gauge transformations give the same expectation values for gauge-invariant observables, and corresponding to this, quantum states related by gauge transformations give the same expectation values for gauge-invariant operators in the operator formalism. 
\subsection{Strings and interactions}\label{sec:string_interaction}
Below, we consider U($N$) gauge theory, because the expressions are slightly simpler than for the case of SU($N$). 

A closed string is created by the Wilson loop 
\begin{align}
\hat{W}={\rm Tr}(\hat{U}_{j,\hat{n}}\hat{U}_{k,\hat{n}+\hat{j}}\cdots)\, , 
\end{align}
where the product of $\hat{U}$'s in the trace is taken along a closed path. This is a counterpart of \eqref{sec;closed_string_singlet} in the Kogut-Susskind formulation. 
Suppose the loop does not have self-intersection, in the sense that no link appears more than once. Then, $\hat{W}\ket{E=0}$ is an energy eigenstate, and the energy is proportional to the length $La$ (equivalently, the number of links in the Wilson loop $L$), as we can see as follows. First, because $\ket{E=0}$ is annihilated by $\hat{E}$, we have
\begin{align}
&
\sum_\alpha\left(\hat{E}_{j,\vec{n}}^{\alpha}\right)^2\hat{W}\ket{E=0}
\nonumber\\
&=
\sum_\alpha\left[
\hat{E}_{j,\vec{n}}^{\alpha},\left[\hat{E}_{j,\vec{n}}^{\alpha},\hat{W}\right]
\right]\ket{E=0}\, . 
\end{align}
$\hat{E}_{j,\vec{n}}^{\alpha}$ commutes with $\hat{W}$ unless the latter contains $\hat{U}_{j,\vec{n}}$ or $\hat{U}^\dagger_{j,\vec{n}}$. When $\hat{U}_{j,\vec{n}}$ is contained, 
\begin{align}
&
\sum_\alpha\left[
\hat{E}_{j,\vec{n}}^{\alpha},\left[\hat{E}_{j,\vec{n}}^{\alpha},\hat{W}\right]
\right]
\nonumber\\
&\qquad=
\sum_\alpha{\rm Tr}\left(\left[
\hat{E}_{j,\vec{n}}^{\alpha},\left[\hat{E}_{j,\vec{n}}^{\alpha},\hat{U}_{j,\hat{n}}\right]\right]\hat{U}_{k,\hat{n}+\hat{j}}\cdots\right)
\nonumber\\
&\qquad=
(a^{-2}g)^2\sum_\alpha{\rm Tr}\left(\tau^\alpha\tau^\alpha\hat{U}_{j,\hat{n}}\hat{U}_{k,\hat{n}+\hat{j}}\cdots\right)
\nonumber\\
&\qquad=
a^{-4}g^2N{\rm Tr}\left(\hat{U}_{j,\hat{n}}\hat{U}_{k,\hat{n}+\hat{j}}\cdots\right)
\nonumber\\
&\qquad=
a^{-4}g^2N\hat{W}\, . 
\end{align}
The same holds when $\hat{U}^\dagger_{j,\hat{n}}$ is in the loop. 
Therefore, 
\begin{align}
\hat{H}_{\rm E}\hat{W}\ket{0}=L\cdot\frac{g^2N}{2a}\hat{W}\ket{0}\, . 
\end{align}
The same holds for any multi-string state, including closed or open strings, as long as there is no intersection. Note that $g^2N$ is the 't Hooft coupling, which is fixed in the 't Hooft large-$N$ limit. The interpretation is clear: the energy of a string is its length $La$ times the string tension $\frac{g^2N}{2a^2}$. 

Next, let us consider the case with intersections. Suppose a link  $\hat{U}_{j,\vec{n}}$ appears twice in the loop, while all other links appear only once. (Generalization to generic situation is straightforward.) Let us write such a loop as
${\rm Tr}(\hat{V}_1\hat{U}_{j,\vec{n}}\hat{V}_2\hat{U}_{j,\vec{n}})$. 
Then, 
\begin{align}
\lefteqn{
\sum_\alpha\left[
\hat{E}_{j,\vec{n}}^{\alpha},\left[\hat{E}_{j,\vec{n}}^{\alpha},{\rm Tr}(\hat{V}_1\hat{U}_{j,\vec{n}}\hat{V}_2\hat{U}_{j,\vec{n}})\right]
\right]
}
\nonumber\\
&=
2a^{-4}g^2N\textrm{Tr}(\hat{V}_1\hat{U}_{j,\vec{n}}\hat{V}_2\hat{U}_{j,\vec{n}})
\nonumber\\
&\quad
+
2\sum_\alpha{\rm Tr}(\hat{V}_1[\hat{E}_{j,\vec{n}}^{\alpha},\hat{U}_{j,\vec{n}}]\hat{V}_2[\hat{E}_{j,\vec{n}}^{\alpha},\hat{U}_{j,\vec{n}}])
\nonumber\\
&=
2a^{-4}g^2N\textrm{Tr}(\hat{V}_1\hat{U}_{j,\vec{n}}\hat{V}_2\hat{U}_{j,\vec{n}})
\nonumber\\
&\quad
+
2a^{-4}g^2\sum_\alpha
{\rm Tr}(\hat{V}_1\tau^\alpha\hat{U}_{j,\vec{n}}\hat{V}_2\tau^\alpha\hat{U}_{j,\vec{n}})
\nonumber\\
&=
2a^{-4}g^2N\textrm{Tr}(\hat{V}_1\hat{U}_{j,\vec{n}}\hat{V}_2\hat{U}_{j,\vec{n}})
\nonumber\\
&\quad
+
2a^{-4}g^2
{\rm Tr}(\hat{V}_1\hat{U}_{j,\vec{n}}){\rm Tr}(\hat{V}_2\hat{U}_{j,\vec{n}})\, . 
\label{eq:KS-string-splitting}
\end{align}
Therefore, 
\begin{align}
&
\hat{H}_E\left({\rm Tr}(\hat{V}_1\hat{U}_{j,\vec{n}}\hat{V}_2\hat{U}_{j,\vec{n}})\ket{E=0}\right)
\nonumber\\
\qquad
&=
L\cdot\frac{g^2N}{2a}{\rm Tr}(\hat{V}_1\hat{U}_{j,\vec{n}}\hat{V}_2\hat{U}_{j,\vec{n}})\ket{E=0}
\nonumber\\
&\quad
+\frac{g^2}{a}\textrm{Tr}(\hat{V}_1\hat{U}_{j,\vec{n}}){\rm Tr}(\hat{V}_2\hat{U}_{j,\vec{n}})\ket{E=0}\, .
\end{align}
The second term can be understood as the splitting of a string into two strings. Note that the second term is suppressed by a factor of $1/N$ compared to the first term. Similarly, two strings can be joined at an intersection to form one string, 
\begin{align}
&
\hat{H}_E\left({\rm Tr}(\hat{V}_1\hat{U}_{j,\vec{n}}){\rm Tr}(\hat{V}_2\hat{U}_{j,\vec{n}})|E=0\rangle\right)
\nonumber\\
\qquad
&=
L\cdot\frac{g^2N}{2a}\textrm{Tr}(\hat{V}_1\hat{U}_{j,\vec{n}}){\rm Tr}(\hat{V}_2\hat{U}_{j,\vec{n}})\ket{E=0}
\nonumber\\
&\quad
+
\frac{g^2}{a}\textrm{Tr}(\hat{V}_1\hat{U}_{j,\vec{n}}\hat{V}_2\hat{U}_{j,\vec{n}})\ket{E=0}\, .
\end{align}
In general, such splitting and joining can take place at any intersection. 

We can also use the counterpart of the non-singlet loop \eqref{sec;closed_string_nonsinglet}, which is defined by 
\begin{align}
\hat{W}^{\rm (non\mathchar`-singlet)}\equiv N^l\hat{U}_{j_1,\vec{x};a_1a_2}\hat{U}_{j_2,\vec{x}+\hat{j}_1;a_2a_3}\cdots\hat{U}_{j_l,\vec{x}-\hat{j}_l;a_la_1}\, . 
\end{align}
Straightforward computations show that exactly the same results hold: for non-intersecting strings, $\hat{H}_E$ counts the number of links; when strings intersect, joining/splitting takes place; and numerical factors are also the same. 

One difference, which may look troublesome at first glance, is that $\hat{U}_{j,\vec{n}}^{ab}\hat{U}^{\dagger bc}_{j,\vec{n}}$ is not $\delta^{ac}$ if a sum over $b$ is not taken. In general, $\hat{W}^{\rm (non\mathchar`-singlet)}$ with and without whiskers (Figure~\ref{fig:whiskers}) becomes identical after the projection to the singlet, and hence, their difference is a redundant state that is in $\mathrm{Ker}(\hat{\mathcal{P}})$. When the expectation value is taken using gauge-invariant states like the ground state, whiskers becomes $\textbf{1}$.

\begin{figure}[htbp]
\begin{center}
  \includegraphics[width=8cm]{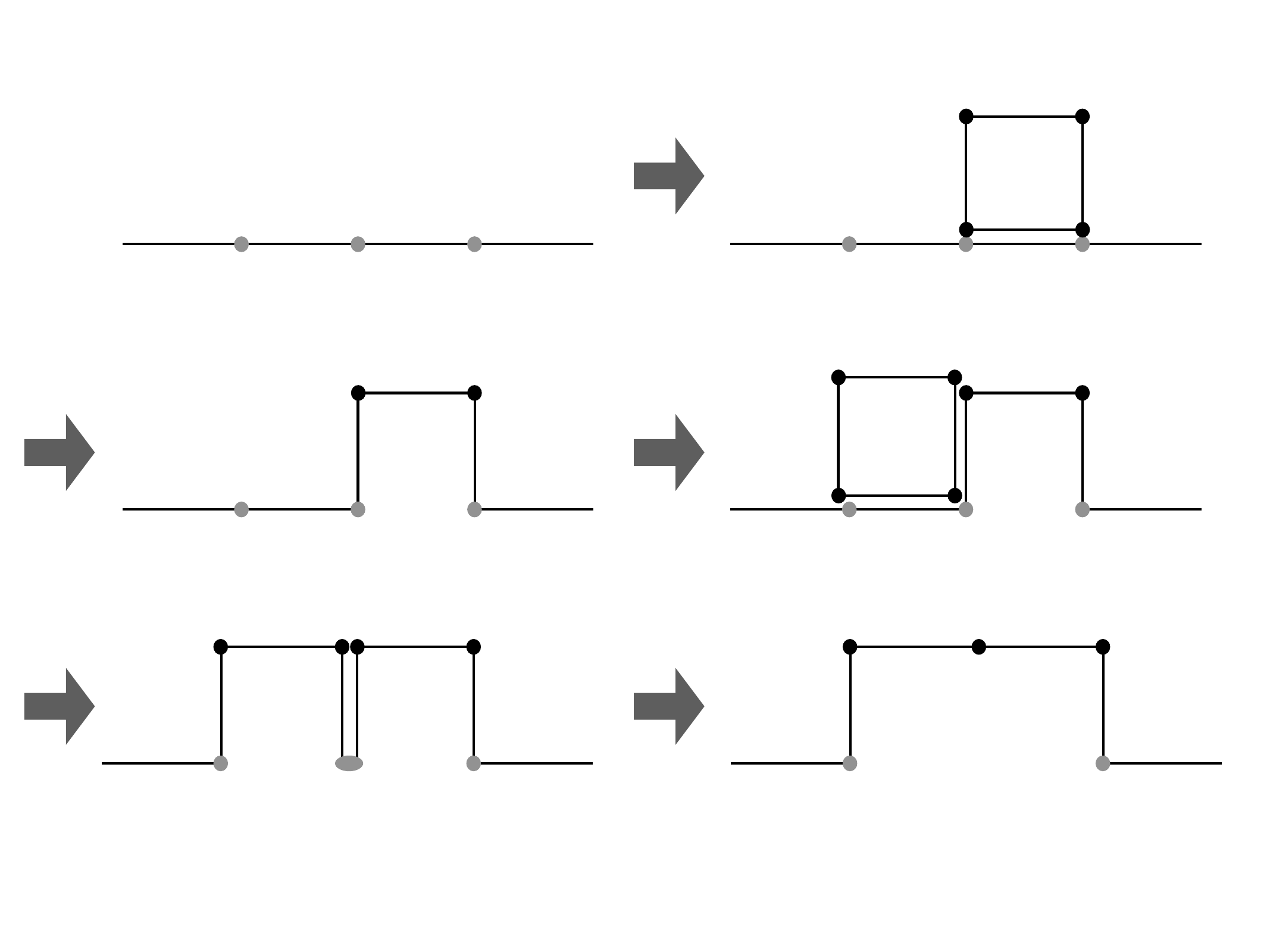}
\end{center}  
  \caption{Interaction between non-singlet string and plaquettes in the Kogut-Susskind Hamiltonian, which are, by construction, singlets. Straight lines are link variables (string bits), and black and grey circles stand for indices which are summed (from 1 to $N$) and not summed. 
  }\label{fig:string-plaquette-interaction}
\end{figure}

Note that, even if the initial state is non-singlet, singlet states can also emerge through the interaction between plaquettes and strings in the initial state, as we can see in Figure~\ref{fig:string-plaquette-interaction}. 

In the above, we used the electric field defined from the infinitesimal transformation acting from the left of the link variable. We could also use the one acting from the right, and the result is the same anyway; see Supplementary Material~\ref{sec:KS-EL-vs-ER}. 

\section{Review of orbifold lattice formulation}\label{sec:orbifold_lattice}
In this section, we review the orbifold lattice Hamiltonian approach for quantum simulations of gauge theory~\cite{Buser:2020cvn,Bergner:2024qjl,Halimeh:2024bth,Halimeh:2025ivn,Bergner:2025zkj}. 
For concreteness, we consider $3+1$-d pure SU($N$) Yang-Mills theory, formulated on the extended Hilbert space. 
\subsection{Operators and commutation relations}
The orbifold lattice Hamiltonian~\cite{Kaplan:2002wv,Buser:2020cvn,Bergner:2024qjl} can be written in terms of the complex link variables $Z_{j,\vec{n}}$ ($j=1,2,3$), and their canonical conjugates $P_{j,\vec{n}}$. 

Note that $\bar{Z}_{j,\vec{n}}$ and $\bar{P}_{j,\vec{n}}$ stand for Hermitian conjugates of $N\times N$ matrices, i.e., 
\begin{align}
\bar{Z}_{j,\vec{n};ab}=(Z_{j,\vec{n};ba})^\ast\, , 
\qquad
\bar{P}_{j,\vec{n};ab}=(P_{j,\vec{n};ba})^\ast\, . 
\end{align}
We do not use dagger $\dagger$ because we save it for conjugate operators acting on the Hilbert space. For the operators this implies  
\begin{align}
\hat{\bar{Z}}_{j,\vec{n};ab}=\left(\hat{Z}_{j,\vec{n};ba}\right)^\dagger\, , 
\qquad
\hat{\bar{P}}_{j,\vec{n};ab}=\left(\hat{P}_{j,\vec{n};ba}\right)^\dagger\, ,  
\end{align}
The canonical commutation relation is
\begin{align}
    \left[
    \hat{Z}_{j,\vec{n};ab},\hat{\bar{P}}_{k,\vec{n}';cd}
    \right]
    =
    \left[
    \hat{Z}_{j,\vec{n};ab},\left(\hat{P}_{k,\vec{n}';dc}\right)^\dagger
    \right]
    =\mathrm{i}\delta_{jk}\delta_{\vec{n}\vec{n}'}\delta_{ad}\delta_{bc}\, . 
\end{align}
In case one prefers real variables, $Z$ can be separated into real and imaginary parts $x$ and $y$, 
\begin{align}
\hat{Z}_{j,\vec{n};ab}
=
\frac{\hat{x}_{j,\vec{n};ab}+\mathrm{i}\hat{y}_{j,\vec{n};ab}}{\sqrt{2}}\, , 
\end{align}
and their conjugate momenta $p^{(x)}$ and $p^{(y)}$, 
\begin{align}
\hat{P}_{j,\vec{n};ab}
=
\frac{\hat{p}^{(x)}_{j,\vec{n};ab}+\mathrm{i}\hat{p}^{(y)}_{j,\vec{n};ab}}{\sqrt{2}}\, , 
\end{align}
that satisfy 
\begin{align}
    \left[
    \hat{x}_{j,\vec{n};ab},\hat{p}^{(x)}_{k,\vec{n}';cd}
    \right]
    =
        \left[
    \hat{y}_{j,\vec{n};ab},\hat{p}^{(y)}_{k,\vec{n}';cd}
    \right]
    =\mathrm{i}\delta_{jk}\delta_{\vec{n}\vec{n}'}\delta_{ad}\delta_{bc}\, , 
\end{align}
while other commutators vanish. 
The orbifold lattice Hamiltonian is designed so that Yang-Mills theory coupled to scalars is obtained when the complex link variable $Z$ is expanded about an appropriate background. 

The complex link variable $Z$ can be interpreted as a product of a positive-definite Hermitian site variable $W$ and a unitary link variable $U$ as~\cite{Unsal:2005yh} 
\begin{align}
Z_{j,\vec{n}}=\sqrt{\frac{a}{2g^2}}W_{j,\vec{n}}U_{j,\vec{n}}\, .  
    \label{eq:polar_decomposition}
\end{align}
They describe an adjoint scalar field $\phi_j$ and gauge field $A_j$, respectively, as $W_{j,\vec{n}}=\exp\left(ag\phi_{j,\vec{n}}\right)$ and $U_{j,\vec{n}}=\exp(\mathrm{i}agA_{j,\vec{n}})$. Intuitively, $W$ and $U$ are the polar coordinates, with the radial coordinate $W$ and angular coordinate $U$, while the $Z$ are the Cartesian coordinates. The key observation~\cite{Buser:2020cvn,Bergner:2024qjl} is that quantization is straightforward in Cartesian coordinates. Indeed, we can use $2N^2$ bosons $\hat{x}_a$ and $\hat{p}_a$ ($a=1,\cdots,2N^2$) that satisfy the canonical commutation relation $[\hat{x}_a,\hat{p}_b]=\delta_{ab}$ to describe the real and imaginary parts of each link variable $Z_{j,\vec{n}}$. This simple fact resolves technical difficulties associated with the Kogut-Susskind Hamiltonian formulation. 
\subsection{Orbifold lattice Hamiltonian}
The orbifold lattice Hamiltonian has been studied as a convenient tool to achieve exponential speedup in quantum simulations of Yang-Mills theory and QCD. 
The orbifold lattice Hamiltonian for pure Yang-Mills theory is~\cite{Kaplan:2002wv,Buser:2020cvn,Bergner:2024qjl}
\begin{align}
\hat{H}
&=
\sum_{\vec{n}}
{\rm Tr}\Biggl(
\sum_{j=1}^3 \hat{P}_{j,\vec{n}} \,  \hat{\bar{P}}_{j,\vec{n}}
\nonumber\\
&\quad
+
\frac{g^2}{2a^3}\left|\sum_{j=1}^3
\left(
\hat{Z}_{j,\vec{n}} \, \hat{\bar{Z}}_{j,\vec{n}} -\hat{\bar{Z}}_{j,\vec{n}-\hat{j}}\hat{Z}_{j,\vec{n}-\hat{j}}
\right)
\right|^2 
\nonumber\\
&\quad
+
\frac{2g^2}{a^3}\sum_{j<k}
\left|
\hat{Z}_{j,\vec{n}} \, \hat{Z}_{k,\vec{n}+\hat{j}}
-
\hat{Z}_{k,\vec{n}} \, \hat{Z}_{j,\vec{n}+\hat{k}}
\right|^2
 \Biggl)
 \nonumber\\
&\quad
 + 
 \hat{H}^{\rm (mass)}\, , 
 \label{eq:orbifold_Hamiltonian}
\end{align}
where $\hat{H}^{\rm (mass)}$ is chosen appropriately depending on the details of the strategy. Ref.~\cite{Bergner:2025zkj} used 
\begin{align}
\hat{H}^{\rm (mass)}
&\equiv
\frac{m^2g^2}{2a}\sum_{\vec{n}}\sum_{j=1}^3
{\rm Tr}
\left|\hat{Z}_{j,\vec{n}}\hat{\bar{Z}}_{j,\vec{n}} -\frac{a}{2g^2}\cdot\textbf{1}_N\right|^2
\nonumber\\
&\,  
+
\frac{m^2_{\rm U(1)}a}{2g^2}
\sum_{\vec{n}}\sum_{j=1}^3
\left|
    \left(\frac{a}{2g^2}\right)^{-N/2}\det(\hat{Z}_{j,\vec{n}})-1
    \right|^2\, . 
    \label{eq:orbifold_mass_term}
\end{align}
With this choice, in the large-mass limit $m^2,m^2_{\rm U(1)}\to\infty$, the complex links reduce to unitary links as $Z_{j,\vec{n}}=\sqrt{\frac{a}{2g^2}}W_{j,\vec{n}}U_{j,\vec{n}}\to\sqrt{\frac{a}{2g^2}}U_{j,\vec{n}}$, $\det U_{j,\vec{n}}$ becomes 1, and the orbifold lattice Hamiltonian reduces to the SU($N$) Kogut-Susskind Hamiltonian.
Away from the Kogut-Susskind limit, the orbifold lattice Hamiltonian describes Yang-Mills theory coupled to heavy scalars that do not affect low-energy physics. (Note that the last term in \eqref{eq:orbifold_mass_term} is not important, because the U(1) part decouples from the SU($N$) part.)

The orbifold lattice Hamiltonian and the commutation relation are invariant under the local U($N$) transformation
\begin{align}
\hat{Z}_{j,\vec{n}}\, ,\  \hat{P}_{j,\vec{n}}
\quad\rightarrow\quad
    \Omega^{-1}_{\vec{n}}
     \hat{Z}_{j,\vec{n}}
     \Omega_{\vec{n}+\hat{j}}\, ,\  
     \Omega^{-1}_{\vec{n}}
     \hat{P}_{j,\vec{n}}
     \Omega_{\vec{n}+\hat{j}}\, , 
\end{align}
where $\Omega_{\vec{n}}$ is an $N\times N$ unitary matrix. We can gauge only SU($N$)~\cite{Bergner:2024qjl}. Note that $W_{j,\vec{n}}$ and $U_{j,\vec{n}}$ transform as an adjoint scalar on a site and unitary link variables, respectively~\cite{Unsal:2005yh} : 
\begin{align}
\hat{W}_{j,\vec{n}}\, ,\  \hat{U}_{j,\vec{n}}
\quad\rightarrow\quad
    \Omega^{-1}_{\vec{n}}
     \hat{W}_{j,\vec{n}}
     \Omega_{\vec{n}}\, ,\  
     \Omega^{-1}_{\vec{n}}
     \hat{U}_{j,\vec{n}}
     \Omega_{\vec{n}+\hat{j}}\, . 
\end{align}

\subsection{Strings and interactions}
We can use singlet and non-singlet version of the Wilson loops $\hat{W}^{\rm (singlet)}_{\rm closed}$ and $\hat{W}^{\rm (non\mathchar`-singlet)}_{\rm closed}$ defined in \eqref{sec;closed_string_singlet} and \eqref{sec;closed_string_nonsinglet}, respectively. They describe QCD strings coupled to scalars. 

It is easy to see that joining and splitting of strings takes place as in the Kogut-Susskind formulation. Indeed, similar to \eqref{eq:KS-string-splitting} for the Kogut-Susskind formulation, there are splitting processes such as 
\begin{align}
&
\sum_{a,b}\left[
\hat{P}_{j,\vec{n},ab},\left[\hat{\bar{P}}_{j,\vec{n},ba}^{\alpha},{\rm Tr}(\hat{V}_1\hat{Z}_{j,\vec{n}}\hat{V}_2\hat{\bar{Z}}_{j,\vec{n}})\right]
\right]
\nonumber\\
&\qquad=
-
\left({\rm Tr}\hat{V}_1\right)
\left({\rm Tr}\hat{V}_2\right)\, . 
\end{align}
Still, some care is needed to see a precise correspondence to the Kogut-Susskind formulation, as we see below. 

Unlike the electric field operator $\hat{E}$ in the Kogut-Susskind formulation, the momentum operators $\hat{P}$ and $\hat{\bar{P}}$ annihilates links, or string bits, $\hat{Z}$ and $\hat{\bar{Z}}$. Therefore, the kinetic term $\mathrm{Tr}(\hat{P}\hat{\bar{P}})$ itself does not count the number of links. It has to be combined with the mass term $\hat{H}_{\rm mass}$, so that links are close to unitary in the low-energy states. The same is true for the `whiskers' (Figure~\ref{fig:whiskers}). When the mass is large and scalars are suppressed, loop operators with and without whiskers are indistinguishable as long as they act on the low-energy states.  

\subsection{Truncated Hamiltonian in terms of Pauli strings}
Unlike in the naive Kogut-Susskind approach, the orbifold lattice Hamiltonian can easily be truncated to finite dimension and be expressed in terms of Pauli strings~\cite{Buser:2020cvn,Bergner:2024qjl}. 

Link variables in the orbifold lattice can be described like the standard bosons $\hat{x}_a$ and $\hat{p}_a$ ($a=1,2,\cdots,2N^2n_{\ell}$), where $n_{\ell}$ is the number of links (for the periodic boundary condition, the number of lattice sites times the spatial dimensions). 
The extended Hilbert space can be expressed using either the coordinate basis or the momentum basis. The coordinate basis is defined as 
\begin{align}
\mathcal{H}_{\rm ext}
=
\otimes_a\mathcal{H}_a\, , 
\qquad
\mathcal{H}_a
=
\mathrm{Span}\{\ket{x_a}|x_a\in\mathbb{R}\}\, , 
\end{align}
where 
\begin{align}
\hat{x}_a\ket{x_a}
=
x_a\ket{x_a}\, . 
\end{align}
The momentum basis is defined using the momentum eigenstate, 
\begin{align}
\hat{p}_a\ket{p_a}
=
p_a\ket{p_a}\, . 
\end{align}
The coordinate basis and momentum basis are related by Fourier transform. 

To put this Hilbert space on a digital quantum computer, we can introduce a cutoff for each boson coordinate $x_a$. 
We allocate $Q$ qubits to each boson. Each $x_a$ may assume $\Lambda=2^Q$ distinct values. 
For practical purposes, it proves advantageous to implement periodic boundary conditions $x+2R\sim x$. 
In this framework, a convenient choice is 
\begin{align}
   \delta_x=\frac{2R}{\Lambda} 
\end{align}
and 
\begin{align}
\hat{x}_a
=
-
\delta_x\cdot
\left(
\frac{\sigma_{z,1}}{2}
+
2\cdot\frac{\sigma_{z,2}}{2}
+
\cdots
+
2^{Q-1}\cdot\frac{\sigma_{z,Q}}{2}
\right)\, . 
\label{eq:x_with_Z}
\end{align}
Here, $\sigma_{z,j}$ represents the Pauli $\sigma_z$ operator acting on the $j$-th qubit assigned to the $a$-th boson, and Pauli $\sigma_z$ operators defined as  
\begin{align}
    \sigma_{z} \equiv \ket{0}\bra{0} - \ket{1}\bra{1}\, .  
\end{align}
Consequently, $\hat{x}_{a}$ takes values $\pm\frac{\delta_x}{2}$, $\pm\frac{3\delta_x}{2}$, ..., $\pm\frac{(\Lambda-1)\delta_x}{2}$.
In the same way, in the momentum basis, we have 
\begin{align}
\hat{p}_a
=
-
\delta_p\cdot
\left(
\frac{\sigma_{z,1}}{2}
+
2\cdot\frac{\sigma_{z,2}}{2}
+
\cdots
+
2^{Q-1}\cdot\frac{\sigma_{z,Q}}{2}
\right)\, ,  
\label{eq:p_with_Z}
\end{align}
where
\begin{align}
\delta_p
=
\frac{\pi}{R}\, . 
\end{align}
Note that the coordinate basis and momentum basis are related, at the discretized level, by a Fourier transform, which can be performed efficiently on a digital quantum computer. We use $\hat{F}$ to denote the Fourier transform from the coordinate basis to the momentum basis. This is a product of Fourier transforms for each boson, which are denoted by $\hat{F}_a$: $\hat{F}=\prod_a\hat{F}_a$. 

Numerous polynomial couplings exist in the form $\hat{x}_{a}\otimes\hat{x}_{b}\otimes\cdots$. 
In the coordinate basis, each $\hat{x}$ comprises a weighted sum of $\sigma_{z,1}$,..., $\sigma_{z,Q}$. 
Therefore, couplings in the orbifold lattice Hamiltonian are written as weighted sums of Pauli strings consisting of Pauli $\sigma_z$'s. 

\section{Potential near-term demonstrations on quantum devices}\label{appendix:near-term-demonstration}
Though our simulation protocols are theoretically justified to scale well, it is always good to have experimental demonstrations on quantum devices.
In the near future, we would want to focus on the minimal setups. The smallest possible example would be the single-site lattice --- whose large-$N$ version is known as the Eguchi-Kawai model~\cite{Eguchi:1982nm} and provides us with a practically useful tool to study lattice gauge theory via large-$N$ volume reduction~\cite{Eguchi:1982nm,Kovtun:2003hr} --- with $d$ links (that corresponds to the spatial dimension $d$) and SU(2) gauge group. The value of  $d$ can be 1 or 2. While $d=2$ is needed for nontrivial interactions, $d=1$ is enough for a demonstration purpose. 

Although the non-singlet description should be the default (Sec.~\ref{sec:nonsinglet_simulations}), there are situations where a \emph{local} singlet projection is unavoidable --- most notably when a non-singlet operator acts on a non-singlet state (Sec.~\ref{sec:when_to_project}). Because such cases involve only a few links, the one- or two-link demonstrations below are not toy exercises: they are precisely the setup in which the projection would actually be used within a large-scale simulation. This is the sense in which these minimal experiments are scalable.

As we saw in Sec.~\ref{sec:numerical_test}, we can represent each link and ancilla state $\ket{\xi}$ using four bosons, and $Q=4$ ($\Lambda=2^Q=16$) is reasonably good for low-energy spectrum. If we do not use the OAA to increase the success rate of the post selection for the singlet projection, we need only one ancilla state $\ket{Z}_2$, which requires four boson; hence we need $4(d+2)Q=16(d+2)$ qubits. Taking $d=1$ or $2$, we get $48$ or $64$ qubits. We can also try SU(2) Hermitian $d$-matrix model (see e.g.,~\cite{Rinaldi:2021jbg}), where dynamical variables are SU(2) Hermitian matrices consisting of 3 bosons; then we need $3d$ bosons for dynamical degrees of freedom, $3$ bosons for the ancilla state analogous to $\ket{Z}_2$, and 4 bosons for $\xi$, and hence total $(3d+7)Q$ qubits. Taking $Q=4$ and $d=1$ or $2$, we get $40$ or $52$ qubits.

These numbers are promising but one would want something even smaller to have demonstrations as soon as possible. For that purpose, projection for 
$n_\ell$ links of U(1) theory embedded into $\mathbb{R}^2$, described in Supplementary Material~\ref{sec:projection_numerical_test_U1}, may be useful. Having $\ket{Z}_1$, $\ket{Z}_2$, and $\mathbb{\xi}$, we need only $2(n_\ell+2)$ bosons ($2n_\ell$ for $\ket{Z}_1$, two for $\ket{Z}_2$ and two for $\mathbb{\xi}$), and $2(n_\ell+2)Q$ qubits. 

For U(1) and $\mathbb{Z}_n$ theories, we can simplify the setup further. The key point is that we do not have to introduce the scalar to truncate U(1) efficiently, because it is a compactification of $\mathbb{R}$, and $\mathbb{Z}_n$ is precisely the truncated version of U(1). Using this property, we can reduce the qubit requirement 50\%, down to $(n_\ell+2)Q$ qubits. (Note that we need to use the argument $\theta$ instead of $e^{\mathrm{i}\theta}$ in this case, and a sum $\theta+\theta'$ instead of a product $e^{\mathrm{i}\theta}\cdot e^{\mathrm{i}\theta'}$. Note also that,  to have interactions, we need to add matter fields. See Supplementary Materials~\ref{sec:U(1)-and-Z_N} for details.)

Finally, let us note that the simplest version of the singlet-projection protocol, translated into the $\mathbb{Z}_2$ gauge theory, requires only $n_\ell+1$ qubits; we do not need an ancilla state $\ket{Z}_2$ in this case. Let $b=0,1$ corresponds to link variable $e^{\pi\mathrm{i}b}=\pm 1$.  
The Haar random distribution on $\mathbb{Z}_2$ is $\frac{\ket{0}+\ket{1}}{\sqrt{2}}=\hat{h}\ket{0}$, where $\hat{h}$ is the Hadamard gate. Therefore, $\ket{\mathcal{G}}=\frac{\ket{0}_\xi+\ket{1}_\xi}{\sqrt{2}}$ can be prepared easily. 
For example, let us take the case of $n_{\ell}=2$. Suppose the initial configuration was $\sum_{b_1,b_2}c_{b_1,b_2}\ket{b_1}_{1}\ket{b_2}_{2}$. Furthermore, suppose that two links share one lattice point and transform by the $\mathbb{Z}_2$ transformation as $\ket{b_1}_{1}\ket{b_2}_{2}\to\ket{b_1\oplus 1}_{1}\ket{b_2\oplus 1}_{2}$. In this case, we can obtain the counterpart of $\ket{Z^{(\xi)}}$ (i.e., $\ket{b\oplus\xi}$) using a CNOT gate (CX gate) that uses $\ket{\xi}$ as a control qubit. Therefore, the singlet projection goes as follows:
\begin{itemize}
  \item \begin{minipage}[t]{0.95\linewidth}
    \raggedright
    We tensor $\ket{0}_\xi$ to this state to obtain $\sum_{b_1,b_2}c_{b_1,b_2}\ket{b_1}_{1}\ket{b_2}_{2}\ket{0}_\xi$
  \end{minipage}

  \item \begin{minipage}[t]{0.95\linewidth}
    \raggedright
Act the Hadamard gate on $\ket{0}_\xi$ to obtain $\sum_{b_1,b_2}c_{b_1,b_2}\ket{b_1}_{1}\ket{b_2}_{2}\frac{\ket{0}_\xi+\ket{1}_\xi}{\sqrt{2}}$
 \end{minipage}

   \item \begin{minipage}[t]{0.95\linewidth}
    \raggedright
Act ${\rm CX}_{\xi,1}$ to obtain $\sum_{b_1,b_2}c_{b_1,b_2}\ket{b_2}_{2}\frac{\ket{b_1}_{1}\ket{0}_\xi+\ket{b_1\oplus 1}_{1}\ket{1}_\xi}{\sqrt{2}}$
 \end{minipage}

  \item \begin{minipage}[t]{0.95\linewidth}
    \raggedright 
Act ${\rm CX}_{\xi,2}$ to obtain $\sum_{b_1,b_2}c_{b_1,b_2}\frac{\ket{b_1}_{1}\ket{b_2}_{2}\ket{0}_\xi+\ket{b_1\oplus 1}_{1}\ket{b_2\oplus 1}_{2}\ket{1}_\xi}{\sqrt{2}}$
 \end{minipage}

   \item \begin{minipage}[t]{0.95\linewidth}
    \raggedright
\item{\leavevmode\raggedright
Project to $\ket{+}_\xi$ to obtain $\sum_{b_1,b_2}c_{b_1,b_2}\frac{\ket{b_1}_{1}\ket{b_2}_{2}+\ket{b_1\oplus 1}_{1}\ket{b_2\oplus 1}_{2}}{\sqrt{2}}$}
 \end{minipage}
\end{itemize}

As far as the time evolution is concerned, we can even use \textit{harmonic oscillator} for a demonstration --- if we apply the universal framework to the harmonic oscillator, what we get is the same universal kind of circuit~\cite{Halimeh:2024bth}! The concrete circuit has been described already in ref.~\cite{Klco:2018zqz}, though its fundamental importance has not been appreciated back then. A possible bottleneck is the precision of quantum Fourier transform at large values of $Q$, and harmonic oscillator is a perfect setup to study it comparing it with analytic results. More generally, one-boson system with a $p$ -th order polynomial such as the double-well potential ($p=4$). The gate requirement is $O(Q^p)$ for CNOT gates and one-qubit rotations, and $O(Q^2)$ for quantum Fourier transform.

\bibliography{ref}

@article{Dalmonte:2016alw,
    author = "Dalmonte, M. and Montangero, S.",
    title = "{Lattice gauge theory simulations in the quantum information era}",
    eprint = "1602.03776",
    archivePrefix = "arXiv",
    primaryClass = "cond-mat.quant-gas",
    reportNumber = "CONT.-PHYS.-(2016), 1-25",
    doi = "10.1080/00107514.2016.1151199",
    journal = "Contemp. Phys.",
    volume = "57",
    pages = "388",
    year = "2016"
}

@article{Banuls:2019bmf,
    author = "Ba{\~n}uls, M. C. and others",
    title = "{Simulating Lattice Gauge Theories within Quantum Technologies}",
    eprint = "1911.00003",
    archivePrefix = "arXiv",
    primaryClass = "quant-ph",
    doi = "10.1140/epjd/e2020-100571-8",
    journal = "Eur. Phys. J. D",
    volume = "74",
    number = "8",
    pages = "165",
    year = "2020"
}

@article{Zohar:2015hwa,
    author = "Zohar, Erez and Cirac, J. Ignacio and Reznik, Benni",
    title = "{Quantum Simulations of Lattice Gauge Theories using Ultracold Atoms in Optical Lattices}",
    eprint = "1503.02312",
    archivePrefix = "arXiv",
    primaryClass = "quant-ph",
    doi = "10.1088/0034-4885/79/1/014401",
    journal = "Rept. Prog. Phys.",
    volume = "79",
    number = "1",
    pages = "014401",
    year = "2016"
}

@article{Banerjee:2012pg,
    author = "Banerjee, D. and Dalmonte, M. and Muller, M. and Rico, E. and Stebler, P. and Wiese, U. -J. and Zoller, P.",
    title = "{Atomic Quantum Simulation of Dynamical Gauge Fields coupled to Fermionic Matter: From String Breaking to Evolution after a Quench}",
    eprint = "1205.6366",
    archivePrefix = "arXiv",
    primaryClass = "cond-mat.quant-gas",
    doi = "10.1103/PhysRevLett.109.175302",
    journal = "Phys. Rev. Lett.",
    volume = "109",
    pages = "175302",
    year = "2012"
}

@article{Elitzur:1989nr,
    author = "Elitzur, Shmuel and Moore, Gregory W. and Schwimmer, Adam and Seiberg, Nathan",
    title = "{Remarks on the Canonical Quantization of the Chern-Simons-Witten Theory}",
    reportNumber = "IASSNS-HEP-89/20",
    doi = "10.1016/0550-3213(89)90436-7",
    journal = "Nucl. Phys. B",
    volume = "326",
    pages = "108--134",
    year = "1989"
}

@article{Mou:2025iiu,
    author = "Mou, Zong-Gang and Chakraborty, Bipasha",
    title = "{Scalable quantum computation of Quantum Electrodynamics beyond one spatial dimension}",
    eprint = "2510.27668",
    archivePrefix = "arXiv",
    primaryClass = "hep-lat",
    month = "10",
    year = "2025"
}

@article{Fliss:2025kzi,
    author = "Fliss, Jackson R. and Frenkel, Alexander and Hartnoll, Sean A. and Soni, Ronak M.",
    title = "{Minimal areas from entangled matrices}",
    eprint = "2408.05274",
    archivePrefix = "arXiv",
    primaryClass = "hep-th",
    doi = "10.21468/SciPostPhys.18.6.171",
    journal = "SciPost Phys.",
    volume = "18",
    number = "6",
    pages = "171",
    year = "2025"
}

@article{Gautam:2022akq,
    author = "Gautam, Vaibhav and Hanada, Masanori and Jevicki, Antal and Peng, Cheng",
    title = "{Matrix entanglement}",
    eprint = "2204.06472",
    archivePrefix = "arXiv",
    primaryClass = "hep-th",
    reportNumber = "DMUS-MP-22/03",
    doi = "10.1007/JHEP01(2023)003",
    journal = "JHEP",
    volume = "01",
    pages = "003",
    year = "2023"
}

@article{Chakraborty:2025veu,
    author = "Chakraborty, Abhijit and Lewis, Randy and Muschik, Christine A.",
    title = "{Charge-singlet measurement toolbox}",
    eprint = "2510.08718",
    archivePrefix = "arXiv",
    primaryClass = "quant-ph",
    doi = "10.1103/353y-scg1",
    journal = "Phys. Rev. Res.",
    volume = "7",
    number = "4",
    pages = "043162",
    year = "2025"
}

@article{Hanada:2021ipb,
    author = "Hanada, Masanori",
    title = "{Bulk geometry in gauge/gravity duality and color degrees of freedom}",
    eprint = "2102.08982",
    archivePrefix = "arXiv",
    primaryClass = "hep-th",
    reportNumber = "DMUS-MP-21/02",
    doi = "10.1103/PhysRevD.103.106007",
    journal = "Phys. Rev. D",
    volume = "103",
    number = "10",
    pages = "106007",
    year = "2021"
}

@article{Gautam:2024zsj,
    author = "Gautam, Vaibhav and Hanada, Masanori and Jevicki, Antal",
    title = "{Operator algebra, quantum entanglement, and emergent geometry from matrix degrees of freedom}",
    eprint = "2406.13364",
    archivePrefix = "arXiv",
    primaryClass = "hep-th",
    doi = "10.1007/JHEP01(2025)019",
    journal = "JHEP",
    volume = "01",
    pages = "019",
    year = "2025"
}

@article{Than:2024zaj,
    author = "Than, Anton T. and others",
    title = "{The phase diagram of quantum chromodynamics in one dimension on a quantum computer}",
    eprint = "2501.00579",
    archivePrefix = "arXiv",
    primaryClass = "quant-ph",
    doi = "10.1038/s41467-025-65198-w",
    journal = "Nature Commun.",
    volume = "16",
    number = "1",
    pages = "10288",
    year = "2025"
}

@article{Rajput:2021trn,
    author = "Rajput, Abhishek and Roggero, Alessandro and Wiebe, Nathan",
    title = "{Quantum error correction with gauge symmetries}",
    eprint = "2112.05186",
    archivePrefix = "arXiv",
    primaryClass = "quant-ph",
    reportNumber = "IQuS@UW-21-014",
    doi = "10.1038/s41534-023-00706-8",
    journal = "npj Quantum Inf.",
    volume = "9",
    number = "1",
    pages = "41",
    year = "2023"
}

@article{Yoder:2014sls,
    author = "Yoder, Theodore J. and Low, Guang Hao and Chuang, Isaac L.",
    title = "{Fixed-Point Quantum Search with an Optimal Number of Queries}",
    eprint = "1409.3305",
    archivePrefix = "arXiv",
    primaryClass = "quant-ph",
    doi = "10.1103/PhysRevLett.113.210501",
    journal = "Phys. Rev. Lett.",
    volume = "113",
    number = "21",
    pages = "210501",
    year = "2014"
}

@inproceedings{Gilyen:2018khw,
    author = "Gily{\'e}n, Andr{\'a}s and Su, Yuan and Low, Guang Hao and Wiebe, Nathan",
    title = "{Quantum singular value transformation and beyond: exponential improvements for quantum matrix arithmetics}",
    booktitle = "{51st Annual ACM SIGACT Symposium on Theory of Computing}",
    eprint = "1806.01838",
    archivePrefix = "arXiv",
    primaryClass = "quant-ph",
    doi = "10.1145/3313276.3316366",
    month = "6",
    year = "2018"
}

@article{Childs:2012gwh,
    author = "Childs, Andrew M. and Wiebe, Nathan",
    title = "{Hamiltonian Simulation Using Linear Combinations of Unitary Operations}",
    eprint = "1202.5822",
    archivePrefix = "arXiv",
    primaryClass = "quant-ph",
    doi = "10.26421/QIC12.11-12-1",
    journal = "Quant. Inf. Comput.",
    volume = "12",
    number = "11",
    pages = "0901--0924",
    year = "2012"
}

@article{Stryker:2018efp,
    author = "Stryker, Jesse R.",
    title = "{Oracles for Gauss's law on digital quantum computers}",
    eprint = "1812.01617",
    archivePrefix = "arXiv",
    primaryClass = "quant-ph",
    reportNumber = "INT-PUB-18-061",
    doi = "10.1103/PhysRevA.99.042301",
    journal = "Phys. Rev. A",
    volume = "99",
    number = "4",
    pages = "042301",
    year = "2019"
}

@article{Salam:1968rm,
    author = "Salam, Abdus",
    title = "{Weak and Electromagnetic Interactions}",
    doi = "10.1142/9789812795915_0034",
    journal = "Conf. Proc. C",
    volume = "680519",
    pages = "367--377",
    year = "1968"
}

@article{Kovtun:2003hr,
    author = "Kovtun, Pavel and Unsal, Mithat and Yaffe, Laurence G.",
    title = "{Nonperturbative equivalences among large N(c) gauge theories with adjoint and bifundamental matter fields}",
    eprint = "hep-th/0311098",
    archivePrefix = "arXiv",
    reportNumber = "UW-PT-03-25, INT-PUB-03-18",
    doi = "10.1088/1126-6708/2003/12/034",
    journal = "JHEP",
    volume = "12",
    pages = "034",
    year = "2003"
}

@article{Eguchi:1982nm,
    author = "Eguchi, Tohru and Kawai, Hikaru",
    title = "{Reduction of Dynamical Degrees of Freedom in the Large N Gauge Theory}",
    reportNumber = "UT-378-TOKYO",
    doi = "10.1103/PhysRevLett.48.1063",
    journal = "Phys. Rev. Lett.",
    volume = "48",
    pages = "1063",
    year = "1982"
}

@article{Yao:2025uxz,
    author = "Yao, Xiaojun",
    title = "{On Quantum Simulation of QED in Coulomb Gauge}",
    eprint = "2507.01089",
    archivePrefix = "arXiv",
    primaryClass = "quant-ph",
    reportNumber = "IQuS@UW-21-106",
    month = "7",
    year = "2025"
}

@article{Hanada:2022pps,
    author = "Hanada, Masanori and Liu, Junyu and Rinaldi, Enrico and Tezuka, Masaki",
    title = "{Estimating truncation effects of quantum bosonic systems using sampling algorithms}",
    eprint = "2212.08546",
    archivePrefix = "arXiv",
    primaryClass = "quant-ph",
    reportNumber = "RIKEN-iTHEMS-Report-23",
    doi = "10.1088/2632-2153/ad035c",
    journal = "Mach. Learn. Sci. Tech.",
    volume = "4",
    number = "4",
    pages = "045021",
    year = "2023"
}

@article{Creutz:1980zw,
    author = "Creutz, M.",
    title = "{Monte Carlo Study of Quantized SU(2) Gauge Theory}",
    doi = "10.1103/PhysRevD.21.2308",
    journal = "Phys. Rev. D",
    volume = "21",
    pages = "2308--2315",
    year = "1980"
}

@article{Maldacena:2023acv,
    author = "Maldacena, Juan",
    title = "{A simple quantum system that describes a black hole}",
    eprint = "2303.11534",
    archivePrefix = "arXiv",
    primaryClass = "hep-th",
    month = "3",
    year = "2023"
}

@article{Jordan:2012xnu,
    author = "Jordan, Stephen P. and Lee, Keith S. M. and Preskill, John",
    title = "{Quantum Algorithms for Quantum Field Theories}",
    eprint = "1111.3633",
    archivePrefix = "arXiv",
    primaryClass = "quant-ph",
    doi = "10.1126/science.1217069",
    journal = "Science",
    volume = "336",
    pages = "1130--1133",
    year = "2012"
}

@article{Feynman:1953zz,
    author = "Feynman, R. P.",
    title = "{Atomic Theory of Liquid Helium Near Absolute Zero}",
    doi = "10.1103/PhysRev.91.1301",
    journal = "Phys. Rev.",
    volume = "91",
    pages = "1301--1308",
    year = "1953"
}

@article{Unsal:2005yh,
    author = "Unsal, Mithat",
    title = "{Compact gauge fields for supersymmetric lattices}",
    eprint = "hep-lat/0504016",
    archivePrefix = "arXiv",
    reportNumber = "BUHEP-05-07",
    doi = "10.1088/1126-6708/2005/11/013",
    journal = "JHEP",
    volume = "11",
    pages = "013",
    year = "2005"
}

@article{Bergner:2025zkj,
    author = "Bergner, Georg and Hanada, Masanori",
    title = "{Exponential speedup in quantum simulation of Kogut-Susskind Hamiltonian via orbifold lattice}",
    eprint = "2506.00755",
    archivePrefix = "arXiv",
    primaryClass = "quant-ph",
    month = "5",
    year = "2025"
}

@article{Bose:1924,
  author  = {Bose, Satyendra Nath},
  title   = {Planck’s Law and the Hypothesis of Light Quanta},
  journal = {Zeitschrift f\"{u}r Physik},
  volume  = {26},
  pages   = {178--181},
  year    = {1924},
  doi     = {10.1007/BF01327326},
}

@article{Einstein:1924,
  author    = {Einstein, Albert},
  title     = {{Quantentheorie} des einatomigen idealen {Gases}. {Erster} {Teil}},
  journal   = {Sitzungsberichte der Preu\ss ischen Akademie der Wissenschaften, Physikalisch-Mathematische Klasse},
  pages     = {261--267},
  year      = {1924}
}

@article{Einstein:1925,
  author    = {Einstein, Albert},
  title     = {{Quantentheorie} des einatomigen idealen {Gases}. {Zweiter} {Teil}},
  journal   = {Sitzungsberichte der Preu\ss ischen Akademie der Wissenschaften, Physikalisch-Mathematische Klasse},
  pages     = {3--14},
  year      = {1925}
}

@article{Weyl:1929,
  author  = {Weyl, Hermann},
  title   = {Electron and Gravitation. I},
  journal = {Zeitschrift f\"{u}r Physik},
  volume  = {56},
  pages   = {330--352},
  year    = {1929},
  doi     = {10.1007/BF01339504},
}

@book{Weinberg:1996kr,
    author = "Weinberg, Steven",
    title = "{The quantum theory of fields. Vol. 2: Modern applications}",
    doi = "10.1017/CBO9781139644174",
    isbn = "978-1-139-63247-8, 978-0-521-67054-8, 978-0-521-55002-4",
    publisher = "Cambridge University Press",
    month = "8",
    year = "2013"
}

@article{Maldacena:1997re,
    author = "Maldacena, Juan Martin",
    title = "{The Large $N$ limit of superconformal field theories and supergravity}",
    eprint = "hep-th/9711200",
    archivePrefix = "arXiv",
    reportNumber = "HUTP-97-A097, HUTP-98-A097",
    doi = "10.4310/ATMP.1998.v2.n2.a1",
    journal = "Adv. Theor. Math. Phys.",
    volume = "2",
    pages = "231--252",
    year = "1998"
}

@article{tHooft:1971qjg,
    author = "'t Hooft, Gerard",
    editor = "Taylor, J. C.",
    title = "{Renormalizable Lagrangians for Massive Yang-Mills Fields}",
    doi = "10.1016/0550-3213(71)90139-8",
    journal = "Nucl. Phys. B",
    volume = "35",
    pages = "167--188",
    year = "1971"
}

@article{tHooft:1972tcz,
    author = "'t Hooft, Gerard and Veltman, M. J. G.",
    title = "{Regularization and Renormalization of Gauge Fields}",
    doi = "10.1016/0550-3213(72)90279-9",
    journal = "Nucl. Phys. B",
    volume = "44",
    pages = "189--213",
    year = "1972"
}

@article{Weinberg:1967tq,
    author = "Weinberg, Steven",
    title = "{A Model of Leptons}",
    doi = "10.1103/PhysRevLett.19.1264",
    journal = "Phys. Rev. Lett.",
    volume = "19",
    pages = "1264--1266",
    year = "1967"
}

@article{Glashow:1961tr,
    author = "Glashow, S. L.",
    title = "{Partial Symmetries of Weak Interactions}",
    doi = "10.1016/0029-5582(61)90469-2",
    journal = "Nucl. Phys.",
    volume = "22",
    pages = "579--588",
    year = "1961"
}

@article{Politzer:1973fx,
    author = "Politzer, H. David",
    editor = "Taylor, J. C.",
    title = "{Reliable Perturbative Results for Strong Interactions?}",
    doi = "10.1103/PhysRevLett.30.1346",
    journal = "Phys. Rev. Lett.",
    volume = "30",
    pages = "1346--1349",
    year = "1973"
}

@article{Gross:1973id,
    author = "Gross, David J. and Wilczek, Frank",
    editor = "Taylor, J. C.",
    title = "{Ultraviolet Behavior of Nonabelian Gauge Theories}",
    doi = "10.1103/PhysRevLett.30.1343",
    journal = "Phys. Rev. Lett.",
    volume = "30",
    pages = "1343--1346",
    year = "1973"
}

@article{Han:1965pf,
    author = "Han, M. Y. and Nambu, Yoichiro",
    editor = "Eguchi, T.",
    title = "{Three Triplet Model with Double SU(3) Symmetry}",
    doi = "10.1103/PhysRev.139.B1006",
    journal = "Phys. Rev.",
    volume = "139",
    pages = "B1006--B1010",
    year = "1965"
}

@article{Fritzsch:1973pi,
    author = "Fritzsch, H. and Gell-Mann, Murray and Leutwyler, H.",
    title = "{Advantages of the Color Octet Gluon Picture}",
    reportNumber = "CALT-68-409",
    doi = "10.1016/0370-2693(73)90625-4",
    journal = "Phys. Lett. B",
    volume = "47",
    pages = "365--368",
    year = "1973"
}

@article{Yang:1954ek,
    author = "Yang, Chen-Ning and Mills, Robert L.",
    editor = "Hsu, Jong-Ping and Fine, D.",
    title = "{Conservation of Isotopic Spin and Isotopic Gauge Invariance}",
    doi = "10.1103/PhysRev.96.191",
    journal = "Phys. Rev.",
    volume = "96",
    pages = "191--195",
    year = "1954"
}

@article{Halimeh:2024bth,
    author = "{Halimeh, Jad C. and Hanada, Masanori and Matsuura, Shunji and Nori, Franco and Rinaldi, Enrico and Sch\"{a}fer, Andreas}",
    title = "{A universal framework for the quantum simulation of Yang-Mills theory}",
    eprint = "2411.13161",
    archivePrefix = "arXiv",
    primaryClass = "quant-ph",
    reportNumber = "RIKEN-iTHEMS-Report-24",
    year = "2024"
}

@article{Susskind:1979up,
    author = "Susskind, Leonard",
    title = "{Lattice Models of Quark Confinement at High Temperature}",
    doi = "10.1103/PhysRevD.20.2610",
    journal = "Phys. Rev. D",
    volume = "20",
    pages = "2610--2618",
    year = "1979"
}

@article{Wilson:1974sk,
    author = "Wilson, Kenneth G.",
    editor = "Taylor, J. C.",
    title = "{Confinement of Quarks}",
    reportNumber = "CLNS-262",
    doi = "10.1103/PhysRevD.10.2445",
    journal = "Phys. Rev. D",
    volume = "10",
    pages = "2445--2459",
    year = "1974"
}

@article{Buser:2020cvn,
    author = "Buser, Alexander J. and Gharibyan, Hrant and Hanada, Masanori and Honda, Masazumi and Liu, Junyu",
    title = "{Quantum simulation of gauge theory via orbifold lattice}",
    eprint = "2011.06576",
    archivePrefix = "arXiv",
    primaryClass = "hep-th",
    reportNumber = "CALT-TH-2020-024, DMUS-MP-20/11, YITP-20-121",
    doi = "10.1007/JHEP09(2021)034",
    journal = "JHEP",
    volume = "09",
    pages = "034",
    year = "2021"
}

@article{Hanada:2020uvt,
    author = "Hanada, Masanori and Shimada, Hidehiko and Wintergerst, Nico",
    title = "{Color confinement and Bose-Einstein condensation}",
    eprint = "2001.10459",
    archivePrefix = "arXiv",
    primaryClass = "hep-th",
    doi = "10.1007/JHEP08(2021)039",
    journal = "JHEP",
    volume = "08",
    pages = "039",
    year = "2021"
}

@inproceedings{Berry:2013tiy,
    author = "Berry, Dominic W. and Childs, Andrew M. and Cleve, Richard and Kothari, Robin and Somma, Rolando D.",
    title = "{Exponential improvement in precision for simulating sparse Hamiltonians}",
    booktitle = "{46th annual ACM symposium on Theory of computing}",
    eprint = "1312.1414",
    archivePrefix = "arXiv",
    primaryClass = "quant-ph",
    doi = "10.1145/2591796.2591854",
    month = "12",
    year = "2013"
}

@article{Gharibyan:2020bab,
    author = "Gharibyan, Hrant and Hanada, Masanori and Honda, Masazumi and Liu, Junyu",
    title = "{Toward simulating Superstring/M-theory on a quantum computer}",
    eprint = "2011.06573",
    archivePrefix = "arXiv",
    primaryClass = "hep-th",
    reportNumber = "CALT-TH-2020-013, DMUS-MP-20/10, YITP-20-146",
    month = "11",
    year = "2020"
}

@article{Kogut:1974ag,
    author = "Kogut, John B. and Susskind, Leonard",
    title = "{Hamiltonian Formulation of Wilson's Lattice Gauge Theories}",
    reportNumber = "Print-74-1186 (CORNELL)",
    doi = "10.1103/PhysRevD.11.395",
    journal = "Phys. Rev. D",
    volume = "11",
    pages = "395--408",
    year = "1975"
}

@article{Kaplan:2002wv,
    author = "Kaplan, David B. and Katz, Emmanuel and Unsal, Mithat",
    title = "{Supersymmetry on a spatial lattice}",
    eprint = "hep-lat/0206019",
    archivePrefix = "arXiv",
    reportNumber = "INT-PUB-02-40, UW-PT-02-14",
    doi = "10.1088/1126-6708/2003/05/037",
    journal = "JHEP",
    volume = "05",
    pages = "037",
    year = "2003"
}

@article{Klco:2018zqz,
    author = "Klco, Natalie and Savage, Martin J.",
    title = "{Digitization of scalar fields for quantum computing}",
    eprint = "1808.10378",
    archivePrefix = "arXiv",
    primaryClass = "quant-ph",
    reportNumber = "INT-PUB-18-044",
    doi = "10.1103/PhysRevA.99.052335",
    journal = "Phys. Rev. A",
    volume = "99",
    number = "5",
    pages = "052335",
    year = "2019"
}

@article{Gautam:2022exf,
    author = "Gautam, Vaibhav and Hanada, Masanori and Holden, Jack and Rinaldi, Enrico",
    title = "{Linear confinement in the partially-deconfined phase}",
    eprint = "2208.14402",
    archivePrefix = "arXiv",
    primaryClass = "hep-th",
    reportNumber = "DMUS-MP-22/14, RIKEN-iTHEMS-Report-22",
    doi = "10.1007/JHEP03(2023)195",
    journal = "JHEP",
    volume = "03",
    pages = "195",
    year = "2023"
}

@article{Hanada:2025yzx,
    author = "Hanada, Masanori and Matsuura, Shunji and Mendicelli, Emanuele and Rinaldi, Enrico",
    title = "{Exponential improvement in quantum simulations of bosons}",
    eprint = "2505.02553",
    archivePrefix = "arXiv",
    primaryClass = "quant-ph",
    reportNumber = "RIKEN-iTHEMS-Report-25",
    month = "5",
    year = "2025"
}

@article{Bergner:2024qjl,
    author = "Bergner, Georg and Hanada, Masanori and Rinaldi, Enrico and Schafer, Andreas",
    title = "{Toward QCD on quantum computer: orbifold lattice approach}",
    eprint = "2401.12045",
    archivePrefix = "arXiv",
    primaryClass = "hep-th",
    reportNumber = "RIKEN-iTHEMS-Report-24",
    doi = "10.1007/JHEP05(2024)234",
    journal = "JHEP",
    volume = "05",
    pages = "234",
    year = "2024"
}

@article{Rinaldi:2021jbg,
    author = "Rinaldi, Enrico and Han, Xizhi and Hassan, Mohammad and Feng, Yuan and Nori, Franco and McGuigan, Michael and Hanada, Masanori",
    title = "{Matrix-Model Simulations Using Quantum Computing, Deep Learning, and Lattice Monte Carlo}",
    eprint = "2108.02942",
    archivePrefix = "arXiv",
    primaryClass = "quant-ph",
    reportNumber = "RIKEN-iTHEMS-Report-21, DMUS-MP-21/10",
    doi = "10.1103/PRXQuantum.3.010324",
    journal = "PRX Quantum",
    volume = "3",
    number = "1",
    pages = "010324",
    year = "2022"
}

@book{Peskin:1995ev,
    author = "Peskin, Michael E. and Schroeder, Daniel V.",
    title = "{An Introduction to quantum field theory}",
    doi = "10.1201/9780429503559",
    isbn = "978-0-201-50397-5, 978-0-429-50355-9, 978-0-429-49417-8",
    publisher = "Addison-Wesley",
    address = "Reading, USA",
    year = "1995"
}

@article{Kugo:1979gm,
    author = "Kugo, Taichiro and Ojima, Izumi",
    title = "{Local Covariant Operator Formalism of Nonabelian Gauge Theories and Quark Confinement Problem}",
    reportNumber = "KUNS-493",
    doi = "10.1143/PTPS.66.1",
    journal = "Prog. Theor. Phys. Suppl.",
    volume = "66",
    pages = "1--130",
    year = "1979"
}

@article{Halimeh:2025ivn,
    author = "Halimeh, Jad C. and Hanada, Masanori and Matsuura, Shunji",
    title = "{Universal framework with exponential speedup for the quantum simulation of quantum field theories including QCD}",
    eprint = "2506.18966",
    archivePrefix = "arXiv",
    primaryClass = "quant-ph",
    reportNumber = "RIKEN-iTHEMS-Report-25",
    month = "6",
    year = "2025"
}
\bibliographystyle{utphys}
\end{document}


\renewcommand{\appendixname}{Supplementary Material}
\renewcommand{\thesection}{S\arabic{section}}
\renewcommand{\theequation}{S\arabic{equation}}
\renewcommand{\thefigure}{S\arabic{figure}}
\renewcommand{\thetable}{S\arabic{table}}

\widetext

\section*{Supplementary Materials for ``Gauge Symmetry in Quantum Simulation --- Redundancy Ia All You Need"}
\section{Review of Yang-Mills theory and SU($N$) gauge symmetry}\label{sec:gauge_theory_review}
Yang-Mills theory is a special kind of gauge theory. Below, we provide a brief introduction to gauge theory putting emphasis on Yang-Mills theories. 
\subsection{Gauge Symmetry}
Gauge theory, gauge symmetry, global symmetry, local symmetry, etc, are somewhat confusing terms. 
For example, according to Wikipedia, 
\textit{In physics, a gauge theory is a type of field theory in which the Lagrangian, and hence the dynamics of the system itself, does not change under local transformations according to certain smooth families of operations (Lie groups).} (from Wikipedia, on September 27, 2025). This is a common misleading way to introduce ``gauge theory".    

In fact, gauge theory can be defined in a manner more suitable for quantum simulation based on canonical quantization, and such a reformulation enables us to define gauge theory based on discrete symmetry as well. 

Historically, gauge symmetry was recognized in electromagnetism~\cite{Weyl:1929}. Electromagnetism is U(1) gauge theory and the Lagrangian is invariant under local U(1) transformation. Note that this is a statement in classical mechanics. In the same manner, Yang-Mills theory is invariant under local SU($N$) transformation at the classical level. Physically meaningful quantities --- electric field and magnetic field --- are invariant under gauge transformation, and hence, field configurations connected by gauge transformation are physically equivalent. In practice, we can take a gauge-fixing condition convenient for the problems under consideration. 

If we try to quantize a theory with a massless vector field in a naive manner keeping the Lorentz invariance, a negative-norm mode appears from the time component of the vector $A_t$ as we saw in the introduction. This problem can be avoided for gauge fields because the negative mode can be gauged away. This is the reason Yang-Mills theory plays an important role in particle physics: A vector boson that mediates force can be introduced without having negative-norm modes~\cite{Peskin:1995ev,Weinberg:1996kr}. 

A straightforward way to quantize a theory respecting local U(1) or SU($N$) symmetry and Lorentz invariance is to perform Feynman path integral with a measure invariant under those transformations. As we show in Supplementary Material~\ref{sec:path-integral-and-Hamiltonian} for a concrete example, we can rewrite path integral expressions as those in canonical quantization in the $A_t=0$ gauge, and then, we can see that there is a residual gauge symmetry that is time-independent, and the states related by this residual gauge symmetry should be identified. 

Now we can give another characterization of gauge theory: Suppose there is a symmetry $G$ acting on Hilbert space, and the Hamiltonian is invariant under the action of $G$. Then, we identify the states in the Hilbert space that are connected by the action of $G$. Meaning this, we say `$G$ is a gauge symmetry' or `$G$ is gauged'~\footnote{
For such a restriction to make sense, Hamiltonian must be invariant under the action of $G$, because otherwise the singlet constraint is not compatible with Hamiltonian time evolution.
}. 
This characterization is valid both for continuous and discrete symmetries. 

Let us see a few concrete examples. 

\subsection{Identical bosons (S$_N$ gauge theory)}
The gauge group can be discrete. In fact, the history of non-Abelian gauge theory dates back to 1920s; the oldest example is the system of identical bosons studied by Bose and Einstein~\cite{Bose:1924,Einstein:1924,Einstein:1925}.

Let us consider the Hamiltonian of $N$ identical bosons in the harmonic potential in $\mathbb{R}^3$:
\begin{align}
\hat{H}
=
\frac{1}{2}\sum_{i=1}^N\left(
\hat{p}_{xi}^2
+
\hat{x}_{i}^2
+
\hat{p}_{yi}^2
+
\hat{y}_{i}^2
+
\hat{p}_{zi}^2
+
\hat{z}_{i}^2
\right).  
\end{align}

This Hamiltonian is invariant under the action of the S$_N$ permutation group, $i\to\sigma(i)$ ($\sigma\in\textrm{S}_N$). 
That $N$ bosons are ``identical" or ``indistinguishable" means that this S$_N$ symmetry is gauged. 

It is convenient to use the Fock states to describe the system. 
We use $\vec{n}_i=(n_{x,i},n_{y,i},n_{z,i})$ to denote the excitation level of the $i$-th boson. 
Then, the Fock states are defined as
\begin{align}
\ket{\vec{n}_1,\vec{n}_2,\cdots,\vec{n}_N}
\equiv
\prod_{j=x,y,z}
\frac{\hat{a}_{1j}^{\dagger n_{1j}}}{\sqrt{n_{1j}!}}
\frac{\hat{a}_{2j}^{\dagger n_{2j}}}{\sqrt{n_{2j}!}}
\cdots
\frac{\hat{a}_{Nj}^{\dagger n_{Nj}}}{\sqrt{n_{Nj}!}}
\ket{0}.
\end{align} 
These states span the extended Hilbert space $\mathcal{H}_{\textrm{ext}}$:
\begin{align}
    \mathcal{H}_{\rm ext}
    =
    \mathrm{Span}\{\ket{\vec{n}_1,\cdots,\vec{n}_N}\, |\, n_{a,j}\ge 0\}
    \nonumber
\end{align}
Here, $a=1,2,\cdots,N$ and $j=x,y,z$. 
The Fock vacuum $\ket{\vec{0},\cdots,\vec{0}}$ is the ground state. We denote the gauge-invariant subspace of the extended Hilbert space by $\mathcal{H}_{\rm inv}$. The permutation operator $\hat{\sigma}$ acts on $\mathcal{H}_{\rm ext}$ as
\begin{align}
\hat{\sigma}\ket{\vec{n}_1,\vec{n}_2,\cdots,\vec{n}_N}
=
\ket{\vec{n}_{\sigma(1)},\vec{n}_{\sigma(2)},\cdots,\vec{n}_{\sigma(N)}}\, . 
\end{align} 
All states related by permutations should be identified. Note that the Hamiltonian is $\mathrm{S}_N$-invariant, and hence permutation and Hamiltonian time evolution commute with each other. Therefore, if the redundancy is properly fixed at the initial time, there is no need to worry at later time.  
One way to remove the redundancy is to use the gauge-invariant states, or equivalently, totally symmetric states, which form the gauge-invariant Hilbert space $\mathcal{H}_{\rm inv}$:
\begin{align}
    \mathcal{H}_{\rm inv}
    =
    \mathrm{Span}\left\{\sum_{\sigma\in\mathrm{S}_N}\hat{\sigma}\ket{\vec{n}_1,\cdots,\vec{n}_N}\, |\, n_{a,j}\ge 0\right\}\, . 
\end{align}
Note that $\sum_{\sigma\in\mathrm{S}_N}\hat{\sigma}$ is a projector from $\mathcal{H}_{\rm ext}$ to $\mathcal{H}_{\rm inv}$. 

The canonical partition function at temperature $T=\beta^{-1}$ can be written in two ways, as
\begin{align}
Z(\beta)
=
\textrm{Tr}_{\mathcal{H}_{\rm inv}}\left(
e^{-\beta\hat{H}}
\right)
\end{align}
and
\begin{align}
Z(\beta)
=
\frac{1}{N!}
\sum_{\sigma\in\mathrm{S}_N}\textrm{Tr}_{\mathcal{H}_{\rm ext}}\left(
\hat{\sigma} e^{-\beta\hat{H}}
\right).   
\label{identical_boson_partition_function}
\end{align}
After the Euclidean time evolution, the projection operator $\frac{1}{N!}\sum_{\sigma\in\mathrm{S}_N}\hat{\sigma}$ is inserted. Because all states by permutations are indistinguishable, we simply take into account all of them and take the average. In the path-integral formulation, there is a sum over twisted boundary conditions associated with the $\mathrm{S}_N$ gauge symmetry~\cite{Feynman:1953zz}. 

The symmetric group S$_N$ is non-Abelian. Einstein considered the large-$N$ limit of this theory, and hence, he was the first person who studied the large-$N$ limit of non-Abelian gauge theory. What he discovered was fantastic --- Bose-Einstein condensation~\cite{Einstein:1925}.
\subsection{Electromagnetism (U(1) gauge theory)}
Let us start with electromagnetism, U(1) gauge theory in $\mathbb{R}^{1,3}$. We start with the classical theory. The field strength $F_{\mu\nu}$ can be written as
\begin{align}
F_{\mu\nu}
=
\partial_\mu A_\nu
-
\partial_\nu A_\mu
\end{align}
by using the vector potential, or equivalently the gauge field, $A_\mu$. We can also write it as 
$F_{\mu\nu}=\mathrm{i}[D_\mu,D_\nu]$, where $D_\mu=\partial_\mu-\mathrm{i}A_\mu$ is the gauge-covariant derivative. 
The field strengths are related to the electric field $E_j$ and magnetic field $B_j$ as $E_j=F_{0j}$ and $B_1=F_{23}$, $B_2=F_{31}$, $B_3=F_{12}$. They are invariant under the `gauge transformation' $A_\mu\to A'_\mu=A_\mu+\partial_\mu\Lambda$. Therefore, the gauge field is redundant, i.e., different configurations of the gauge field related by the gauge transformation describe the same electromagnetic fields. 

Now we consider the quantization in the $A_t=0$ gauge. Roughly speaking, the dynamical degrees of freedom are spatial gauge fields $A_j$, and the conjugate momenta are the electric fields $E_j=\dot{A}_j$. They are promoted to operators $\hat{A}_j$ and $\hat{E}_j$ which satisfy the canonical commutation relation. 
(A precise treatment requires the lattice Hamiltonian discussed in Supplementary Material~\ref{sec:Kogut_Susskind}, \ref{sec:orbifold_lattice}, and \ref{sec:path-integral-and-Hamiltonian}.)

The equation of motion for $A_t$ leads to $\sum_{i=1}^3\partial_jE_j=0$, i.e., the Gauss-law constraint. As we will derive in Supplementary Material~\ref{sec:path-integral-and-Hamiltonian}, the counterpart in the quantum theory in the $A_t=0$ gauge is the gauge redundancy at each time slice, which corresponds to the residual gauge symmetry in the classical theory, i.e., time-independent gauge transformations.

\subsection{Yang-Mills theory (SU($N$) gauge theory)}
In this section, we introduce the SU($N$) Yang-Mills theory on $\mathbb{R}^{1,d}$. This is a natural generalization of the U(1) gauge theory described above. The Lagrangian is 
\begin{align}
L=\frac{1}{4}\int d^dx {\rm Tr}F_{\mu\nu}^2\, ,
\end{align}
where $F_{\mu\nu}$ is the field strength defined by 
\begin{align}
F_{\mu\nu}
=
\mathrm{i}[D_\mu,D_\nu]
=
\partial_\mu A_\nu
-
\partial_\nu A_\mu
-
\mathrm{i}g[A_\mu,A_\nu]\, . 
\end{align}
Here, $A_\mu(t,\vec{x})$ is the gauge field, which is $N\times N$ traceless Hermitian, and $D_\mu=\partial_\mu-\mathrm{i}A_\mu$ is the gauge covariant derivative. 

The gauge transformation is $D_\mu\to D'_\mu = e^{\mathrm{i}\Lambda}D_\mu e^{-\mathrm{i}\Lambda}$, where $D'_\mu=\partial_\mu-\mathrm{i}A'_\mu$, and $F_{\mu\nu}\to F'_{\mu\nu}=\mathrm{i}[D'_\mu,D'_\nu]=e^{\mathrm{i}\Lambda}F_{\mu\nu} e^{-\mathrm{i}\Lambda}$. (The same expressions hold for the U(1) theory as well; in that case $F_{\mu\nu}$ and $\Lambda$ are not matrices but ordinary numbers, and hence, $F'_{\mu\nu}=F_{\mu\nu}$.) Because of the trace cyclicity, ${\rm Tr}F_{\mu\nu}^2$ is invariant under this gauge transformation.

The equation of motion for $A_0$ is $\sum_{j=1}^3D_jE_j=0$, where $D_j E_k\equiv\partial_jE_k-\mathrm{i}[A_j,E_k]$. This is the generalization of Gauss's law. In quantum theory, the Gauss-law constraint appears in the path-integral formulation by imposing the gauge condition $A_t=0$. In the operator formulation, the counterpart of Gauss's law is that the states that are connected by the (residual) gauge transformation are equivalent. This can be seen from \eqref{identical_boson_partition_function} derived in Supplementary Material~\ref{sec:path-integral-and-Hamiltonian}, which is essentially the same as \eqref{YM_partition_function}.

\section{Hamiltonian formulation in temporal gauge and path-integral formulation}\label{sec:path-integral-and-Hamiltonian}
In this Supplementary Material, we show how the path integral formulation and the Hamiltonian formulation in the $A_t=0$ gauge are related to each other. We will consider the case of the orbifold lattice. For Kogut-Susskind, see ref.~\cite{Gautam:2022exf}. 

As a starter, let us consider the canonical partition function. 
We only have to repeat a famous computation of Feynman's, adding gauge field. 
We split Euclidean time evolution into $K$ steps, and send $K$ to infinity. In addition, we insert $\textbf{1}=\hat{\Omega}_{(k)}^{-1}\hat{\Omega}_{(k)}$ at each step:
\begin{align}
Z(T)
&=
\frac{1}{\textrm{vol}\mathcal{G}}
\int\left(\prod_{k=1}^Kd\Omega_k\right)
{\rm Tr}_{{\cal H}_{\rm ext}}
\Bigl(
\hat{\Omega}_{(K)}
e^{-\frac{H(\hat{P},\hat{Z})}{TK}}
\hat{\Omega}_{(K-1)}^{-1}\hat{\Omega}_{(K-1)}
e^{-\frac{H(\hat{P},\hat{Z})}{TK}}
\hat{\Omega}_{(K-2)}^{-1}\hat{\Omega}_{(K-2)}
\cdots
\hat{\Omega}_{(1)}^{-1}\hat{\Omega}_{(1)}
e^{-\frac{H(\hat{P},\hat{Z})}{TK}}
\Bigl)
\nonumber\\
&=
\frac{1}{[{\rm volU}(N)]^K}
\int\left(\prod_{k=1}^Kd\Omega_{(k)}\right)
\int\left(\prod_{k=1}^KdZ_{(k)}\right)
\bra{Z_{(K)}}
\hat{\Omega}_{(K)}
e^{-\frac{H(\hat{P},\hat{Z})}{TK}}
\hat{\Omega}_{(K-1)}^{-1}
\ket{Z_{(K-1)}}
\nonumber\\
&
\qquad\qquad\qquad
\times
\bra{Z_{(K-1)}}
\hat{\Omega}_{(K-1)}
e^{-\frac{H(\hat{P},\hat{Z})}{TK}}
\hat{\Omega}_{(K-2)}^{-1}
\ket{Z_{(K-2)}}
\times
\cdots
\times
\bra{Z_{(1)}}
\hat{\Omega}_{(1)}
e^{-\frac{H(\hat{P},\hat{Z})}{TK}}
\ket{Z_{(K)}}\, . 
\end{align}
For $H(\hat{P},\hat{Z})=\mathrm{Tr}\hat{P}\hat{\bar{P}}+V(\hat{Z})$, we can rewrite each term in the product as
\begin{align}
&
\bra{Z_{(k)}}
\hat{\Omega}_{(k)}
e^{-\frac{H(\hat{P},\hat{Z})}{TK}}
\hat{\Omega}_{(k-1)}^{-1}
\ket{Z_{(k-1)}}
\nonumber\\
&\quad=
\bra{Z_{(k)}^{(\Omega_{(k)})}}
e^{-\frac{H(\hat{P},\hat{Z})}{TK}}
\ket{Z_{(k-1)}^{(\Omega_{(k-1)})}}
\nonumber\\
&\quad=
e^{-\frac{1}{TK}V(Z_{(k-1)}^{(\Omega_{(k-1)})})}
\bra{Z_{(k)}^{(\Omega_{(k)})}}
e^{-\frac{1}{TK}{\rm Tr}\hat{P}\hat{\bar{P}}}
\ket{Z_{(k-1)}^{(\Omega_{(k-1)})}}
\nonumber\\
&\quad=
e^{-\frac{1}{TK}V(Z_{(k-1)})}
\int dP
\bra{Z_{(k)}^{(\Omega_{(k)})}}
\ket{P}\left\langle P\Bigl|
e^{-\frac{1}{TK}{\rm Tr}\hat{P}\hat{\bar{P}}}
\Bigl| Z_{(k-1)}^{(\Omega_{(k-1)})}\right\rangle
\nonumber\\
&\quad=
e^{-\frac{1}{TK}V(Z_{(k-1)})}
\int dP e^{-\frac{1}{TK}{\rm Tr}P\bar{P}}
\exp\Bigl(
\mathrm{i}\mathrm{Tr}\Bigl[P\left(\bar{Z}_{(k)}^{(\Omega_{(k)})}-\bar{Z}_{(k-1)}^{(\Omega_{(k-1)})}\right)
+
\bar{P}\left(Z_{(k)}^{(\Omega_{(k)})}-Z_{(k-1)}^{(\Omega_{(k-1)})}\right)
\Bigl]\Bigl)
\nonumber\\
&\quad=
e^{-\frac{1}{TK}V(Z_{(k-1)})}
\exp\Bigl(
-TK\mathrm{Tr}\left|Z_{(k)}^{(\Omega_{(k)})}-Z_{(k-1)}^{(\Omega_{(k-1)})}\right|^2
\Bigl)
\nonumber\\
&\quad=
e^{-\frac{1}{TK}V(Z_{(k-1)})}
\exp\Bigl(
-TK\mathrm{Tr}\left|Z_{(k)}^{(\Omega_{(k-1)}^{-1}\Omega_{(k)})}-Z_{(k-1)}\right|^2
\Bigl)
\nonumber\\
&\quad\simeq
\exp\Bigl(
-\frac{1}{TK}V(Z_{(k-1)})
-
\frac{1}{TK}\mathrm{Tr}\left|D_tZ_{(k)}\right|^2
\Bigl)\, , 
\end{align}
where $\simeq$ becomes $=$ as $K\to\infty$. 
Here we used
\begin{align}
\Omega^{-1}_{(k-1)}\Omega_{(k)}\equiv e^{-\mathrm{i}A_{(k)}/(KT)}, 
\end{align}
and
\begin{align}
Z_{(k)}^{(\Omega_{(k-1)}^{-1}\Omega_{(k)})}-Z_{(k-1)}
\simeq
\frac{D_tZ_{(k)}}{KT}\, . 
\end{align}
Therefore, in the limit $K\to\infty$, we obtain the Euclidean path integral, 
\begin{align}
Z(T)
&=
\int [dA] [dZ]\exp\left(-\int dt L[D_tZ,Z]\right)\, , 
\end{align}
where $L$ is the Euclidean action, 
\begin{align}
L[D_tZ,Z]
=
\mathrm{Tr}\left(
(D_tZ)(D_t\bar{Z})
+
V(Z)
\right)\, . 
\end{align}
Note that $\Omega_{(K)}$ is the Polyakov loop. 

For the Hamiltonian time evolution, we start with
\begin{align}
\bra{\Psi_{\rm final}}
\hat{\mathcal{P}}
\exp\left(
-\mathrm{i}\hat{H}(t_{\rm final}-t_{\rm initial})
\right)
\ket{\Psi_{\rm initial}}\, , 
\end{align}
where $\hat{\mathcal{P}}=\frac{1}{{\rm Vol}G}\int_Gd\Omega \hat{\Omega}$. Then, we can repeat essentially the same computations above and show that this is the same as the Lorentzian path integral
\begin{align}
&
\int [dA] [dZ]\exp\left(\mathrm{i}\int_{t_{\rm initial}}^{t_{\rm final}} dt L[D_tZ,Z]\right)
\nonumber\\
&\quad\times
\Psi_{\rm final}^\ast(Z(t_{\rm final}))\Psi_{\rm initial}(Z(t_{\rm initial}))\, . 
\end{align}
\section{Review of the universal framework for quantum simulation of Yang-Mills theory}\label{sec:universal_framework}
In this Supplementary Material, we review a universal simulation framework for Yang-Mills theory~\cite{Halimeh:2024bth}\footnote{
The simulation protocols described below are not new and have long been applied to simpler systems such as the harmonic oscillator and scalar quantum field theory~\cite{Jordan:2012xnu,Klco:2018zqz}. More recently, they have been applied to quantum electrodynamics~\cite{Yao:2025uxz,Mou:2025iiu}. The nontrivial observation~\cite{Halimeh:2024bth} is that this elementary, textbook-style approach suffices for practically all systems of interest, including Yang-Mills theory and QCD for arbitrary gauge groups and spacetime dimensions.
}. A key observation is that the orbifold lattice Hamiltonian takes a simple and universal form common to various systems including matrix models~\cite{Gharibyan:2020bab,Maldacena:2023acv}, scalar quantum field theories~\cite{Jordan:2012xnu}, and even a harmonic oscillator. Specifically, the Hamiltonian can be written as 
\begin{align}
\hat{H}
=
\frac{1}{2}\sum_a\hat{p}_a^2
+
V(\hat{x})\, , 
\end{align}
with the canonical commutation relations
\begin{align}
[\hat{x}_a,\hat{p}_b]
=
\mathrm{i}\delta_{ab}\, , 
\qquad
[\hat{x}_a,\hat{x}_b]
=
[\hat{p}_a,\hat{p}_b]
=
0\, . 
\end{align}
For the orbifold lattice, the real and imaginary parts of each matrix entries of $\hat{Z}$'s (resp., $\hat{P}$'s) corresponds to $\hat{x}$'s (resp., $\hat{p}$'s). 
The potential $V$ is a quartic polynomial. If we add U(1) mass to the SU(3) theory to take the SU($N$) Kogut-Susskind limit, $V$ is a polynomial of degree six. Below, we assume $V$ is a polynomial of degree $n$, and we do not specify the value of $n$. 
\subsection{Hamiltonian time evolution via Trotterization}\label{sec:Trotter_cost_piece}
For simplicity, we assume that $V$ is a polynomial of degree $n$. It is straightforward to generalize our analysis to other polynomial potentials. 

We treat the momentum part and the interaction part separately. 
Because $[\hat{p}_j,\hat{p}_k]=0$, the momentum part factorizes as 
\begin{align}
\exp\left(-\textrm{i}\theta\sum_j\hat{p}^2_j\right)
=
\prod_{j}\exp\left(-\textrm{i}\theta\hat{p}^2_j\right)\, . 
\end{align}
The potential part factorizes as well.
Schematically,
\begin{align}
\prod_{j_1,\cdots,j_n}\exp\left(-\textrm{i}\theta \mathcal{C}_{j_1\cdots j_n}\, 
\hat{x}_{j_1}\cdots\hat{x}_{j_n}\right)\, , 
\end{align}
where coefficients $\mathcal{C}_{j_1\cdots j_n}$ define the potential. 

Using \eqref{eq:x_with_Z} and \eqref{eq:p_with_Z}, we can write them schematically as 
\begin{align}
    \exp\left(-\mathrm{i}\theta\hat{p}^2\right)
    =
    \prod_{a<b}
    \exp\left(-\mathrm{i}\theta \mathcal{C}''_{ab}\sigma_{z,a}\sigma_{z,b}\right)\, . 
\end{align}
and
\begin{align}
&
\exp\left(-\textrm{i}\theta \mathcal{C}_{j_1\cdots j_n}\, 
\hat{x}_{j_1}\cdots\hat{x}_{j_n}\right)
\nonumber\\
&\quad=
\prod_{a_1\cdots a_n}\exp\left(-\textrm{i}\theta \mathcal{C}'_{a_1\cdots a_n}\sigma_{z,a_1}\cdots \sigma_{z,a_n}\right)\, . 
\end{align}

To simplify these operators, we use the CNOT gate with the control qubit $a$ and target qubit $b$, which is defined by 
\begin{align}
    C_{a,b}\ket{n_a}_a\ket{n_b}_b
    =
    \ket{n_a}_a\ket{n_a\oplus n_b}_b\, .
\end{align}
Using this and the definition of $\sigma_z$, 
\begin{align}
    \sigma_{z,a}\ket{n_a}=(-1)^{n_a}\ket{n_a}\, , 
\end{align}
we can show 
\begin{align}
    Z_{a_1}Z_{a_2}\cdots Z_{a_n} 
    =
    \left(\prod_{i=1}^{n-1}C_{a_i,a_n}\right)
    \cdot Z_{a_n}\cdot
    \left(\prod_{i=1}^{n-1}C_{a_i,a_n}\right)\, , 
\end{align}  
\begin{align}      
    &\exp\left(-\mathrm{i}\theta\mathcal{C}'_{a_1\cdots a_n} Z_{a_1}Z_{a_2}\cdots Z_{a_n} \right)
    =
    \left(\prod_{i=1}^{n-1}C_{a_i,a_n}\right)
    \cdot \exp\left(-\mathrm{i}\theta\mathcal{C}'_{a_1\cdots a_n} Z_{a_n}\right)\cdot
   \left(\prod_{i=1}^{n-1}C_{a_i,a_n}\right)\, .
\end{align}  
The decomposition into CNOT gates is not unique. For example, we can also use 
\begin{align}
&
    Z_{a_1}Z_{a_2}\cdots Z_{a_n}
     =
    \left(C_{a_1,a_2}\cdots C_{a_{n-1},a_n}\right)
    \cdot Z_{a_n}\cdot
    \left(C_{a_{n-1},a_n}\cdots C_{a_1,a_2}\right)\, .
\end{align}

The gate complexity for simulating the Hamiltonian $\hat{H} = \hat{H}_{\text{kin}} + \hat{H}_{\text{pot}}$ via Trotterization is as follows. The first-order Trotter approximation $e^{-i\hat{H}t} \approx (e^{-i\hat{H}_{\text{kin}}\Delta t} e^{-i\hat{H}_{\text{pot}}\Delta t})^r$ has error scaling as $\mathcal{O}(r^{-1}t^2\|\hat{H}\|^2)$, requiring $r = \mathcal{O}(t^2\|\hat{H}\|^2/\epsilon)$ Trotter steps to achieve target error $\epsilon$. The second-order symmetric splitting $e^{-i\hat{H}t} \approx (e^{-i\hat{H}_{\text{kin}}\Delta t/2} e^{-i\hat{H}_{\text{pot}}\Delta t} e^{-i\hat{H}_{\text{kin}}\Delta t/2})^r$ improves this to $r = \mathcal{O}(t^{3/2}\|\hat{H}\|^{3/2}/\epsilon^{1/2})$ steps due to its $\mathcal{O}(\Delta t^3)$ error per step. 

Each Trotter step requires quantum Fourier transformations to switch between coordinate and momentum bases. The kinetic term $\hat{H}_{\text{kin}}$ is diagonal in the momentum basis, necessitating one forward QFT to transform from coordinate to momentum representation, followed by an inverse QFT to return to the coordinate basis. For the orbifold lattice, with $Q$ qubits per real variable and $n_{\text{link}} \times 2N^2$ real variables in total (where $n_{\text{link}} = 3V$ is the number of links for three spatial dimensions, each being an $N \times N$ complex matrix), each QFT requires $\mathcal{O}(Q^2)$ gates per variable, giving a total QFT cost of $\mathcal{O}(n_{\text{link}} N^2 Q^2)$ gates per Trotter step. The potential term $\hat{H}_{\text{pot}}$, being diagonal in the coordinate basis, requires $\mathcal{O}(n_{\text{link}} N^4 Q^4)$ gates per step using CNOT ladder decompositions for the quartic interactions~\footnote{
If $m^2_{\rm U(1)}$ in \eqref{eq:orbifold_mass_term} were not set to zero, the cost would be $\mathcal{O}(n_{\text{link}} (N!)^2 Q^{2N})$.
}. Therefore, the total gate complexity of first-order Trotterization is $\mathcal{O}(t^2\|\hat{H}\|^2 n_{\text{link}}(N^2 Q^2 + N^4 Q^4)/\epsilon)$, and of second-order Trotterization is $\mathcal{O}(t^{3/2}\|\hat{H}\|^{3/2} n_{\text{link}}(N^2 Q^2 + N^4 Q^4)/\epsilon^{1/2})$.

Note that this is an overly pessimistic estimate that considers the entire energy spectrum, including the high-energy modes whose energy diverges as truncation is removed, and which are irrelevant for low-energy dynamics that describes the continuum limit. Because the largest energy scale we consider is either $a^{-1}$, $m$, or $m_{\rm U(1)}$, and for a precise description of the states below these energy scales, we do not have to scale the Trotter step size with $||\hat{H}||$~\footnote{Total energy of the system can increase with the volume if the energy density is fixed, however we would not scale the Trotter step with volume as long as we care about the error rate per volume.}. See Sec.~\ref{sec:Trotter_error_numerical_test} for numerical experiments which confirm this point. 
\subsubsection{Adding gauge-penalty term}\label{sec:gauge_penalty_implementation}
As explained in Sec.~\ref{sec:penalty_G^2}, a simple way to remove gauge non-singlets from the low-energy spectrum is to add a penalty to the non-singlet states. 
Using non-compact variables, the generators of U($N$) transformation acting on the link variable from left and right are defined as
\begin{align}
\hat{\mathcal{E}}_{j,\vec{n}}^{(\mathrm{L})ab}
&=
\sum_c
\left(
\mathrm{i}\hat{Z}_{j,\vec{n}}^{ac}\cdot\hat{\bar{P}}^{cb}_{j,\vec{n}}
-
\mathrm{i}\hat{\bar{Z}}^{cb}_{j,\vec{n}}\cdot\hat{P}^{ac}_{j,\vec{n}}
\right)
\end{align} 
and
\begin{align}
\hat{\mathcal{E}}_{j,\vec{n}}^{(\mathrm{R})ab}
&=
\sum_c
\left(
\mathrm{i}\hat{Z}^{cb}_{j,\vec{n}}\cdot\hat{\bar{P}}^{ac}_{j,\vec{n}}
-
\mathrm{i}\hat{\bar{Z}}^{ac}_{j,\vec{n}}\cdot\hat{P}^{cb}_{j,\vec{n}}
\right)\, . 
\end{align} 
We can use U($N$) generators $\tau_\alpha$ ($\alpha=1,\cdots,N^2$) to extract the generators of U($N$) as 
\begin{align}
    \hat{\mathcal{E}}^{(\mathrm{L})\alpha}
    =
    \mathrm{Tr}\left(
    \tau_\alpha
    \hat{\mathcal{E}}^{(\mathrm{L})}
    \right)\, , 
    \qquad
        \hat{\mathcal{E}}^{(\mathrm{R})\alpha}
    =
    \mathrm{Tr}\left(
    \tau_\alpha
    \hat{\mathcal{E}}^{(\mathrm{R})}
    \right)\, .      
\end{align}
The generators of SU($N$) gauge transformation is 
\begin{align}
\hat{G}^\alpha_{\vec{n}}
=
\sum_{j=1}^3
\left(
-\hat{\mathcal{E}}^{(\mathrm{L})\alpha}_{j,\vec{n}}
+
\hat{\mathcal{E}}^{(\mathrm{R})\alpha}_{j,\vec{n}-\hat{j}}
\right)\, . 
\end{align}
By adding a penalty term proportional to $\sum_{\alpha,\vec{n}}\left(\hat{G}_{\vec{n}}^\alpha\right)^2$ to the Hamiltonian, we can remove gauge non-singlets. 
At each site $\vec{n}$, we have
\begin{align}
\sum_{\alpha}\left(\hat{G}_{\vec{n}}^\alpha\right)^2
&=
\sum_{j,j'=1}^3
\left(
-\hat{\mathcal{E}}^{(\mathrm{L})\alpha}_{j,\vec{n}}
+
\hat{\mathcal{E}}^{(\mathrm{R})\alpha}_{j,\vec{n}-\hat{j}}
\right)
\left(
-\hat{\mathcal{E}}^{(\mathrm{L})\alpha}_{j',\vec{n}}
+
\hat{\mathcal{E}}^{(\mathrm{R})\alpha}_{j',\vec{n}-\hat{j}}
\right)
\nonumber\\
&=
\sum_j
\left(
\textcolor{red}{
\left(\hat{\mathcal{E}}^{(\mathrm{L})\alpha}_{j,\vec{n}}\right)^2
}
+
\textcolor{red}{
\left(\hat{\mathcal{E}}^{(\mathrm{R})\alpha}_{j,\vec{n}-\hat{j}}\right)^2
}
-
\textcolor{blue}{
2\hat{\mathcal{E}}^{(\mathrm{L})\alpha}_{j,\vec{n}}
\hat{\mathcal{E}}^{(\mathrm{R})\alpha}_{j,\vec{n}-\hat{j}}
}
\right)
+
\sum_{j\neq j'}
\textcolor{blue}{
\left(
-\hat{\mathcal{E}}^{(\mathrm{L})\alpha}_{j,\vec{n}}
+
\hat{\mathcal{E}}^{(\mathrm{R})\alpha}_{j,\vec{n}-\hat{j}}
\right)
\left(
-\hat{\mathcal{E}}^{(\mathrm{L})\alpha}_{j',\vec{n}}
+
\hat{\mathcal{E}}^{(\mathrm{R})\alpha}_{j',\vec{n}-\hat{j}}
\right)
}\, . 
\label{eq:G^2-penalty-term}
\end{align}

As we mentioned above, for the orbifold lattice, the real and imaginary parts of each matrix entry of $\hat{Z}$'s (resp., $\hat{P}$'s) correspond to $\hat{x}$'s (resp., $\hat{p}$'s). 
Terms written in blue can be written as a sum of terms of the form $\hat{p}_a\hat{x}_b\hat{p}_c\hat{x}_d$, where $a$, $b$, $c$, and $d$ are all different. In this case, we can use the momentum basis for $\hat{p}_a$ and $\hat{p}_c$, and the coordinate basis for $\hat{x}_b$ and $\hat{x}_d$. In this basis, we can write $\hat{p}_a\hat{x}_b\hat{p}_c\hat{x}_d$ in the following form:
\begin{align}
\hat{p}_a\hat{x}_b\hat{p}_c\hat{x}_d
=
\sum_{pqrs}\mathcal{C}'_{pqrs}Z_pZ_qZ_rZ_s\, . 
\end{align}
(Note that $Z$ here means Pauli $Z$ operator, which is different from $\hat{Z}_{j,\vec{n}}$.) 
Therefore, we can use the method we introduced above for $\hat{p}^2$ and $\hat{x}_1\cdots\hat{x}_n$. 

The terms in \eqref{eq:G^2-penalty-term} shown in red  contains terms like $\hat{p}_a\hat{x}_b\hat{p}_c\hat{x}_a$, for which the same method is not applicable. To resolve this issue and build a simple circuit, we note that $\left(\hat{\mathcal{E}}^{(\mathrm{L})\alpha}_{j,\vec{n}}\right)^2$ and $\left(\hat{\mathcal{E}}^{(\mathrm{R})\alpha}_{j,\vec{n}}\right)^2$ are the Laplacian on the SU($N$) group manifold. This manifold is embedded into $\mathbb{C}^{N^2}=\mathbb{R}^{2N^2}$, and hence, the  Laplacian on this space, $\mathrm{Tr}(\hat{P}_{j,\vec{n}}\hat{\bar{P}}_{j,\vec{n}})$, can be written as the Laplacian on SU($N$) and that on the rest of the degrees of freedom. The latter takes large values and decouple from the low-energy spectrum when the mass parameters in    \eqref{eq:orbifold_mass_term} are large. Taking a normalization factor into account, we can replace both $\left(\hat{\mathcal{E}}^{(\mathrm{L})\alpha}_{j,\vec{n}}\right)^2$ and $\left(\hat{\mathcal{E}}^{(\mathrm{R})\alpha}_{j,\vec{n}}\right)^2$ with $\frac{a}{2g^2}\mathrm{Tr}(\hat{P}_{j,\vec{n}}\hat{\bar{P}}_{j,\vec{n}})$. This can be absorbed into the kinetic term of the Hamiltonian.

\subsection{Linear Combination of Unitaries}\label{sec:orbifold_block_encoding}
As reviewed in Supplementary Material~\ref{Sec:algorithms}, convenient algorithms can be used when an operator of interest can be written as Linear Combination of Unitaries (LCU), which is straightforward for the orbifold lattice Hamiltonian. 
Using \eqref{eq:x_with_Z} and \eqref{eq:p_with_Z}, the Hamiltonian is written in the coordinate basis in terms of Pauli strings $\hat{\Pi}_i\sim\sigma_z\otimes\sigma_z\otimes\cdots\otimes\sigma_z$, $\hat{\tilde{\Pi}}_{j}\sim\sigma_z\otimes\sigma_z$ and the quantum Fourier transforms $\hat{F}$ as
\begin{align}
\hat{H}
=
\sum_{i} \alpha_{i} \hat{\Pi}_{i}
+
\sum_{j} \tilde{\alpha}_{j}
\hat{F}^\dagger \hat{\tilde{\Pi}}_{j}\hat{F}
, \qquad \alpha_{i}>0,\quad \tilde{\alpha}_{j} > 0 
\end{align}
The first and second terms on the right-hand side are the potential $V(\hat{x}_1,\cdots\hat{x}_K)$ and kinetic term $\frac{1}{2}\sum_a\hat{p}_a^2$, respectively. 

The operator $\mathcal{T}^{(\xi)}$ defined and used in Sec.~\ref{sec:singlet_simulation} for gauge-singlet projection can be expressed as an LCU order-by-order in the Taylor expansion of the exponential. Each term can be written as 
\begin{align}    \hat{F}^\dagger\hat{\tilde{\Pi}}\hat{F}\hat{\Pi}\, . 
\end{align}

\section{Potentially useful quantum algorithms}\label{Sec:algorithms}

\subsection{Review of Block Encoding via Linear Combination of Unitaries}
For a given state $\ket{\Phi}$ (say the ground state), we want to obtain $\frac{\hat{O}\ket{\Phi}}{||\hat{O}\ket{\Phi}||}$, where $\hat{O}$ is written as a linear combination of unitaries:
\begin{align}
    \hat{O}
    =
    \sum_{i}
    \alpha_i\hat{\Pi}_i\, ,
    \qquad\alpha_i>0\, . 
\end{align}
Here, each $\hat{\Pi}_i$ is a unitary operator which is often, but not necessarily, a Pauli string. 
Note that $\hat{O}$ is not necessarily unitary or Hermitian. For this purpose, a block encoding via Linear Combination of Unitaries (LCU), which is reviewed below, can be used. 

We introduce ancilla states $\ket{i}_{\rm ancilla}$ ($i=1,2,\cdots$) and construct the unitary operator $\hat{U}$ from control-$\hat{\Pi}_i$ operators,  
\begin{align}
\hat{U} 
=
\sum_i
\left(
\left(\ket{i}\bra{i}\right)_{\rm ancilla}
\otimes
\hat{\Pi}_i
\right)\, . 
\end{align}
We prepare a state $\ket{\alpha}$ defined by  
\begin{align}
\ket{\alpha}  
=  
\sum_i\sqrt{\frac{\alpha_i}{\sum_j\alpha_j}}\ket{i}_{\rm ancilla}\, .  
\end{align}
To prepare $\ket{G}$, we used a unitary transformation $\hat{A}$ acting on the ancilla states that satisfies
\begin{align}
    \hat{A}\ket{1}_{\rm ancilla}=\ket{\alpha}\, . 
\end{align}
Then, we act with $\hat{U}$ on $\ket{\alpha}\ket{\Phi}$:
\begin{align}
\hat{U}\left(\ket{\alpha}\ket{\Phi}\right)
=
\sum_i\sqrt{\frac{\alpha_i}{\sum_j\alpha_j}}\ket{i}_{\rm ancilla}\left(\hat{\Pi}_i\ket{\Phi}\right)\, . 
\end{align}
Now we act with $\hat{A}^\dagger$ on this state, and measure the ancilla states. When the measured state is $\ket{1}_{\rm ancilla}$, the state is, up to normalization, 
\begin{align}
\ket{1}_{\rm ancilla}
\bra{1}_{\rm ancilla}\hat{A}^\dagger\hat{U}\left(\ket{\alpha}\ket{\Phi}\right)
&=
\ket{1}_{\rm ancilla}
\bra{\alpha}\hat{A}^\dagger\hat{U}\left(\ket{\alpha}\ket{\Phi}\right)
\nonumber\\
&=
\frac{1}{\sum_j\alpha_j}
\ket{1}_{\rm ancilla}
\hat{O}\ket{\Phi}\, . 
\end{align}
Ignoring the ancilla qubits and taking into account the normalization factor, we have $\frac{\hat{O}\ket{\Phi}}{||\hat{O}\ket{\Phi}||}$, which we wanted. 

To increase the success rate of post selection, Oblivious Amplitude Amplification~\cite{Berry:2013tiy} can be used.

\subsection{Product of operator-valued $N\times N$ matrices}\label{Sec:algorithms-matrix-multiplication}
Suppose that we want to act with a Wilson loop operator
\begin{align}
&
\hat{W}\equiv{\rm Tr}(\hat{Z}_{j_1,\vec{x}}\hat{Z}_{j_2,\vec{x}+\hat{j}_1}\cdots\hat{Z}_{j_l,\vec{x}-\hat{j}_l})
\nonumber\\
&=
\sum_{a_1,\cdots,a_l}
\hat{Z}_{j_1,\vec{x};a_1a_2}\hat{Z}_{j_2,\vec{x}+\hat{j}_1;a_2a_3}\cdots\hat{Z}_{j_l,\vec{x}-\hat{j}_l;a_la_1}\, .
\end{align}
on $\ket{\rm VAC}$ to get a gauge-invariant state. Naively, we need to take a sum of $N^l$ terms. This is problematic because we need to take $l$ large as the lattice spacing becomes small to keep the physical size of the loop finite.

One way to avoid such an increase is to introduce ancillary states $\ket{i}_{\vec{x}}$ ($i=1,2,\cdots,N$) at each site $\vec{x}$ and define 
\begin{align}
    \hat{\mathcal{Z}}^{(n)}_{j,\vec{x}}
    \equiv
    \sum_{k,l=1}^N
    \ket{k}_{\vec{x}}
    \bra{l}_{\vec{x}+\hat{j}}
    e^{2\pi\mathrm{i}n(k-l)/N}   
    \hat{Z}_{j,\vec{x};kl}
\end{align}
for $n=1,2,\cdots,N$. 
Then, 
\begin{align}
&
\hat{\mathcal{Z}}^{(n)}_{j_1,\vec{x}}\hat{\mathcal{Z}}^{(n)}_{j_2,\vec{x}+\hat{j}_1}\cdots\hat{\mathcal{Z}}^{(n)}_{j_l,\vec{x}-\hat{j}_l}
=
\sum_{k,l=1}^N
    \ket{k}_{\vec{x}}
    \bra{l}_{\vec{x}}
    e^{2\pi\mathrm{i}n(k-l)/N}       
    \left(
\hat{Z}_{j_1,\vec{x}}\hat{Z}_{j_2,\vec{x}+\hat{j}_1}\cdots\hat{Z}_{j_l,\vec{x}-\hat{j}_l}
\right)_{kl}\, , 
\end{align}
and
\begin{align}
&\frac{1}{N}
\sum_{n=1}^N
\hat{\mathcal{Z}}^{(n)}_{j_1,\vec{x}}\hat{\mathcal{Z}}^{(n)}_{j_2,\vec{x}+\hat{j}_1}\cdots\hat{\mathcal{Z}}^{(n)}_{j_l,\vec{x}-\hat{j}_l}
=
\sum_{k=1}^N
    \ket{k}_{\vec{x}}
    \bra{k}_{\vec{x}}     
    \left(
\hat{Z}_{j_1,\vec{x}}\hat{Z}_{j_2,\vec{x}+\hat{j}_1}\cdots\hat{Z}_{j_l,\vec{x}-\hat{j}_l}
\right)_{kk}\, . 
\end{align}
Applying a Fourier transform to the ancillary states and choose the zero-momentum sector, we obtain $\mathrm{Tr}\left(
\hat{Z}_{j_1,\vec{x}}\hat{Z}_{j_2,\vec{x}+\hat{j}_1}\cdots\hat{Z}_{j_l,\vec{x}-\hat{j}_l}
\right)$. 

\section{Supplementary material on quantum simulations with singlet constraint}\label{Supplementary_Material:singlet_simulation}

\subsection{Detailed derivation}
We will provide some details of the computations discussed in Sec.~\ref{sec:singlet_simulation}.

We use $Z_{j,\vec{n}}^{(\Omega)}\equiv \Omega_{\vec{n}}Z_{j,\vec{n}}\Omega^{-1}_{\vec{n}+\hat{j}}$ to denote the gauge transformation of $Z_{j,\vec{n}}$ by $\Omega\in\mathrm{SU}(N)$. A gauge singlet state can be written as
\begin{align}
    \ket{F}\equiv\int\mathrm{d}Z\, F(Z)\ket{Z}
\end{align}
with 
\begin{align}
    F(Z^{(\Omega)}) = F(Z)\, . 
\end{align}
We want to obtain such a state from  
\begin{align}
    \ket{f}\equiv\int\mathrm{d}Z\, f(Z)\ket{Z}
\end{align}
which is not necessarily a gauge singlet. 

For this purpose, we introduce a complex matrix $\xi_{\vec{n}}\in\mathbb{C}^{N\times N}$ and ancilla states $\ket{\xi}=\otimes_{\vec{n}}\ket{\xi_{\vec{n}}}$. These ancilla states can be truncated straightforwardly, as we did for the complex link variables. The operator $\hat{\xi}_{\vec{n}}$ can be defined as a sum of Pauli Z gates, as \eqref{eq:x_with_Z}. We prepare a state
\begin{align}
\ket{\mathcal{G}}=\int \mathrm{d}\xi\varphi(\xi)\ket{\xi}
\end{align}
in such a way that $\ket{\mathcal{G}}$ is a linear combination of $\xi_{\vec{n}}\in\mathrm{SU}(N)$ with Haar measure. 
Equivalently, $\int \varphi(\xi)^2\mathrm{d}\xi...$ is the integral over gauge transformations with the Haar measure. 
Then, by using 
\begin{align}
    \hat{Z}^{(\xi)}_{j,\vec{n}}
    \equiv
    \hat{\xi}_{\vec{n}}
    \hat{Z}_{j,\vec{n}}
    \hat{\xi}^\dagger_{\vec{n}+\hat{j}}\, , 
\end{align}
we define an operator $\hat{\mathcal{T}}^{\prime(\xi)}$ that transforms $\ket{Z}$ to $\ket{Z^{(\xi)}}$, where $Z^{(\xi)}=\xi Z\xi^\dagger$, as 
\begin{align}
\hat{\mathcal{T}}^{\prime(\xi)}
\equiv
\prod_{j,\vec{n}}
\hat{\mathcal{T}}^{\prime(\xi)}_{j,\vec{n}}
&\equiv
    \coordorder{
    \exp\left(
\mathrm{i}\sum_{j,\vec{n}}\mathrm{Tr}\left[\hat{P}_{j,\vec{n}}(\hat{\bar{Z}}_{j,\vec{n}}-\hat{\bar{Z}}^{(\xi)}_{j,\vec{n}})
    +
    \hat{\bar{P}}_{j,\vec{n}}(\hat{Z}_{j,\vec{n}}-\hat{Z}^{(\xi)}_{j,\vec{n}})\right]
    \right)}
    \nonumber\\
    &=
    \coordorder{\exp\left(\mathrm{Tr}(\mathrm{i}\bar{P}_{j,\vec{n}}(Z_{j,\vec{n}}-Z_{j,\vec{n}}^{(\xi)}))\right)}
\times
\coordorder{\exp\left(\mathrm{Tr}(\mathrm{i}P_{j,\vec{n}}(\bar{Z}_{j,\vec{n}}-\bar{Z}_{j,\vec{n}}^{(\xi)}))\right)}\, , 
    \label{eq:singlet_projection_T_operator}
\end{align}
where $\coordorder{\ }$ stands for the ordering convention, ``momentum operators come to the left, coordinate operators come to the right". We used this symbol because this is analogous to the normal ordering $:\ :$ (creation operators come to the left, annihilation operators come to the right). 
Indeed, we can see that 
\begin{align}
\bra{P}\hat{\mathcal{T}}^{\prime(\xi)}
\ket{Z}\ket{\xi}
&=
\bra{P}\ket{Z}\cdot\ket{\xi}
\, \cdot
\exp\left(
    \mathrm{i}\sum_{j,\vec{n}}\mathrm{Tr}
    \left[
    P_{j,\vec{n}}(\bar{Z}_{j,\vec{n}}-\bar{Z}^{(\xi)}_{j,\vec{n}})
    +
    \bar{P}_{j,\vec{n}}(Z_{j,\vec{n}}-Z^{(\xi)}_{j,\vec{n}})\right]
    \right)
    \nonumber\\
&=\ket{\xi}\cdot\exp\left(
    -\mathrm{i}\sum_{j,\vec{n}}\mathrm{Tr}
    \left[
    P_{j,\vec{n}}\bar{Z}^{(\xi)}_{j,\vec{n}}
    +
    \bar{P}_{j,\vec{n}}Z^{(\xi)}_{j,\vec{n}}\right]
    \right)
    \nonumber\\    
&
=
\bra{P}\ket{Z^{(\xi)}}\cdot\ket{\xi}\    
\end{align}
    
\begin{align}
    \hat{\mathcal{T}}^{\prime(\xi)}
    \ket{Z}\ket{\xi}
    &=
    \int \mathrm{d}P\ket{P}\bra{P}\hat{\mathcal{T}}^{(\xi)}
    \ket{Z}\ket{\xi}
    \nonumber\\
    &=    
    \int \mathrm{d}P\ket{P}\bra{P}\ket{Z^{(\xi)}}\ket{\xi}
    \nonumber\\
    &=    
    \ket{Z^{(\xi)}}\ket{\xi}\, .  
\end{align}
By using this operator, we have 
\begin{align}
\hat{\mathcal{T}}^{\prime(\xi)}\ket{Z}\ket{\mathcal{G}}
=
\int\mathrm{d}\xi\varphi(\xi)\ket{Z^{(\xi)}}\ket{\xi}
\end{align}
and
\begin{align}
\bra{\mathcal{G}}
\hat{\mathcal{T}}^{\prime(\xi)}
    \ket{Z}\ket{\mathcal{G}}
    &=
    \int\mathrm{d}\xi\varphi(\xi)^2\ket{Z^{(\xi)}}
    \nonumber\\
    &=
    \int\mathrm{d}\Omega\ket{Z^{(\Omega)}}\, .
\end{align}
Therefore, 
\begin{align}
\bra{\mathcal{G}}
\hat{\mathcal{T}}^{\prime(\xi)}
    \ket{f}\ket{\mathcal{G}}
    &=
    \int\mathrm{d}\Omega\int\mathrm{d}Z\,  f(Z)\ket{Z^{(\Omega)}}
    \nonumber\\
    &=
    \int\mathrm{d}Z\int\mathrm{d}\Omega\, f(Z^{(\Omega^{-1})})\ket{Z}
    \nonumber\\
    &\equiv
    \int\mathrm{d}Z\, F(Z)\ket{Z}\, .   
\end{align}

Now we discuss the implementation of $\hat{\mathcal{T}}^{\prime(\xi)}_{j,\vec{n}}$ --- which turns out to be hard! The simplest construction is to approximate the exponential by a truncated Taylor series, namely $e^x\simeq\sum_{j=0}^n\frac{x^n}{n!}$. Then, we have a sum of products of $\hat{P}$, $\hat{Z}$, and $\hat{\xi}$. For the $n$-th order term, we need 
\begin{align}
\left(
\prod_{\ell=1}^n\hat{\bar{P}}_{a_\ell b_\ell}
\right)
\left(
\prod_{\ell=1}^n(\hat{Z}-\hat{Z}^{(\xi)})_{b_\ell a_\ell}
\right)\, . 
\end{align}
Fourier transform is needed only twice, before and after applying $\hat{\bar{P}}$. A potential issue is that the number of terms grows quickly with the truncation level $n$. 

To avoid this issue, we can use the following trick. Firstly, we create $\int\mathrm{d}Z\ket{Z}_1\ket{Z}_2\ket{\xi}$ from $\int\mathrm{d}Z\ket{Z}_1\ket{\xi}$. Here, we used a subscript to distinguish two copies. This straightforward because $\ket{Z}$ is binary; we just tensor $\ket{0,0,\cdots,0}_2$ and act CNOT gates. Now, we define a new operator:
\begin{align}
\hat{\mathcal{T}}^{(\xi)}
\equiv
    \exp\left(
\mathrm{i}\mathrm{Tr}\left[\hat{P}_1(\hat{\bar{Z}}_2-\hat{\bar{Z}}_2^{(\xi)})
    +
    \hat{\bar{P}}_1(\hat{Z}_2-\hat{Z}^{(\xi)}_2)\right]
    \right)\, . 
    \label{eq:singlet_projection_T_prime_operator}
\end{align}
Here, $\hat{P}_1$ and $\hat{\bar{P}}_1$ acts on $\ket{Z}_1$, $\hat{Z}_2$ and $\hat{\bar{Z}}_2$ acts on $\ket{Z}_2$, and $\hat{Z}_2^{(\xi)}$ is made of $\hat{Z}_2$ and $\hat{\xi}$. 
All operators inside the exponential commute with each other, and hence we do not need the ordering symbol $ \coordorder{\ }$. 
Then, we can see that   
\begin{align}
    \hat{\mathcal{T}}^{(\xi)}
    \ket{Z}_1\ket{Z}_2\ket{\xi}
&=
\int\mathrm{d}P\ket{P}_1\bra{P}_1
 \hat{\mathcal{T}}^{(\xi)}
    \ket{Z}_1\ket{Z}_2\ket{\xi}
    \nonumber\\
    &=
    \int\mathrm{d}P
    \exp\left(
    \mathrm{i}\mathrm{Tr}
    \left[
    P(\bar{Z}-\bar{Z}^{(\xi)})
    +
    \bar{P}(Z-Z^{(\xi)})\right]
    \right)
    \ket{P}_1\bra{P}_1\ket{Z}_1
    \ket{Z}_2\ket{\xi}
    \nonumber\\
    &=
      \int\mathrm{d}P
    \exp\left(
    -\mathrm{i}\mathrm{Tr}
    \left[
    P\bar{Z}^{(\xi)}
    +
    \bar{P}Z^{(\xi)}\right]
    \right)
    \ket{P}_1
    \ket{Z}_2\ket{\xi}
    \nonumber\\
    &= \ket{Z^{(\xi)}}_1\ket{Z}_2\ket{\xi}. 
\end{align}
The ancilla state $\ket{Z}_2$ is eliminated using a protocol we discussed in the main text. 
\subsubsection{Construction of $\ket{\mathcal{G}}$}
To construct this projection explicitly, we introduce a complex matrix $\xi_{\vec{n}}\in\mathbb{C}^{N\times N}$ and ancilla states $\ket{\xi}=\otimes_{\vec{n}}\ket{\xi_{\vec{n}}}$, and prepare a state $\ket{\mathcal{G}}=\int \mathrm{d}\xi\varphi(\xi)\ket{\xi}$ in such a way that $\ket{\mathcal{G}}$ is a linear combination of $\xi_{\vec{n}}\in\mathrm{SU}(N)$ with Haar measure. 
Equivalently, $\varphi(\xi)$ is SU($N$)-invariant and vanishes quickly as $\xi$ departs from $\mathrm{SU}(N)\in\mathbb{C}^{N\times N}$. 

There can be many ways to build $\ket{\mathcal{G}}$. One way is to multiply $\exp(-C\, \mathrm{Tr}(\hat{\xi}\hat{\bar{\xi}}-\textbf{1})^2)$ and $\exp(-C'|\det\hat{\xi}-1|^2)$ to a state proportional to $\sum_\xi\ket{\xi}$, where $C$ and $C'$ are large positive numbers. This is straightforward by using the Linear Combination of Unitaries (LCU; see Supplementary Material~\ref{Sec:algorithms}). Specifically, we can use relations like
\begin{align}
    \exp(-C\mathrm{Tr}(\hat{\xi}\hat{\bar{\xi}}-\textbf{1})^2)
    =
    \lim_{n\to\infty}
    \left(
    1-\frac{1}{n}\cdot C\mathrm{Tr}(\hat{\xi}\hat{\bar{\xi}}-\textbf{1})^2
    \right)^n\, . 
\end{align}
The multiplication of $1-\frac{1}{n}\cdot C\mathrm{Tr}(\hat{\xi}\hat{\bar{\xi}}-\textbf{1})^2$ via LCU is straightforward because $\hat{\xi}$ and $\hat{\bar{\xi}}$ have simple Pauli-Z decompositions, and we can simply repeat it $n$ times. The multiplication of $\exp(-C'|\det\hat{\xi}-1|^2)$ can be realized in a similar manner. 
\subsection{Numerical demonstration for U(1)}\label{sec:projection_numerical_test_U1}
At a finite truncation level, we need to use 
\begin{align}
&
  \ket{Z^{(\xi)}}_{\rm truncated}
   \equiv
    \frac{1}{\sqrt{\mathrm{dim}\mathcal{H}}}\sum_P
    \exp\left(
    -\mathrm{i}\sum_{j,\vec{n}}\mathrm{Tr}
    \left[
    P_{j,\vec{n}}\bar{Z}^{(\xi)}_{j,\vec{n}}
    +
    \bar{P}_{j,\vec{n}}Z^{(\xi)}_{j,\vec{n}}\right]
    \right)
    \ket{P}\, . 
\end{align}
An important question is if 
\begin{align}
    \ket{Z}_{\rm proj.}
    \equiv
    \sum_\xi\varphi(\xi)    \ket{Z^{(\xi)}}_{\rm truncated}
\end{align}
is ``close to an SU($N$)-singlet". We used a quotation symbol because the SU($N$) symmetry is explicitly broken by the truncation. A reasonable measure of the truncation effect is given by 
\begin{align}
    \bra{Z}_{\rm proj.}
    \hat{G}^2_{\rm truncated}
    \ket{Z}_{\rm proj.}
\end{align}
where $\hat{G}_{\rm truncated}$ is the truncated version of the SU($N$) generators. 

Numerical evaluation of this quantity requires a large cost. As a tractable case, let us consider U$(1)$ embedded into $\mathbb{C}=\mathbb{R}^2$, and just one link.   
The Hilbert space is a truncated version of $\mathrm{Span}\{\ket{z}\, |\, z\in\mathbb{C}\}$. We take $z=(x_1+\mathrm{i}x_2)/\sqrt{2}$, and restrict $x_1$ and $x_2$ to $\pm\delta_x/2,\pm 3\delta_x/2,\cdots,(\Lambda-1)\delta_x/2$. The dimension of the truncated Hilbert space is $\Lambda^2$. In the same way, we introduce $\xi\in\mathbb{C}$, so that $z$ is transformed to $z^{(\xi)}\equiv\xi z$. 

The U(1) generator is
\begin{align}
\hat{G}
=
\hat{x}\hat{p}_y-\hat{y}\hat{p}_x\, . 
\end{align} 
We choose $\varphi(\xi)$ in such a way that $\sum_\xi\varphi(\xi)\ket{\xi}$ gives the ground state of the Hamiltonian similar to \eqref{n-sphere-model}, specifically
\begin{align}
\hat{H}_\xi
=
\frac{1}{2}\hat{p}_{\xi^{\rm (R)}}^2
+
\frac{1}{2}\hat{p}_{\xi^{\rm (I)}}^2
+
\frac{m^2}{8}\left((\hat{\xi}^{\rm (R)})^2+(\hat{\xi}^{\rm (I)})^2-1\right)^2\, , 
\end{align}
where $\xi=\xi^{\rm (R)}+\mathrm{i}\xi^{\rm (I)}$, $[\hat{\xi}^{\rm (R)},\hat{p}_{\xi^{\rm (R)}}]=[\hat{\xi}^{\rm (I)},\hat{p}_{\xi^{\rm (I)}}]=\mathrm{i}$,
with $m=40$, $R=2$ and $\delta_x=2R/\Lambda=4/\Lambda$.
We truncate $\xi=\xi^{\rm (R)}$ and $\xi^{\rm (I)}$ such that they take the values $\pm\delta_x/2,\pm 3\delta_x/2,\cdots,(\Lambda-1)\delta_x/2$. We take $\Lambda$ to be even, and take $x=\frac{\delta_x}{2}\left(\frac{\Lambda}{2}-1\right)=1-\frac{2}{\Lambda}$ and $y=0$. Then, 
\begin{align}
\ket{x^{(\xi)},y^{(\xi)}}_{\rm truncated}
    &\equiv
    \frac{1}{\Lambda}\sum_{p_x,p_y}
    \exp\left(
    -\mathrm{i}
    \left(
p_x x^{(\xi)}
+
p_y y^{(\xi)}
\right)
    \right)
    \ket{p_x,p_y}
    \nonumber\\
    &=
    \frac{1}{\Lambda}\sum_{p_x,p_y}
    \exp\left(
    -\mathrm{i}
    \left(
1-\frac{2}{\Lambda}
    \right)
    \left[
p_x\xi^{\rm (R)}
+
p_y\xi^{\rm (I)}\right]
    \right)
    \ket{p_x,p_y}\, ,
\end{align}
\begin{align}
\big\langle x,y\big|\hat{G}\big|
x^{(\xi)},y^{(\xi)}\big\rangle_{\rm truncated}
&=
    \frac{1}{\Lambda}\sum_{p_x,p_y}
    \exp\left(
    -\mathrm{i}
    \left(
1-\frac{2}{\Lambda}
    \right)
    \left[
p_x\xi^{\rm (R)}
+
p_y\xi^{\rm (I)}\right]
    \right)
\times
(xp_y-yp_x)
    \exp\left(
    \mathrm{i}(xp_x+yp_y)
    \right)\, , 
\end{align}
\begin{align}
\bra{x,y}\hat{G}\ket{z}_{\rm proj.}
&=
\sum_\xi\varphi(\xi)\cdot
    \frac{1}{\Lambda}\sum_{p_x,p_y}
    \exp\left(
    -\mathrm{i}
    \left(
1-\frac{2}{\Lambda}
    \right)
    \left[
p_x\xi^{\rm (R)}
+
p_y\xi^{\rm (I)}\right]
    \right)
\times
(xp_y-yp_x)
    \exp\left(
    \mathrm{i}(xp_x+yp_y)
    \right)\, , 
\end{align}
and 
\begin{align}
    \bra{z}_{\rm proj.}\hat{G}^2\ket{z}_{\rm proj.}
    =
\sum_{x,y}\Big|\bra{x,y}\hat{G}\ket{z}_{\rm proj.}\Big|^2
\end{align}
Using these expressions, we can compute $\bra{z}_{\rm proj.}\hat{G}^2\ket{z}_{\rm proj.}$ with a cost of order $\Lambda^6$. 

The result is shown in Fig.~\ref{fig:G2_convergence} in the main article. We can see that, as $\Lambda$ increases, $\bra{z}_{\rm proj.}\hat{G}^2\ket{z}_{\rm proj.}$ quickly approaches zero. 

\subsection{U(1) and $\mathbb{Z}_N$ gauge theory without embedding into $\mathbb{R}^2$}\label{sec:U(1)-and-Z_N}
In this section, we show how U(1) and $\mathbb{Z}_N$ ($N=2^Q$) can be realized without embedding into $\mathbb{R}^2$, although such a method does not generalize to SU($N$). We introduce arguments $\theta\in\mathbb{R}$ and $\varphi\in\mathbb{R}$ of $z\in\mathbb{C}$ and $\xi\in\mathbb{C}$:
\begin{align}
z
=
e^{\mathrm{i}\theta}\, , 
\qquad
\xi
=
e^{\mathrm{i}\varphi}\, . 
\end{align}
Instead of $z$ and $\xi$, we use $\theta$ and $\varphi$ as implemented on quantum computer. 
Then, the multiplication of $\xi$ on $z$ is the shift of $\theta$ by $\varphi$. The operator $\hat{\mathcal{T}}(\varphi)$ becomes simplified to 
\begin{equation}
\hat{\mathcal{T}}(\varphi)
=
\prod_{j,\vec{n}}
\exp\left(
\mathrm{i}(\hat{\varphi}_{\vec{n}}-\hat{\varphi}_{\vec{n}+\hat{j}})\hat{p}_{\theta;j,\vec{n}}
\right)\, .  
\end{equation}
For $N=2^Q$, we can use a small modification of \eqref{eq:x_with_Z} to implement $\hat{\theta}$ and $\hat{\xi}$. Specifically, 
\begin{align}
-
\frac{2\pi}{2^Q}\cdot
\left(
\frac{1}{2}
+
\frac{\sigma_{z,1}}{2}
+
2\cdot\frac{\sigma_{z,2}}{2}
+
\cdots
+
2^{Q-1}\cdot\frac{\sigma_{z,Q}}{2}
\right)\, . 
\end{align}
The conjugate momentum operators are defined similarly. 

\section{$\hat{E}_j^{(L)}$ and $\hat{E}_j^{(R)}$}\label{sec:KS-EL-vs-ER}
In Appendix~\ref{sec:Kogut_Susskind}, we used the electric field defined from the infinitesimal transformation acting from the left of the link variable. We could also use the one acting from the right. Let us compare these two options. Explicitly, they are defined by   
\begin{eqnarray}
\left[
\hat{E}^{(L)\alpha}_{j,\vec{n}},
\hat{U}_{k,\vec{n}'}^{pq}
\right]
=
a^{-2}g\delta_{jk}\delta_{\vec{n}\vec{n}'}(\tau_\alpha\hat{U}_{k,\vec{n}'})^{pq}\, 
\end{eqnarray}
and 
\begin{eqnarray}
\left[
\hat{E}^{(R)\alpha}_{j,\vec{n}},
\hat{U}_{k,\vec{n}'}^{pq}
\right]
=
a^{-2}g\delta_{jk}\delta_{\vec{n}\vec{n}'}(\hat{U}_{k,\vec{n}'}\tau_\alpha)^{pq}\, . 
\end{eqnarray}
$\hat{E}^{(L)}$ is the electric field we used so far. We can see that $\sum_\alpha(\hat{E}^{(L)\alpha})^2$ and $\sum_\alpha(\hat{E}^{(R)\alpha})^2$ acts on the string states in exactly the same way. To check it, we only have to check two cases: (i) when they act on one link, and (ii) when they act on two links. For the case (i), we see the agreement immediately, because
\begin{align}
\sum_\alpha\left[
\hat{E}_{j,\vec{n}}^{(L)\alpha},\left[\hat{E}_{j,\vec{n}}^{(L)\alpha},\hat{U}^{pq}_{j,\hat{n}}\right]
\right]
=
(a^{-2}g)^2\sum_\alpha\left(\tau^\alpha\tau^\alpha\hat{U}_{j,\hat{n}}\right)^{pq}
=
a^{-4}g^2N\hat{U}_{j,\hat{n}}^{pq}
\end{align}
and 
\begin{align}
\sum_\alpha\left[
\hat{E}_{j,\vec{n}}^{(R)\alpha},\left[\hat{E}_{j,\vec{n}}^{(R)\alpha},\hat{U}^{pq}_{j,\hat{n}}\right]
\right]
=
(a^{-2}g)^2\sum_\alpha\left(\hat{U}_{j,\hat{n}}\tau^\alpha\tau^\alpha\right)^{pq}
=
a^{-4}g^2N\hat{U}_{j,\hat{n}}^{pq}\, . 
\end{align}
To check (ii), let us consider a loop in which $\hat{U}_{j,\vec{n}}$ appears twice. Then, 
\begin{align}
\lefteqn{
\sum_\alpha\left[
\hat{E}_{j,\vec{n}}^{(L)\alpha},\left[\hat{E}_{j,\vec{n}}^{(L)\alpha},\hat{V}_1^{pq}\hat{U}_{j,\vec{n}}^{qr}\hat{V}_2^{rs}\hat{U}^{sp}_{j,\vec{n}}\right]
\right]
}
\nonumber\\
&=
2a^{-4}g^2N\hat{V}_1^{pq}\hat{U}_{j,\vec{n}}^{qr}\hat{V}_2^{rs}\hat{U}^{sp}_{j,\vec{n}}
+
2\sum_\alpha\hat{V}^{pq}_1[\hat{E}_{j,\vec{n}}^{(L)\alpha},\hat{U}^{qr}_{j,\vec{n}}]\hat{V}^{rs}_2[\hat{E}_{j,\vec{n}}^{(L)\alpha},\hat{U}^{sp}_{j,\vec{n}}]
\nonumber\\
&=
2a^{-4}g^2N\hat{V}_1^{pq}\hat{U}_{j,\vec{n}}^{qr}\hat{V}_2^{rs}\hat{U}^{sp}_{j,\vec{n}}
+
2a^{-4}g^2\sum_\alpha\hat{V}^{pq}_1(\tau^\alpha\hat{U}_{j,\vec{n}})^{qr}\hat{V}^{rs}_2(\tau^\alpha\hat{U}_{j,\vec{n}})^{sp}
\nonumber\\
&=
2a^{-4}g^2
\left(
N\hat{V}_1^{pq}\hat{U}_{j,\vec{n}}^{qr}\hat{V}_2^{rs}\hat{U}^{sp}_{j,\vec{n}}
+
\hat{V}^{pq}_1\hat{U}^{qp}_{j,\vec{n}}\hat{V}^{rs}_2\hat{U}_{j,\vec{n}}^{sr}
\right)
\end{align}
and
\begin{align}
\lefteqn{
\sum_\alpha\left[
\hat{E}_{j,\vec{n}}^{(R)\alpha},\left[\hat{E}_{j,\vec{n}}^{(R)\alpha},\hat{V}_1^{pq}\hat{U}_{j,\vec{n}}^{qr}\hat{V}_2^{rs}\hat{U}^{sp}_{j,\vec{n}}\right]
\right]
}
\nonumber\\
&=
2a^{-4}g^2N\hat{V}_1^{pq}\hat{U}_{j,\vec{n}}^{qr}\hat{V}_2^{rs}\hat{U}^{sp}_{j,\vec{n}}
+
2\sum_\alpha\hat{V}^{pq}_1[\hat{E}_{j,\vec{n}}^{(R)\alpha},\hat{U}^{qr}_{j,\vec{n}}]\hat{V}^{rs}_2[\hat{E}_{j,\vec{n}}^{(R)\alpha},\hat{U}^{sp}_{j,\vec{n}}]
\nonumber\\
&=
2a^{-4}g^2N\hat{V}_1^{pq}\hat{U}_{j,\vec{n}}^{qr}\hat{V}_2^{rs}\hat{U}^{sp}_{j,\vec{n}}
+
2a^{-4}g^2\sum_\alpha\hat{V}^{pq}_1(\hat{U}_{j,\vec{n}}\tau^\alpha)^{qr}\hat{V}^{rs}_2(\hat{U}_{j,\vec{n}}\tau^\alpha)^{sp}
\nonumber\\
&=
2a^{-4}g^2
\left(
N\hat{V}_1^{pq}\hat{U}_{j,\vec{n}}^{qr}\hat{V}_2^{rs}\hat{U}^{sp}_{j,\vec{n}}
+
\hat{V}^{pq}_1\hat{U}^{qp}_{j,\vec{n}}\hat{V}^{rs}_2\hat{U}_{j,\vec{n}}^{sr}
\right)\, . 
\end{align}

\bibliography{ref}